\documentclass[epj]{svjour}
\usepackage{graphics}
\usepackage{graphicx}
\usepackage{fancyhdr}
\usepackage{amsmath,amssymb}
\usepackage{epsfig,bbm}
\newcommand{\nc}{\newcommand}
\nc{\hef}{\ensuremath{^4\mathrm{He}}}
\nc{\het}{\ensuremath{^3\mathrm{He}}}
\nc{\lisx}{\ensuremath{^6\mathrm{Li}}}
\nc{\lisv}{\ensuremath{^7\mathrm{Li}}}
\nc{\bes}{\ensuremath{^7\mathrm{Be}}}
\nc{\beet}{\ensuremath{^8\mathrm{Be}}}
\nc{\ben}{\ensuremath{^9\mathrm{Be}}}
\nc{\dm}{{\rm D}}
\nc{\hefm}{{\rm ^4He}}
\nc{\hetm}{{\rm ^3He}}
\nc{\lisxm}{{\rm ^6Li}}
\nc{\lisvm}{{\rm ^7Li}}
\nc{\besm}{{\rm ^7Be}}
\nc{\beetm}{{\rm ^8Be}}
\nc{\benm}{{\rm ^9Be}}
\nc{\bs}{(N$X^-$)}
\nc{\xm}{$X^-$}
\nc{\xp}{$X^+$}
\nc{\xz}{$X^0$}
\nc{\bex}{(\bes\xm)}
\nc{\bexm}{(\besm X^-)}
\nc{\px}{($p$\xm)}
\nc{\Ox}{\ensuremath{\mathrm{O}}}
\nc{\Fe}{\ensuremath{\mathrm{Fe}}}
\nc{\Hyd}{\ensuremath{\mathrm{H}}}
\nc{\Be}{\ensuremath{\mathrm{Be}}}
\nc{\tauX}{\ensuremath{\tau_{X^-}}}
\nc{\YX}{\ensuremath{Y_{X^-}}}
\nc{\YXdec}{\ensuremath{Y^{\mathrm{dec}}_{X^-}}}
\nc{\Ysldec}{\ensuremath{Y^{\mathrm{dec}}_{{\widetilde{l}_1}}}}
\nc{\Ysldecm}{\ensuremath{Y^{\mathrm{dec}}_{{\widetilde{l}_1^-}}}}
\newcommand{\Order}{{\cal O}}   
\newcommand{\Arg}{\mbox{Arg}}
\newcommand{\meV}{\mathrm{meV}}
\newcommand{\eV}{\mathrm{eV}}
\newcommand{\keV}{\mathrm{keV}}
\newcommand{\MeV}{\mathrm{MeV}}
\newcommand{\GeV}{\mathrm{GeV}}
\newcommand{\TeV}{\mathrm{TeV}}
\newcommand{\Mpc}{\mathrm{Mpc}}
\newcommand{\km}{\mathrm{km}}
\newcommand{\cm}{\mathrm{cm}}
\newcommand{\FS}{\mathrm{FS}}
\newcommand{\seconds}{\mathrm{s}}

\newcommand{\MPl}{\mathrm{M}_{\mathrm{P}}}
\newcommand{\proton}{\mathrm{p}}
\newcommand{\antiproton}{\bar{\mathrm{p}}}
\newcommand{\MET}{E_{T}^{\mathrm{miss}}}
\newcommand{\gravitino}{{\widetilde{G}}}
\newcommand{\axino}{{\widetilde{a}}}
\newcommand{\ax}{\ensuremath{\widetilde{a}}}
\newcommand{\slepton}{{\tilde{l}_1}}
\newcommand{\stau}{{\widetilde{\tau}_1}}
\newcommand{\sleptonR}{\ensuremath{{\tilde{l}_{\mathrm{R}}}}}
\newcommand{\stauR}{\ensuremath{{\widetilde{\tau}_{\mathrm{R}}}}}

\newcommand{\sel}{{\widetilde{e}_1}}
\newcommand{\smu}{{\widetilde{\mu}_1}}
\newcommand{\scalartop}{{\widetilde{t}_1}}
\newcommand{\st}{{\tilde{\tau}_1}}
\newcommand{\bino}{{\widetilde B}}
\newcommand{\Bino}{{\widetilde B}}
\newcommand{\Wino}{{\widetilde W}}
\newcommand{\wino}{{\widetilde W}}
\newcommand{\gluino}{{\widetilde g}}
\newcommand{\neutralino}{{\widetilde \chi}^{0}_{1}}
\newcommand{\sneutrino}{{\widetilde \nu}_{1}}
\newcommand{\chargino}{{\widetilde{\chi}_1^{\pm}}}
\newcommand{\HiggsinoUp}{{\widetilde H}^{0}_{u}}
\newcommand{\HiggsinoDown}{{\widetilde H}^{0}_{d}}
\newcommand{\Higgs}{\mathrm{H}}
\newcommand{\hhiggs}{\ensuremath{{h}^0}}
\newcommand{\Hhiggs}{\ensuremath{{H}^0}}
\newcommand{\Ahiggs}{\ensuremath{{A}^0}}

\newcommand{\bquark}{\ensuremath{\mathrm{b}}}
\newcommand{\antibquark}{\ensuremath{\bar{\mathrm{b}}}}

\newcommand{\mh}{\ensuremath{m_{\hhiggs}}}

\newcommand{\mA}{\ensuremath{m_{\Ahiggs}}}

\newcommand{\mZ}{\ensuremath{M_{\mathrm{Z}}}}

\newcommand{\mgr}{m_{\widetilde{G}}}

\newcommand{\mgravitino}{\mgr}
\newcommand{\mst}{m_{\tilde{\tau}_1}}
\newcommand{\mslepton}{m_{\slepton}}
\newcommand{\mstop}{m_{\scalartop}}
\newcommand{\mneu}{m_{{\widetilde \chi}^{0}_{1}}}
\newcommand{\PQ}{\mathrm{PQ}}

\newcommand{\LSP}{\mathrm{LSP}}
\newcommand{\NLSP}{\mathrm{NLSP}}
\newcommand{\LOSP}{\mathrm{LOSP}}
\newcommand{\NTP}{\mathrm{NTP}}
\newcommand{\TP}{\mathrm{TP}}
\newcommand{\thermal}{\mathrm{therm}}
\newcommand{\equil}{\mathrm{eq}}

\newcommand{\freezeout}{\mathrm{f}}
\newcommand{\CDM}{\mathrm{dm}}

\newcommand{\EM}{\mathrm{em}}
\newcommand{\HAD}{\mathrm{had}}

\newcommand{\rms}{\mathrm{rms}}
\newcommand{\rhotot}{\rho_{\mathrm{tot}}}
\newcommand{\Omegatot}{\Omega_{\mathrm{tot}}}

\newcommand{\OmegaL}{\Omega_{\Lambda}}

\newcommand{\OmegaDM}{\Omega_{\mathrm{dm}}}
\newcommand{\OmegaB}{\Omega_{\mathrm{b}}}
\newcommand{\OmegaGamma}{\Omega_{\gamma}}
\newcommand{\OmegaNu}{\Omega_{\nu}}
\newcommand{\OmegaX}{\Omega_{X}}
\newcommand{\LQCD}{\Lambda_{\mathrm{QCD}}}

\newcommand{\dec}{\mathrm{dec}}
\newcommand{\GUT}{\mathrm{GUT}}

\newcommand{\Reheating}{\mathrm{R}}
\newcommand{\TR}{T_{\Reheating}}

\newcommand{\Tmin}{T_{\mathrm{min}}}
\newcommand{\Tmax}{T_{\mathrm{max}}}
\newcommand{\TRmax}{T_{\Reheating}^{\max}}
\newcommand{\Tosc}{T_{\mathrm{osc}}}
\newcommand{\TL}{T_{\mathrm{low}}}
\newcommand{\Color}{\mathrm{c}}
\newcommand{\Weak}{\mathrm{L}}
\newcommand{\Hypercharge}{\mathrm{Y}}
\newcommand{\Lisix}{{}^6 \mathrm{Li}}
\newcommand{\Hefour}{{}^4 \mathrm{He}}

\newcommand{\taustau}{\tau_{\widetilde{\tau}_1}}

\newcommand{\monetwo}{m_{1/2}}
\newcommand{\mzero}{m_{0}}
\newcommand{\tanb}{\tan{\beta}}
\newcommand{\mgut}{M_\mathrm{GUT}}
\newcommand{\Omegatp}{\Omega_{\widetilde{G}}^{\mathrm{TP}}}
\newcommand{\Omegantp}{\Omega_{\widetilde{G}}^{\mathrm{NTP}}}

\newcommand{\champ}{X^{\! -}}

\newcommand{\deuterium}{\mathrm{D}}
\newcommand{\inflaton}{\mathrm{I}}
\newcommand{\Hinf}{H_{\inflaton}}
\newcommand{\Einf}{E_{\inflaton}}
\newcommand{\epseff}{\epsilon_{\mathrm{eff}}}
\newcommand{\GN}{G_{\rm N}}
\newcommand{\Nnu}{N_{\nu}^{\mathrm{eff}}}
\newcommand{\Laxion}{L_a}
\newcommand{\gagg}{g_{a\gamma\gamma}}
\newcommand{\gaNN}{g_{aNN}}
\newcommand{\gaee}{g_{aee}}

\newcommand{\be}{\begin{equation}}
\newcommand{\ee}{\end{equation}}
\newcommand{\bea}{\begin{eqnarray}}
\newcommand{\eea}{\end{eqnarray}}
\newcommand{\benn}{\begin{displaymath}}
\newcommand{\eenn}{\end{displaymath}}
\newcommand{\beann}{\begin{eqnarray*}}
\newcommand{\eeann}{\end{eqnarray*}}
\begin{document}
\title{Dark Matter Candidates}
\subtitle{Axions, Neutralinos, Gravitinos, and Axinos}
\author{Frank Daniel Steffen\inst{1}
\thanks{\emph{Email:} steffen@mppmu.mpg.de}%
}                     
%
%
\institute{Max-Planck-Institut f\"ur Physik, F\"ohringer Ring 6,
  D-80805 Munich, Germany}
%
\date{\today}
%
%
\abstract{
  The existence of dark matter provides strong evidence for physics
  beyond the Standard Model. Extending the Standard Model with the
  Peccei-Quinn symmetry and/or supersymmetry, compelling dark matter
  candidates appear. For the axion, the neutralino, the gravitino, and
  the axino, I review primordial production mechanisms, cosmological
  and astrophysical constraints, experimental searches, and prospects
  for experimental identification.
} 
%
%
%
\maketitle
%

\section{Introduction}
\label{intro}

Based on numerous cosmological and astrophysical studies, we believe
today that our Universe is flat and thus that the total energy density
is critical, $\rhotot\simeq\rho_c=3H_0^2/8\pi\GN$ or
$\Omegatot\equiv\rhotot/\rho_c\simeq 100\%$, with contributions
of~\cite{Komatsu:2008hk,Amsler:2008zz}
\bea
\OmegaL\simeq 72\%, 
\,\,\,\,\,\,
\OmegaDM\simeq 23\%,
\,\,\,\,\,\,
\OmegaB\simeq 4.6\%,
\nonumber\\
\OmegaGamma\simeq 0.005\%,
\,\,\,\,\,\,
0.1\% \lesssim \OmegaNu \lesssim 1.5\%
\,\,\,\,\,\,
\label{Eq:Composition}
\eea
provided in the form of dark energy, non-baryonic dark matter,
baryons, photons, and neutrinos, respectively. Here
$\Omega_i\equiv\rho_i/\rho_c$,
$H_0$ is the present Hubble expansion rate, and $\GN$ is Newton's
constant.
The density parameter $\OmegaGamma$ is given basically by the photons
of the cosmic microwave background (CMB). The understanding of the
remaining $\Omegatot-\OmegaGamma\simeq 99.995\%$ requires physics
beyond the Standard Model:
\begin{itemize}
\item[(i)] The amount of dark energy $\OmegaL$ could be provided by a
  cosmological constant.
  However, the fact that the cosmological constant or vacuum energy
  inferred from a naive quantum field theoretical estimate exceeds
  $\OmegaL$ by 120 orders of magnitude is the most serious fine-tuning
  problem in physics.
  An alternative explanation of dark energy is a slowly evolving
  scalar field~\cite{Wetterich:1987fm,Ratra:1987rm,Caldwell:1997ii}
  that cannot be part of the Standard Model.
\item[(ii)] The amount of baryons $\OmegaB$---inferred from the CMB
  an\-iso\-tropies, the framework of primordial nucleosynthesis, and
  the observationally inferred abundances of light nuclei---indicates
  a matter-anti\-matter asymmetry that cannot be explained within the
  Standard Model.
\item[(iii)] The amount of neutrinos $\OmegaNu$ is understood to be
  provided in the form of the cosmic neutrino background (C$\nu$B)
  whose detection is a major challenge; cf.~\cite{Ringwald:2004np} and
  references therein. 
  While $\OmegaNu^{m_{\nu_i}=0}\simeq0.003\%$ for three massless
  neutrino species, we know from oscillation experiments that at least
  two neutrinos have small but non-zero masses, $\sum_i
  m_{\nu_i}\gtrsim 0.05~\eV$, which are not part of the Standard
  Model, such that $\OmegaNu\gtrsim 0.1\%$.
  Moreover, since light neutrinos are hot dark matter, which leaves an
  imprint on large scale structure (LSS)~\cite{Hannestad:2007dd}, one
  can extract the cosmological limit
  $\sum_i m_{\nu_i}\lesssim\Order(1~\eV)$
  or equivalently $\OmegaNu\lesssim 1.5\%$~\cite{Lesgourgues:2006nd}.
\item[(iv)] The dominant amount of the non-baryonic dark matter
  density $\OmegaDM$ must reside in one or more species with a
  negligible (thermal) velocity to allow for structure formation.
  Moreover, relying on the astrophysical and cosmological
  considerations that point to the existence of dark matter
  (see~\cite{Bergstrom:2000pn,Bertone:2004pz} and references therein),
  we think that a particle physics candidate for dark matter has to be
  electrically neutral, color neutral, and stable or have a lifetime
  $\tau_{\CDM}$ that is generally larger than the age of the Universe
  today, $t_0 \simeq 14~\mathrm{Gyr}$, i.e.,%
  \footnote{Studies of diffuse x-ray and $\gamma$-ray backgrounds, of
    the cosmic ionization history, and of the CMB point to radiative
    lifetimes of electromagnetically decaying dark matter that are
    orders of magnitude above
    $t_0$~\cite{Pierpaoli:2003rz,Kasuya:2003sm,Chen:2003gz,Zhang:2007zzh,Yuksel:2007dr}.
    Moreover, for dark matter with decays into weakly-interacting
    relativistic particles such as neutrinos, $\tau_{\CDM}\gg t_0$ is
    inferred from studies of LSS, of the
    CMB~\cite{Oguri:2003nn,Takahashi:2003iu,Ichiki:2004vi}, of type Ia
    supernovae (SN)~\cite{Gong:2008gi}, and of contraints on the
    cosmic neutrino flux~\cite{PalomaresRuiz:2007ry}.}
\begin{align}
  \tau_{\CDM}> t_0 \simeq 4.3\times 10^{17}\,\seconds
  \ .
  \label{Eq:lifetimeDM}
\end{align}
  With the standard active neutrinos being too light, such a dark
  matter candidate cannot be found within the Standard Model. Thus,
  one can consider the existence of dark matter as evidence for new
  physics.
\end{itemize}

In this review we focus on dark matter candidates that appear once the
Standard Model is extended with the Peccei--Quinn (PQ) symmetry and/or
supersymmetry (SUSY): the axion, the lightest neutralino, the
gravitino, and the axino.
These hypothetical particles are particularly well motivated:
\begin{itemize}
\item The PQ mechanism allows for an elegant solution of the strong CP
  problem~\cite{Peccei:1977hh,Peccei:1977ur}. In fact, an additional
  global U(1) symmetry---referred to as PQ symmetry---that is broken
  spontaneously at the
  PQ~scale~\cite{Sikivie:1999sy,Raffelt:2006cw,Amsler:2008zz}
$f_a \gtrsim 6 \times 10^8\,\GeV$
can explain the smallness (or vanishing) of the CP violating
$\Theta$-vacuum term in quantum chromodynamics (QCD).  The associated
pseudo-Nambu-Gold\-stone boson is the
axion~\cite{Weinberg:1977ma,Wilczek:1977pj} which becomes massive due
to QCD instanton effects. Indeed, the axion is electrically neutral,
color neutral, and sufficiently long-lived for being a compelling dark
matter candidate~\cite{Preskill:1982cy,Abbott:1982af,Dine:1982ah}.
\item SUSY extensions of the Standard Model are an appealing concept
  because of their remarkable properties, for example, with respect to
  gauge coupling unification, the hierarchy problem, and the embedding
  of
  gravity~\cite{Wess:1992cp,Nilles:1983ge,Haber:1984rc,Martin:1997ns,Drees:2004jm,Baer:2006rs}.
  As superpartners of the Standard Model particles, new particles
  appear including fields that are electrically neutral and color
  neutral.  Since they have not been detected at particle
  accelerators, these sparticles must be heavy or extremely weakly
  interacting.
  Moreover, because of the non-ob\-ser\-va\-tion of reactions that
  violate lepton number $L$ or baryon number $B$, it is often
  assumed---as also in this review---that SUSY theories respect the
  multiplicative quantum number
\begin{align}
  R=(-1)^{3B+L+2S} 
  \ ,
\end{align}
known as R-parity, with $S$ denoting the spin. Since Standard Model
particles and superpartners carry respectively even $(+1)$ and odd
$(-1)$ R-parity, its conservation implies that superpartners can only
be produced or annihilated in pairs and that the lightest
supersymmetric particle (LSP) cannot decay even if it is heavier than
most (or all) of the Standard Model particles.%
\footnote{While R-parity conservation is assumed in this review, its
  violation is a realistic option; see, e.g.,
  \cite{Dreiner:1997uz,Allanach:2007vi,Takayama:2000uz,Buchmuller:2007ui,Ibarra:2007jz}.}
An electrically neutral and color neutral LSP candidate---such as the
lightest neutralino, the gravitino, or the axino---can thus also be a
compelling dark matter candidate.
\end{itemize}

For each dark matter candidate $X$, it is crucial to calculate its
relic density $\OmegaX$ and to compare the result with the dark matter
density $\OmegaDM$, for which a nominal $3\sigma$ range can be
inferred from measurements of the CMB anisotropies by the Wilkinson
Micro\-wave An\-iso\-tropy Probe (WMAP)
satellite~\cite{Spergel:2006hy}
\begin{equation}
        \Omega_{\CDM}^{3\sigma}h^2=0.105^{+0.021}_{-0.030} 
\label{Eq:OmegaDM}
\end{equation} 
with $h=0.73^{+0.04}_{-0.03}$ denoting the Hubble constant in units of
$100~\km\,\Mpc^{-1}\seconds^{-1}$. Note that the nominal $3\sigma$
range is derived assuming a restrictive six-parameter ``vanilla''
mod\-el. A larger range is possible---even with additional data from
other cosmological probes---if the fit is performed in the context of
a more general model that includes other physically motivated
parameters such as non-zero neutrino masses~\cite{Hamann:2006pf}:
$0.094 < \OmegaDM h^2 < 0.136$.
Moreover, there are limits on the present free-streaming velocity
$v_0^X$ of the dark matter candidate $X$ from observations and
simulations of LSS. Indeed, as mentioned above, the dominant part of
$\OmegaDM$ has to have a negligible (thermal) velocity to allow for
structure formation.

Primordial nucleosynthesis---or big-bang nucleosynthesis (BBN)---is
another cosmological probe for the viability of dark matter scenarios
and, more generally, for the viability of models beyond the Standard
Model.
Relying on Standard Model physics, general relativity, and the
baryon-to-photon ratio inferred from the CMB an\-iso\-tro\-pies,
standard BBN (SBBN) predicts the primordial abundances of deuterium,
helium, and lithium in good overall agreement with the observationally
inferred values and provides thereby an important consistency check of
standard
cosmology~\cite{Malaney:1993ah,Sarkar:1995dd,Schramm:1997vs,Olive:1999ij,Serpico:2004gx,Iocco:2008va,Amsler:2008zz}.
This agreement can be affected by new physics, e.g., in the three
following ways.  First, there can be changes in the timing of the
nuclear reactions---caused, e.g., by a change in the Hubble expansion
rate because of an increase of the energy density during BBN
contributed by a new relativistic
species~\cite{Sarkar:1995dd,Mangano:2006ur,Simha:2008zj}.  Second,
non-thermal processes---caused, e.g., by the injection of energetic
Standard Model particles in late decays of heavier particles into an
extremely weakly interacting dark matter candidate---can reprocess the
produced light
nuclei~\cite{Lindley:1984bg,Ellis:1984er,Scherrer:1985zt,Reno:1987qw,Dimopoulos:1988ue,Levitan:1988au,Sigl:1995kk,Jedamzik:1999di,Cyburt:2002uv,Jedamzik:2004er,Kawasaki:2004qu,Jedamzik:2006xz,Kawasaki:2008qe}.
Third, long-lived electromagnetically or strongly interacting
relics---which can occur, e.g., in scenarios with an extremely weakly
interacting dark matter candidate---can form bound states with light
nuclei and thereby lead to a catalysis of nuclear reactions, i.e., to
catalyzed BBN
(CBBN)~\cite{Pospelov:2006sc,Kohri:2006cn,Kaplinghat:2006qr,Cyburt:2006uv,Hamaguchi:2007mp,Bird:2007ge,Kawasaki:2007xb,Takayama:2007du,Jittoh:2007fr,Jedamzik:2007cp,Pradler:2007is,Jedamzik:2007qk,Kusakabe:2007fu,Kusakabe:2007fv,Pospelov:2007js,Kawasaki:2008qe,Pospelov:2008ta}.
We will encounter the associated BBN constraints and other
cosmological and astrophysical constraints explicitly in the
discussions of the dark matter scenarios given below.

The viability of a dark matter scenario is tightly connected to the
early history of the Universe. 
As suggested by the flatness, isotropy, and homogeneity of the
Universe, we assume that its earliest moments were governed by
inflation~\cite{Linde:2005ht,Kolb:1990vq}. The inflationary expansion is
then followed by a phase in which the Universe is reheated.  The
reheating phase repopulates the Universe and provides the initial
conditions for the subsequent radiation-dominated epoch.  We refer to
the reheating temperature $T_{\Reheating}$ as the initial temperature
of this early radiation-dominated epoch of our Universe.
In fact, inflation models can point to $T_{\Reheating}$ well above
$10^{10}~\GeV$~\cite{Linde:2005ht,Kolb:1990vq,Linde:1991km}.
Here one should stress that BBN requires a minimum temperature
of~\cite{Kawasaki:1999na,Kawasaki:2000en,Hannestad:2004px,Ichikawa:2005vw}
\begin{equation}
   T\gtrsim 0.7\!-\!4~\MeV\equiv \Tmin
\label{Eq:Tafter_limit}
\end{equation}
and that any temperature above $\Tmin$ is still speculative since BBN
is currently the deepest reliable probe of the early Universe.
Interestingly, futuristic space-based gra\-vita\-tional-wave detectors
such as the Big Bang Observer (BBO) or the Deci-hertz Interferometer
Gra\-vi\-ta\-tional Wa\-ve Ob\-servatory (DECIGO)~\cite{Seto:2001qf}
could allow for tests of the thermal history before BBN and could even
probe the reheating temperature $\TR$ after
inflation~\cite{Nakayama:2008ip,Nakayama:2008wy}.

Let us now turn to the dark matter candidates one by one:
axions (Sect.~\ref{sec:AxionDM}), neutralinos
(Sect.~\ref{sec:NeutralinoDM}), gravitinos
(Sect.~\ref{sec:GravitinoDM}), and axinos (Sect.~\ref{sec:AxinoDM}).
For each of those candidates, I review primordial production
mechanisms, cosmological/astro\-physical constraints, and prospects of
experimental identification.

\section{Axion Dark Matter}
\label{sec:AxionDM}

In this section we consider the axion which is a promising dark matter
candidate that is not tied to SUSY but that can coexist with any SUSY
dark matter candidate. Let us start this section with a brief review
of the strong CP problem, the PQ mechanism, and a short description of
basic properties of the axion. More detailed discussions can be found,
e.g.,
in~\cite{Kolb:1990vq,Peccei:1998jt,Peccei:2006as,Raffelt:2006cw,Amsler:2008zz,Kim:2008hd}.

The isoscalar $\eta'$ meson being too heavy to qualify as a
pseudo-Nambu-Goldstone boson of a spontaneously broken axial
U(1)$_{\mathrm{A}}$ is the well-known U(1)$_{\mathrm{A}}$ problem of
the strong interactions. This problem has an elegant solution in QCD
based on non-trivial topological
properties~\cite{'tHooft:1976up,'tHooft:1976fv}.  The $\eta'$ meson
mass can be understood as a consequence of the Adler-Bell-Jackiw
anomaly~\cite{Bell:1969ts,Adler:1969gk} receiving contributions from
gauge-field configurations with non-zero topological charge such as
instantons~\cite{Belavin:1975fg}.  The existence of such
configurations implies the additional $\Theta$-vacuum term in the QCD
Lagrangian~\cite{'tHooft:1976up,'tHooft:1976fv,Bell:1969ts,Adler:1969gk}
\begin{align}
  {\cal L}_{\Theta} = \Theta\frac{g_s^2}{32\pi^2}
  G_{\mu\nu}^a\widetilde{G}^{a\mu\nu}
\label{Eq:Theta-Vacuum}
\end{align}
with the strong coupling $g_s$ and the gluon-field-strength tensor
$G_{\mu\nu}^a$ whose dual is given by
$\widetilde{G}^{a}_{\mu\nu}=\epsilon_{\mu\nu\rho\sigma}G^{a\:\!\rho\sigma}\!\!/2$.
The $\Theta$-vacuum term violates the discrete symmetries P, T, and CP
for any value of $\Theta \neq n\pi$ with $n \in \mathbbm{Z}$.  Such
violations have not been observed in strong interactions and
experiments on the electric dipole moment of the neutron give an upper
bound of $|\Theta|<10^{-9}$.  Within QCD, $\Theta=0$ seems natural
based on the observed conservation of those discrete symmetries.
However, once QCD is embedded in the Standard Model of strong and
electroweak interactions with CP violation being an experimental
reality, $\Theta$ and the argument of the determinant of the quark
mass matrix $M$---two {\em a priori} unrelated quantities---must
cancel to an accuracy of $10^{-9}$ according to the upper bound:%
\footnote{If one or more quarks are massless~\cite{Kaplan:1986ru},
  $\bar{\Theta}$ can be rotated away by a chiral rotation;
  cf.~\cite{Peccei:1998jt,Peccei:2006as}.}
\begin{align}
        |\bar{\Theta}| \equiv |\Theta+\Arg\det M|< 10^{-9}
        \ .
\label{Eq:StrongCPProblem}
\end{align}
This fine-tuning problem is the strong CP problem (cf.\
also~\cite{Peccei:1998jt,Peccei:2006as} and references therein).

The elegant solution of the strong CP problem suggested by Peccei and
Quinn~\cite{Peccei:1977hh,Peccei:1977ur} requires a new global chiral
U(1) symmetry---the PQ symmetry U(1)$_{\PQ}$---that is broken
spontaneously at the PQ scale $f_a$. The corresponding
pseudo-Nambu-Goldstone boson is the axion
$a$~\cite{Weinberg:1977ma,Wilczek:1977pj}, which couples to gluons
such that the chiral anomaly in the U(1)$_{\PQ}$ current is
reproduced,
\begin{align}
        {\cal L}_{a gg} 
        = \frac{a}{f_a/N}\,
        \frac{g_s^2}{32\pi^2}\,G_{\mu\nu}^a\widetilde{G}^{a\mu\nu}
        \ ,
\label{Eq:axion-gluon-gluon}
\end{align}
where the color anomaly $N$ of the PQ symmetry depends on the axion
model as discussed below. This interaction
term~(\ref{Eq:axion-gluon-gluon}) together with the vacuum
term~(\ref{Eq:Theta-Vacuum}) for $\Theta \to
\bar{\Theta}=\Theta+\Arg\det M$ provide the axion field with an
effective potential $V_{\rm eff}$ at low energies. This solves the
strong CP problem since the coefficient of the CP violating
$G\widetilde{G}$ term becomes dynamical and vanishes for the value
$\langle a \rangle = - \bar{\Theta}f_a/N$ at which $V_{\rm eff}$ has
its minimum.

\begin{figure}[t!]
  \centerline{\includegraphics[width=0.35\textwidth]{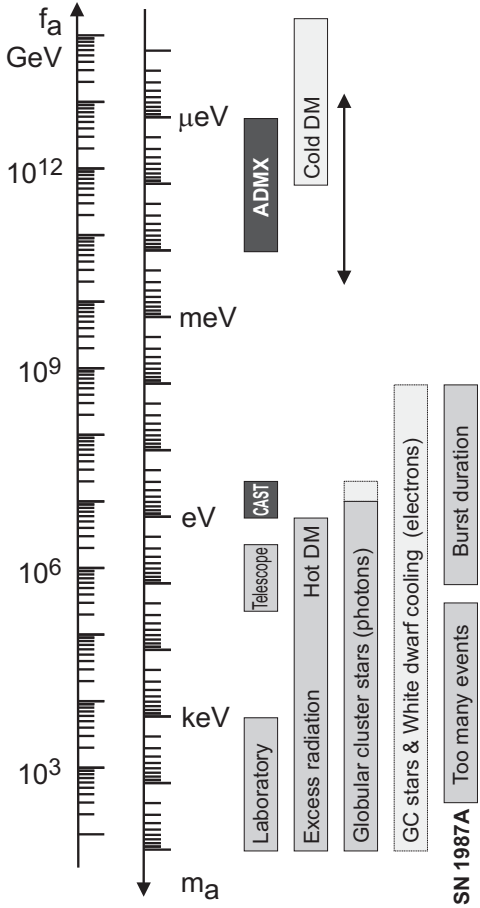}}
  \caption{The PQ scale $f_a$, its relation to the axion mass $m_{a}$
    (for $N=1$) together with axion-search ranges (dark-shaded) and
    exclusion limits from axion searches, cosmological constraints,
    and astrophysical limits (light-shaded and medium-shaded, columns
    from left to right). The vertical line with the two arrowheads
    indicates the $f_a$ range where $\Omega_a$ from the misalignment
    mechanism can provide naturally a sizable fraction of $\OmegaDM$
    for $f_a>\TR$. For a scenario with $f_a<\TR$, $\Omega_a>\OmegaDM$
    in the region shown by the ``Cold DM'' bar.  The other bar in this
    column indicates the region that is excluded by constraints from
    LSS (``Hot DM'') or by a thermal relic axion density
    $\Omega_a^{\thermal}>\OmegaDM$.  The transition region between
    those constraints is excluded by telescope searches for
    $a\to\gamma\gamma$ decays as indicated by the ``Telescope'' bar in
    the leftmost column.  The dark-shaded bars in that column indicate
    the search ranges of the axion-dark-matter-search experiment ADMX
    and of the axion helioscope CAST. The exclusion ranges in the
    rightmost column are inferred from the supernova SN 1987A, i.e.,
    from the observed duration of the emitted neutrino burst (``Burst
    duration'') and from the non-observation of axion-induced events
    (``Too many events''). The second column from the right applies to
    models with a direct axion-electron coupling and is inferred from
    globular cluster (GC) stars and white-dwarf cooling. Also the
    middle column is inferred from GC stars but from the axion-photon
    coupling: While the medium-shaded bar applies to a DFSZ model, the
    exclusion range can be more extended in a KSVZ model as indicated
    by the light-shaded tip. Details are given in the main text.
    From~\cite{Raffelt:2006cw}.}
  \label{Fig:AxionLimits}
\end{figure}

Because of the chiral U(1)$_{\PQ}$ anomaly, the axion receives a mass
from QCD instanton effects~\cite{Bardeen:1977bd,Baluni:1978rf} that
govern the effective axion potential $V_{\rm eff}$ at low energies
\begin{align}
        m_{a}^2 
        =\frac{m_u m_d}{(m_u+m_d)^2}
        \left(\frac{f_{\pi}m_{\pi}}{f_a/N}\right)^2
        \ ,
\label{Eq:m_axion}
\end{align}
where $m_u$ ($m_d$) is the mass of the up (down) quark and
$f_{\pi}\approx 92~\MeV$ and $m_{\pi}=135~\MeV$ are respectively the
decay constant and mass of the pion.
Using $z=m_u/m_d$,
\bea
        &&\!\!\!m_{a} 
        =\frac{\sqrt{z}}{1+z}
        \left(\frac{f_{\pi}m_{\pi}}{f_a/N}\right)
\label{Eq:Axion_Mass}
\\
        &&\!\!\!= 0.60~\meV\,
        \frac{2\,\sqrt{z/0.56}}{1+(z/0.56)}\!
        \left(\frac{10^{10}\,\GeV}{f_a/N}\right)
        \ .
\label{Eq:m_axion_z056}
\eea
This relation is shown for $z=0.56$~\cite{Leutwyler:1996qg} and
$f_a/N\to f_a$ in Fig.~\ref{Fig:AxionLimits} (from
\cite{Raffelt:2006cw});
%
%
note that $z$ could be within 0.3--0.6~\cite{Amsler:2008zz}.

While the original PQ proposal~\cite{Peccei:1977hh,Peccei:1977ur}
assumed $f_a$ to be at the weak scale, axion searches, astrophysical
observations, and cosmological arguments, which are discussed below,
point to a significantly higher value of the PQ scale
(cf.~\cite{Sikivie:1999sy,Raffelt:2006cw,Amsler:2008zz} and references
therein)
\begin{align}
f_a/N \gtrsim 6 \times 10^8\,\GeV \ . 
\label{Eq:faMin}
\end{align}
Accordingly, the interactions of the axion are strongly suppressed and
its mass must be very small, $m_a \lesssim 0.01~\eV$, so that the
axion can be classified as an extremely weakly interacting particle
(EWIP).

Axion interactions are model dependent. The two most popular classes
of phenomenologically viable ``invisible axion'' models are the
hadronic or Kim--Shifman--Vainshtein--Zakharov (KSVZ)
models~\cite{Kim:1979if,Shifman:1979if} and the
Dine--Fischler--Srednicki--Zhitnitskii (DFSZ)
models~\cite{Dine:1981rt,Zhitnitsky:1980tq}. In models of the KSVZ
type, at least one additional heavy quark is introduced which couples
directly to the axion while all other fields do not carry PQ charge.
Thus, the axion interacts with ordinary matter through the anomaly
term from loops of this new heavy quark.  Integrating out the heavy
quark loops, one obtains the effective dimension-5 coupling of axions
to gluons given in~(\ref{Eq:axion-gluon-gluon}) with $N=1$. In
particular, couplings of the axion to Standard Model matter fields are
suppressed by additional loop factors. In the DFSZ schemes, no
additional heavy quarks are introduced. Instead, the Standard Model
matter fields and at least two Higgs doublets carry appropriate PQ
charges such that the axion also couples directly to the Standard
Model fields.  Again, at low energies the axion-gluon
interaction~(\ref{Eq:axion-gluon-gluon}) arises, but now with $N = 6$.
For more details on these axion models, we refer to the
reviews~\cite{Kim:1986ax,Cheng:1987gp,Kim:2008hd}. In addition, axions
with $f_a\sim 10^{16}\,\GeV$ appear also in string theory;
cf.~\cite{Fox:2004kb,Gaillard:2005gj,Svrcek:2006yi}.

For the axion lifetime $\tau_a$ and axion phenomenology, the
axion-photon interaction plays a central role:
\begin{align}
        {\cal L}_{a\gamma\gamma} 
        \,=\, \frac{\gagg}{4}\,
        F_{\mu\nu}\widetilde{F}^{\mu\nu}
\label{Eq:axion-photon-photon}
\end{align}
with the electromagnetic field-strength tensor $F_{\mu\nu}$, its dual
$\widetilde{F}_{\mu\nu}=\epsilon_{\mu\nu\rho\sigma}F^{\rho\sigma}\!\!/2$,
and the coupling constant
\bea
        \gagg 
        &=& \frac{\alpha}{2\pi f_a/N}
        \left(\frac{E}{N}-\frac{2}{3}\frac{4+z}{1+z}\right)
\label{Eq:g-axion-photon-photon_fa}
\\
        &=& \frac{\alpha}{2\pi}
        \frac{1+z}{\sqrt{z}}
        \frac{m_a}{f_{\pi}m_{\pi}}
        \left(\frac{E}{N}-\frac{2}{3}\frac{4+z}{1+z}\right)
        \ ,
\label{Eq:g-axion-photon-photon_ma}
\eea
where both the electromagnetic anomaly $E$ and the color anomaly $N$
of the PQ symmetry depend on the axion model. For example, $E/N=0$ in
a KSVZ model with (an) electrically neutral new heavy quark(s) and
$E/N=8/3$ in DSFZ models. Note that $E/N=2$ is a possibility for which
the axion-photon coupling would be suppressed~\cite{Kaplan:1985dv} due
to cancellations in the bracket
of~(\ref{Eq:g-axion-photon-photon_fa},\ref{Eq:g-axion-photon-photon_ma}).

The lifetime of the axion governed by the decay $a\to\gamma\gamma$
(and the associated rate $\Gamma_{a\to\gamma\gamma}$) is obtained
from~(\ref{Eq:axion-photon-photon}):
\bea
        \tau_a 
        &=& \Gamma_{a\to\gamma\gamma}^{-1}
        \,\,=\,\, \frac{64\pi}{g_{a\gamma\gamma}^2 m_a^3}
\label{Eq:AxionLifetime}\\
        &\simeq& 4.6\times 10^{40}\,\seconds\,
        \left(\frac{E}{N}-1.95\right)^{\!\!-2}
        \left(\frac{f_a/N}{10^{10}\,\GeV}\right)^{\!\!5}
        \ ,
\label{Eq:AxionLifetime_fa}
\eea
where $z=0.56$ was used. Comparing $\tau_a$ for $E/N=0$ with the
present age of the Universe $t_0$ given in~(\ref{Eq:lifetimeDM}), one
finds $\tau_a\gtrsim t_0$ for $f_a/N\gtrsim 3\times 10^5\,\GeV$ or
equivalently $m_a\lesssim 20~\eV$. Thus, the axion is stable on
cosmological time scales for the $f_a/N$ values~(\ref{Eq:faMin}) that
are favored by cosmological and astrophysical constraints.

\subsection{Primordial Origin}
\label{sec:AxionProduction}

Axion dark matter can have different origins and different properties
depending on the cosmic history and on the value of $f_a/N$. 
A remarkable generic property of the axion is the fact that it is
massless for $T\gtrsim 1~\GeV\gtrsim\LQCD$ and that it acquires mass
only through instanton effects for $T\lesssim \LQCD$ such that $m_a$
is given by~(\ref{Eq:Axion_Mass}) for $T\to 0$.

The most straightforward situation is encountered when $f_a/N$ is
sufficiently small so that the axion couples sufficiently strong for
being in thermal equilibrium with the early primordial plasma.  The
axion will then decouple (or freeze out) at a temperature
$T_{\freezeout}\gg m_a$ as a thermal relic
\begin{align}
        \Omega_a^{\thermal} h^2 
        = 0.077
        \left(\frac{10}{g_{*S}(T_{\freezeout})}\right) 
        \left(\frac{m_a}{10~\eV}\right)
\label{Eq:OmegaAxionThermal}
\end{align}
where $g_{*S}$ denotes the number of effectively massless degrees of
freedom such that the entropy density reads $s=2\pi^2 g_{*S} T^3/45$.
For $f_a/N\lesssim 3\times 10^7\,\GeV$ (corresponding to $m_a\gtrsim
0.2~\eV$), the axion is a thermal relic that decouples after the
quark--hadron transition, $T_{\freezeout}\lesssim 150~\MeV$, where
$\pi\pi\leftrightarrow\pi a$ 
is a generic process that keeps the axions in thermal equilibrium
before their decoupling~\cite{Chang:1993gm,Hannestad:2005df}.%
\footnote{Before the quark--hadron transition, axions can be kept in
  thermal equilibrium by QCD reactions that involve the axion-gluon
  interaction~(\ref{Eq:axion-gluon-gluon}) with a sufficiently small
  $f_a/N$~\cite{Masso:2002np,Sikivie:2006ni}.}
Accordingly, the dark matter constraint $\Omega_a\leq\OmegaDM$ implies
$m_a\lesssim 18~\eV$, where also $\tau_a\gtrsim t_0$ is satisfied as
discussed above. However, such axions are hot dark matter so that
$\Omega_a$ can only be a minor fraction of $\OmegaDM$. In fact,
observations of LSS imply a restrictive limit of $m_a\lesssim
1.02~\eV$ (95\%
CL)~\cite{Hannestad:2005df,Hannestad:2007dd,Hannestad:2008js}, which
is indicated as ``Hot DM'' in Fig.~\ref{Fig:AxionLimits} (second
column from the left).

For large values of $f_a/N$ such that axions are never in thermal
equilibrium with the primordial plasma, $\Omega_a$ becomes sensitive
to the earliest cosmological history.
In this limit, the simplest setting is encountered when the
spontaneous breaking of the PQ symmetry occurs before inflation
leading to a reheating temperature of $\TR<f_a$ so that no PQ symmetry
restoration takes place during inflation or in the course of
reheating.
In this setting, axionic strings can be neglected since they have been
diluted by inflation and cannot be produced later on.
Moreover, the initial misalignment angle $\Theta_i$ of the axion with
respect to the CP-conserving position (i.e., its position in the
mexican hat potential) is everywhere (basically) the same in our
observable patch of the Universe.
Since $\Theta_i$ is typically not at the minimum of the effective
potential (i.e., not at the minimum of the tilted mexican hat)
generated by instanton effects at $T\sim\LQCD$, the field starts to
oscillate coherently around its minimum for $m_a(\Tosc)\simeq 3
H(\Tosc)$.  Once $m_a$ takes on its $T$-independent
value~(\ref{Eq:m_axion},\ref{Eq:Axion_Mass}), this axion condensate
behaves as cold dark
matter~\cite{Preskill:1982cy,Abbott:1982af,Dine:1982ah}
with a relic density that is governed by the initial misalignment
angle $-\pi<\Theta_i\leq\pi$~\cite{Beltran:2006sq,Sikivie:2006ni}:%
\begin{align}
        \Omega_a h^2 \sim 0.15\,\xi\, 
        f(\Theta_i^2)\,
        \Theta_i^2
        \left(\frac{f_a/N}{10^{12}\,\GeV}\right)^{\! 7/6}
\label{Eq:OmegaAxionMisalignment}
\end{align}
with $\xi=\Order(1)$ parametrizing theoretical uncertainties related,
e.g., to details of the quark--hadron transition and of the
$T$-dependence of $m_a$ (cf.~\cite{Bae:2008ue}) and $f(\Theta_i^2)$
accounting for anharmonicity at sizable $\Theta_i$; $f(\Theta_i^2)\to
1$ for $\Theta_i^2\to 0$.
For $10^{10}\,\GeV\lesssim f_a/N \lesssim 10^{13}\,\GeV$, this
``misalignment mechanism'' can provide {\em naturally}, i.e., for
$\Theta_i = \Order(1)$, a sizable contribution $\Omega_a$ to
$\OmegaDM$, which is indicated by the vertical line with the two
arrowheads in Fig.~\ref{Fig:AxionLimits}.
However, with any value of $|\Theta_i|\in[0,\pi]$ being equally
probable, there is always the possibility of $|\Theta_i|\approx 0$ in
our patch of the
Universe.%
\footnote{In fact, as a matter of principle, it is impossible to
  calculate the precise value of $\Omega_a$ from first principles in
  this setting.}
In fact, for a finely tuned $|\Theta_i|<10^{-2}$, this setting allows
for the high values of $f_a\sim 10^{16}\,\GeV$ suggested by concepts
of grand unification and string
theory~\cite{Fox:2004kb,Gaillard:2005gj,Svrcek:2006yi}.
With the original motivation for the axion residing in a solution of
the fine-tuning problem~(\ref{Eq:StrongCPProblem}), a finely tuned
$\Theta_i$ is somewhat unpleasant but can be associated with anthropic
considerations~\cite{Tegmark:2005dy,Hertzberg:2008wr}.
Note that it would actually be more correct to replace $\Theta_i^2$
in~(\ref{Eq:OmegaAxionMisalignment}) by $\langle\Theta_i^2\rangle$ at
$T\sim\LQCD$, which cannot be arbitrarily small 
in the case of inflation with energy scale $\Einf$ or Hubble scale
$\Hinf=8\pi\GN\Einf^4/3$ defined when those modes exit the horizon
that reenter the horizon
today~\cite{Linde:1990yj,Turner:1990uz,Lyth:1991ub,Lyth:1992tw,Fox:2004kb,Beltran:2006sq},
\begin{align}
  \langle\Theta_i^2\rangle
  \gtrsim 
  30 
  \left(\frac{\Hinf}{2\pi f_a/N}\right)^{\!2}
        \ ,
\label{Eq:ThetaMin}
\end{align}
due to fluctuations in the axion field from de Sitter quantum
fluctuations during the inflationary epoch.

Let us now consider the case in which the spontaneous breaking of the
PQ symmetry occurs after inflation, $\TR>f_a$, again for large values
of $f_a/N$ such that axions are never in thermal equilibrium with the
primordial plasma.
After spontaneous breaking of the PQ symmetry at $T\sim f_a$, our
observable patch of the Universe will consist of many smaller patches,
each of which with an arbitrary value of $\Theta_i\in(-\pi,\pi]$, and
of an associated network of axionic strings and domain walls.
Indeed, since a uniform distribution of $\Theta_i$ is provided with
$\langle\Theta_i^2\rangle=\pi^2/3$, the axion relic density from the
misalignment mechanism becomes independent of $\Theta_i$:
\begin{align}
        \Omega_a h^2 \sim 0.6 \,\xi 
        \left(\frac{f_a/N}{10^{12}\,\GeV}\right)^{\! 7/6}
        \ ,
\label{Eq:OmegaAxionMisalignment_ThetaRMS}
\end{align}
which includes the anharmonic correction factor
$f(\Theta_i^{\mathrm{rms}}=\pi/\sqrt{3})=1.2$.  The dark matter
constraint $\Omega_a\leq\OmegaDM$ thus implies, e.g., for a moderate
$\xi\simeq 0.4$ the limit $f_a/N\lesssim 6\times 10^{11}\,\GeV$
($m_a\gtrsim 10^{-5}\,\eV$) as indicated in Fig.~\ref{Fig:AxionLimits}
(second column from the left).
This limit is conservative since there can be additional sizable
contributions to $\Omega_a$ from decays of axionic strings, domain
walls, and non-zero momentum modes of the axion field.
For a discussion of those contributions and potential domain wall
problems, see~\cite{Sikivie:2006ni} and references therein.

\subsection{Cosmological Constraints}
\label{sec:AxionCosmoConstraints}

The thermal relic axions encountered for sufficiently small values of
$f_a/N$ will contribute to the radiation density at the time of BBN in
a way analogous to the case of an extra light neutrino species.
Thereby, these axions will give the following contribution to the
effective number of neutrinos $\Nnu$ at $T\simeq
1~\MeV$~\cite{Chang:1993gm}
\begin{align}
        \Delta \Nnu 
        = \frac{4}{7} \left(\frac{10.75}{g_{*S}(T_{\freezeout})}\right)^{4/3}
        .
\label{Eq:DeltaNnuAxion}
\end{align}
For $T_{\freezeout}>1~\MeV$, $\Delta\Nnu\leq 4/7=0.57$ which is compatible with
the BBN limit~\cite{Mangano:2006ur}
\begin{align}
        \Nnu = 3.1^{+1.4}_{-1.2} 
        \quad 
        (95\%~\mathrm{CL})
        \ ;
\label{Eq:NnuLimitBBN}
\end{align}
see also Fig.~2 of Ref.~\cite{Simha:2008zj}.
However, in the literature also more restrictive limits on
$\Delta\Nnu$ can be found, which can be associated with an $m_a$ bound
that is more restrictive than the limit $m_a\lesssim 1.02~\eV$ (95\%
CL)~\cite{Hannestad:2005df,Hannestad:2007dd,Hannestad:2008js} from
observations of cosmological LSS discussed in the previous section.
For example, the limit $\Delta\Nnu<0.25$ would imply $g_{*S}(T_{\freezeout})>20$
and thereby $f_a\gtrsim 2\times 10^7\,\GeV$ (or $m_a\lesssim 0.3~\eV$)
as can be seen from Fig.~4 of Ref.~\cite{Hannestad:2005df}.

For large $f_a/N$ scenarios with PQ breaking before inflation and
$\TR<f_a$, in which axions are never in thermal equilibrium with the
primordial plasma, axion density fluctuations produced during
inflation lead to iso\-cur\-vature
\cite{Seckel:1985tj,Linde:1985yf,Lyth:1989pb,Turner:1990uz,Linde:1991km,Beltran:2006sq,Hertzberg:2008wr}
and non-Gaussian \cite{Lyth:1991ub,Lyth:1992tx,Kawasaki:2008sn}
effects in the temperature fluctuations of the CMB. Such effects are
constrained by cosmological data.

With the axion field being essentially massless during inflation, de
Sitter quantum fluctuations are imprinted on the axion as on any other
light scalar field.
Since the couplings of the axion are so weak, an isocurvature mode
survives that is uncorrelated to the usual adiabatic mode seeded by
the quantum fluctuations of the inflaton field.
The resulting deviations from adiabaticity could manifest themselves
in the temperature fluctuations of the CMB and could thereby support
the existence of the axion (or, more generally, of at least two light
scalar fields during inflation, one of which giving rise to dark
matter that is never in thermal equilibrium with the primordial
plasma).

So far, no evidence for dark matter isocurvature perturbations has
been found. Using $\alpha_0$ to parameterize the ratio of the power
spectrum of entropy perturbation, $P_S(k)$, to the one of curvature
perturbation, $P_R(k)$, at the pivot wavenumber $k_0=0.002~\Mpc^{-1}$,
\begin{align}
  \frac{\alpha(k_0)}{1-\alpha(k_0)}\equiv\frac{P_S(k_0)}{P_R(k_0)}
  \ ,
\label{Eq:alpha0def}
\end{align}
current limits on axionic entropy perturbations are $\alpha_0<0.16$
(95\% CL) and $\alpha_0<0.067$ (95\% CL) obtained respectively from
a WMAP-only and a WMAP+BAO+SN%
\footnote{Here BAO refers to baryon acoustic oscillations and SN to
  supernovae of type Ia. See~\cite{Komatsu:2008hk} for references and
  details.}
analysis~\cite{Komatsu:2008hk}.
Confronting the $\alpha_0$ limit with the expression derived for an
axion that receives quantum fluctuations with a nearly scale invariant
spectrum during inflation and $N=1$~\cite{Beltran:2006sq},%
\footnote{Note that there are typos in Eqs.~(44) and~(27) of
  Ref.~\cite{Beltran:2006sq}. The corrected prefactor in~(44) is
  $2.4\times 10^{10}$ and the associated limit in~(27) should read
  $1.2\times 10^{16}\,\GeV$. I thank J.\ Lesgourgues for the
  clarification of this point.}
\begin{align}
  \alpha(k) \simeq 
  2.4\times 10^{10}\,
  \frac{\GN\Einf^4 R_a^2}{\pi f_a^2 \langle\Theta_i^2\rangle}
  \ ,
\label{Eq:alpha0Axion}
\end{align}
one obtains cosmological constraints on $f_a$ that depend on $\Einf$
(or $\Hinf$) and on the fraction
$R_a=\Omega_a/\OmegaDM$~\cite{Beltran:2006sq,Hertzberg:2008wr}.

\begin{figure}[t!]
\includegraphics[width=0.48\textwidth,clip=true]{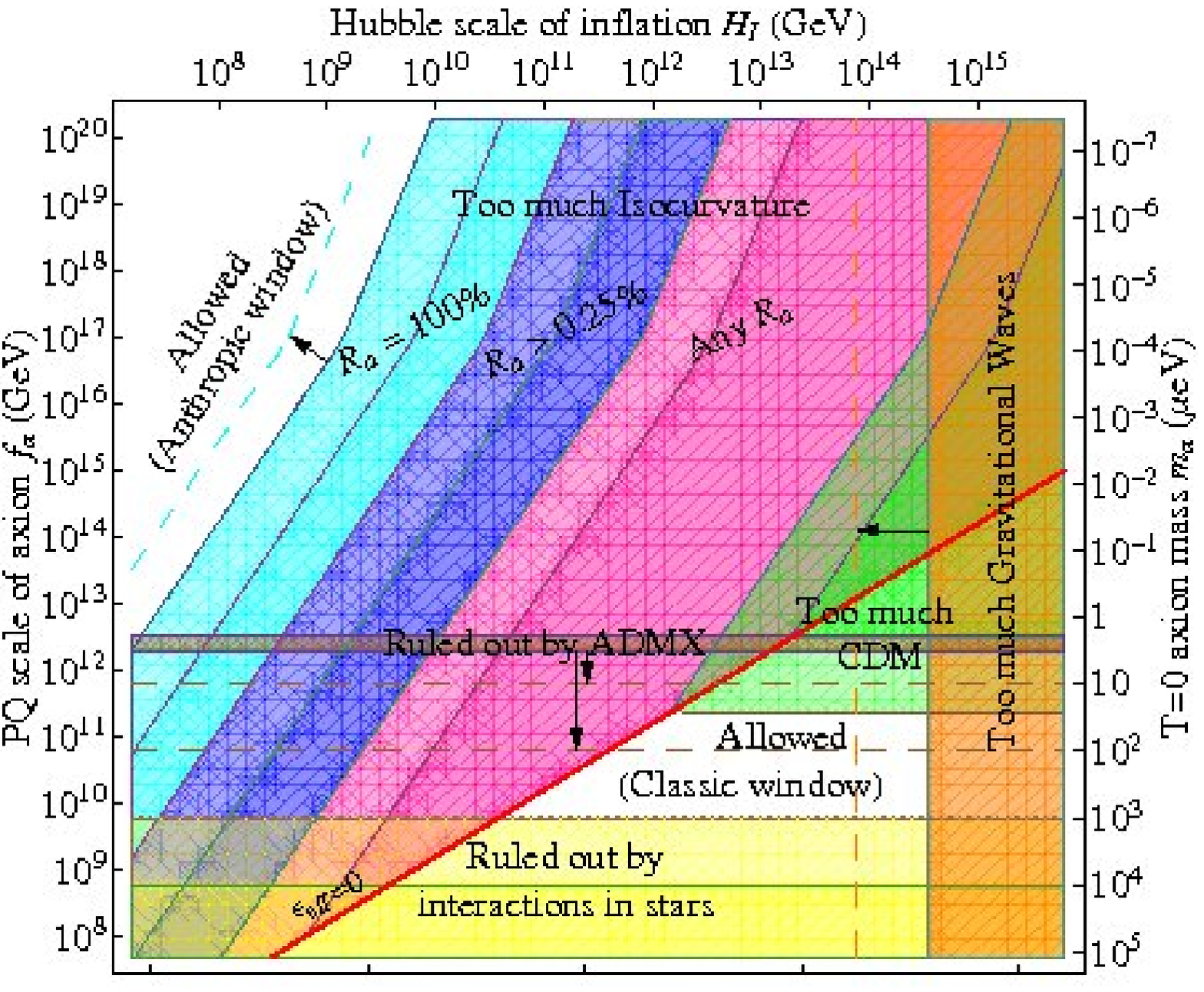}\vspace{-0.05cm}
\includegraphics[width=0.48\textwidth,clip=true]{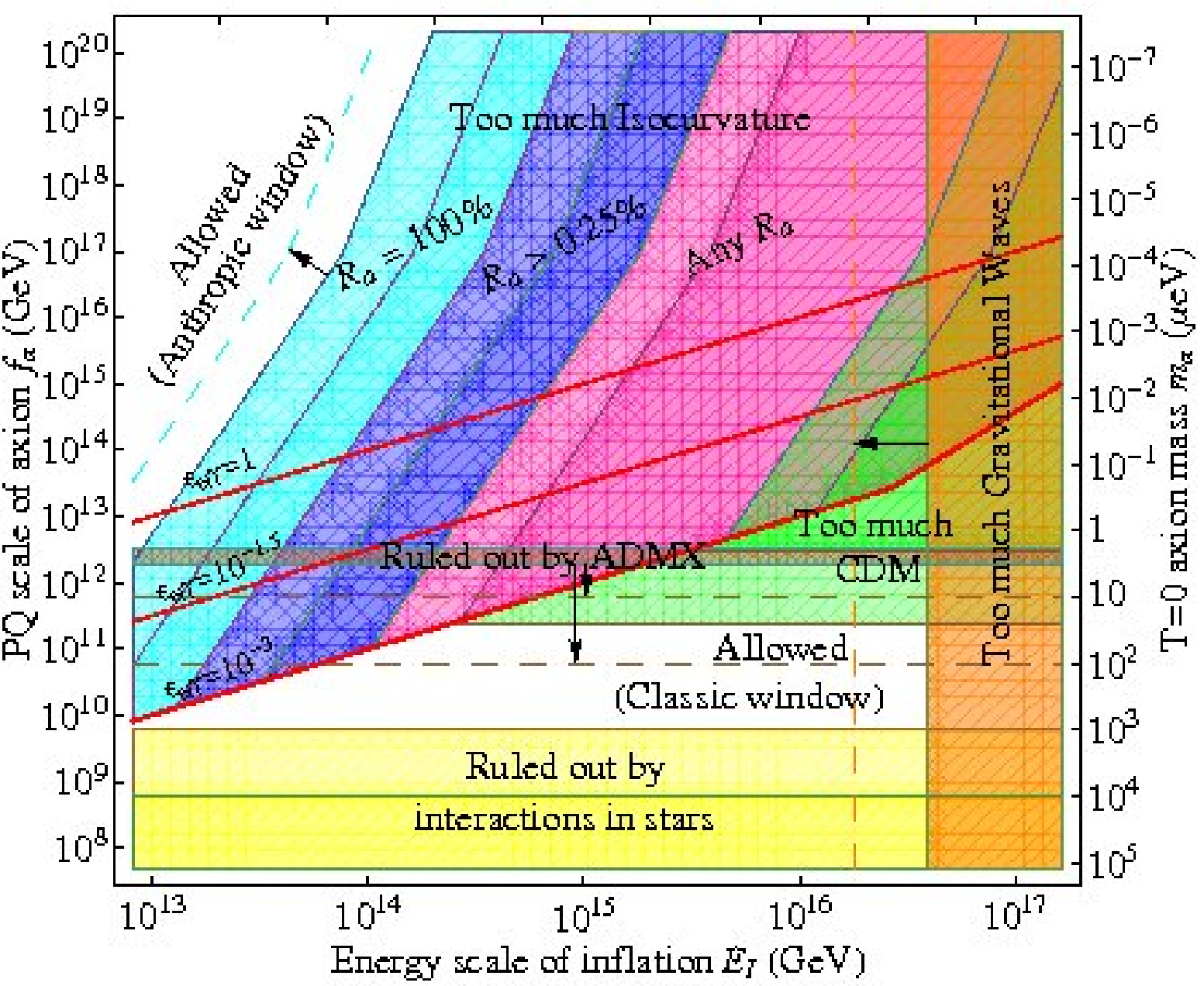}
\caption{Constraints on the PQ scale $f_a$ (for $N=1$) as a function
  of the energy scale of inflation $\Einf$ or Hubble scale
  $\Hinf=8\pi\GN\Einf^4/3$ defined when those modes exit the horizon
  that reenter the horizon today.  The thick solid (red) lines show
  $f_a=\Hinf/2\pi$ (upper panel) and $f_a=\Tmax=\epseff\Einf$ for
  $\epseff=1$, $10^{-1.5}$, $10^{-3}$ (lower panel). Below (above) the
  corresponding line, PQ symmetry restoration does (not) occur
  during/after inflation. The hatched regions above those lines show
  the cosmological constraints from the WMAP+BAO+SN limit
  $\alpha_0<0.067$ (95\% CL) on axionic isocurvature
  perturbations~(\ref{Eq:alpha0Axion}). Those constraints depend on
  $R_a=\Omega_a/\OmegaDM$ as labeled and as indicated by the different
  shadings. The constraints labeled as ``Any $R_a$'' result from the
  lower limit on axion fluctuations due to quantum fluctuations during
  inflation~(\ref{Eq:ThetaMin}). For $\Einf>10^{15}\,\GeV$, this can
  lead to $\Omega_a>\OmegaDM$ as indicated by the (green) region
  labeled as ``Too much CDM''.  The WMAP+BAO+SN limit on the tensor to
  scalar ratio $r<0.2$ (95\% CL) excludes the region with
  $\Einf>3.8\,\times 10^{16}\,\GeV$.  The exclusion ranges from axion
  searches and astrophysical considerations are indicated respectively
  by the dark (brown) horizontal bar and by the light-shaded (yellow)
  region with $f_a<10^9\,\GeV$.  For $f_a>\MPl$, the PQ scale is
  super-Planckian.  More details are given in the main text.
  From~\cite{Hertzberg:2008wr}.}
  \label{Fig:AxionLimitsfaEinflaton}
\end{figure}

Figure~\ref{Fig:AxionLimitsfaEinflaton} (from~\cite{Hertzberg:2008wr})
%
%
shows isocurvature limits inferred from $\alpha_0<0.067$ for $R_a=1$
(cyan) and $R_a\geq 0.25$ (purple) and $\Omega_a\leq\OmegaDM$ limits
(green). The $R_a$ independent limits (pink) are associated with the
lower limit~(\ref{Eq:ThetaMin}).
For each of the constraints that depend
on~(\ref{Eq:OmegaAxionMisalignment}),
a more and a less conservative limit obtained respectively for
$\xi=0.05$ and $\xi=1$ is indicated by the darker and lighter
shadings.
The kink of these limits for $f_a>10^{16}\,\GeV$ results from
$\Tosc\lesssim\LQCD$ and an associated $\Omega_a$ expression with a
different parametric dependence on $f_a/N$ than
in~(\ref{Eq:OmegaAxionMisalignment}) derived for $\Tosc\gtrsim\LQCD$;
for details, see e.g.~\cite{Fox:2004kb,Hertzberg:2008wr}.
The upper (lower) panel shows the case of very inefficient (efficient)
thermalization at the end of inflation. The thick solid (red) lines
indicate $f_a=\Hinf/2\pi$ and $f_a=\Tmax=\epseff\Einf$ for
$\epseff=1$, $10^{-1.5}$, $10^{-3}$. Below (above) the corresponding
line, PQ symmetry restoration does (not) occur during/after inflation.
Obtained from a WMAP+BAO+SN analysis~\cite{Komatsu:2008hk}, the current
limit on the amplitude of primordial gravitational waves expressed in
terms of the tensor to scalar ratio $r<0.2$ (95\% CL) excludes
$\Einf>3.8\,\times 10^{16}\,\GeV$ as indicated by the orange region.
In the region indicated by the horizontal (brown) bar, axion models
with an unsuppressed axion-photon coupling $\gagg$ are excluded by the
axion dark matter search experiment ADMX;
cf.~Fig.~\ref{Fig:AxionLimitsSearches} and discussions in the next
section. The (yellow) region with $f_a<10^9\,\GeV$ is disfavored by
conservative astrophysical constraints;
cf.~Fig.~\ref{Fig:AxionLimits}. The dashed lines indicate targets of
future experiments/observations.

While the quantum fluctuations in $\Theta_i$ during inflation are
Gaussian distributed, the axion-induced temperature fluctuations in
the CMB are governed by fluctuations in the axion number density after
formation of the axion condensate with $n_a\propto \Theta_i^2$ modulo
anharmonic corrections.  Thereby a non-linearity enters that leads to
a non-Gaussian contribution to the CMB temperature
fluctuation~\cite{Lyth:1991ub,Lyth:1992tx,Kawasaki:2008sn}. This may
allow one to probe axions cosmologically even for the case
$\Omega_a\ll\OmegaDM$~\cite{Kawasaki:2008sn}.

If the PQ symmetry is not broken before inflation, there will be no
light axion field present to acquire de Sitter quantum fluctuations
and thus no axionic source of isocurvature perturbations.
Also if the PQ symmetry is restored during inflation, $\Hinf/2\pi>f_a$
(i.e., the typical amplitude of quantum fluctuations exceeds the PQ
scale), or during reheating, $\Tmax>f_a$ (i.e., thermal PQ symmetry
restoration), primordial axionic perturbations will be washed out
which implies negligible axion-induced deviations from adiabaticity.
Moreover, as described in the previous section, the subsequent PQ
breaking will lead to a sample of $\Theta_i$ values with a flat
probability distribution after PQ breaking. Since these $\Theta_i$
values are distributed at random throughout our observable Universe,
spatial axion density variations are expected once the axion acquires
its mass. These variations can then lead to ``axion mini clusters''
containing a sizable contribution to $\OmegaDM$~\cite{Sikivie:2006ni}.

\subsection{Astrophysical Constraints}
\label{sec:AxionAstroConstraints}

Axions could be produced not only in the early Universe but also in
hot and dense astrophysical environments such as those encountered in
ordinary stars, white dwarfs, and supernovae.  The axion luminosity
$\Laxion$ of such sources depends on $f_a$, the relevant axion
production processes (and thereby on the axion couplings and model),
and on the astrophysical understanding/model of the source under
consideration.  A sizable axion luminosity is associated with an
additional energy transport out of the corresponding astrophysical
source. This can affect the behavior/character\-is\-tics of the source
strongly.  Astrophysical studies of stars, white dwarfs, and
supernovae can thus be used to derive constraints on $f_a/N$ or
$m_a$~\cite{Dicus:1978fp,Vysotsky:1978dc,Turner:1989vc,Raffelt:1990yz,Kolb:1990vq,Raffelt:1999tx,Raffelt:2006cw}.

To allow for axion production, the relevant energy/tem\-per\-ature in
the source must be sufficiently large with respect to $m_a$ already
for kinematical reasons. In addition, the interaction strength of the
axion enters in the following two ways: On the one hand, a stronger
axion interaction (smaller $f_a/N$ or larger $m_a$) allows for a more
efficient production in the source. On the other hand, stronger
interactions are also associated with smaller mean free paths in
medium and possibly with rescatterings within the source which can
delay energy emission via the axion channel.

The current status of astrophysical axion bounds is reviewed, e.g., in
Refs.~\cite{Raffelt:2006cw,Amsler:2008zz} and shown in
Fig.~\ref{Fig:AxionLimits} (three rightmost columns). Here only a
brief summary is given.

In stars including our sun, axions can be produced via their coupling
to photons $\gagg$ through the Primakoff process $\gamma+Ze\to
Ze+a$~\cite{Dicus:1978fp,Raffelt:1985nk,Altherr:1993zd,Raffelt:1987np}.
The axionic energy drain (described by $\Laxion\propto\gagg^2$) can
then lead to an enhanced consumption of nuclear fuel within the star
and can thereby, e.g., shorten the lifetime of a star. In this respect
globular clusters (GCs), which are bound systems of a homogeneous
population of low-mass stars, are particularly valuable since they
allow for tests of stellar-evolution theory. In fact, studies of GC
stars point to $\gagg\lesssim
10^{-10}~\GeV^{-1}$~\cite{Raffelt:1996wa,Raffelt:1999tx,Raffelt:2006cw}
which implies $f_a>2.3\times 10^{7}\,\GeV$ in a KSVZ axion model with
$E/N=0$ and $f_a>0.8\times 10^{7}\,\GeV$ in a DFSZ axion model with
$E/N=8/3$ for $z=0.56$.  These limits are indicated by the middle
column in Fig.~\ref{Fig:AxionLimits}. Not shown are the limits
inferred from studies of our sun since they are less restrictive. For
example, helioseismological studies of the sound-speed profile give
the limit $\gagg\lesssim 10^{-9}~\GeV^{-1}$~\cite{Schlattl:1998fz}.
Moreover, $\gagg\lesssim 5\times
10^{-10}~\GeV^{-1}$~\cite{Raffelt:2006cw} is inferred from
measurements of the solar $^8$B neutrino flux which is sensitive to
enhanced nuclear burning~\cite{Schlattl:1998fz}.

Axion production via $\gamma+e^-\to e^-+a$ and $e^-+Ze\to Ze+e^-+a$ is
more efficient than the Primakoff processes in models with a direct
axion-electron coupling $\gaee$. Studies of GC
stars~\cite{Raffelt:1994ry} and of white-dwarf
cooling~\cite{Raffelt:1985nj,Wang:1992gc,Corsico:2001be,Isern:2003xj}
thereby lead to restrictive limits on $\gaee$. For the DSFZ model with
two Higgs doublets carrying PQ charge and the ratio of the associated
Higgs vacuum expectation values (VEVs) given by $\tan\beta$, these
limits imply $f_a<1.3\times 10^9\,\GeV
\cos^2\beta$~\cite{Raffelt:2006cw} which is indicated for
$\cos^2\beta=0.5$ by the second column from the right in
Fig.~\ref{Fig:AxionLimits}.

In core collapse supernovae (SN of type II) that lead to the formation
of a hot---$T=\Order(10~\MeV)$---proto-neutron star, axions can be
produced via their coupling to nucleons $\gaNN$ through axion
bremsstrahlung emission $N+N\to
N+N+a$~\cite{Raffelt:1987yt,Carena:1988kr} in the dense nuclear
medium~\cite{Ellis:1987pk,Raffelt:1987yt,Turner:1987by,Mayle:1987as,Janka:1995ir,Keil:1996ju}.
In fact, in these dense environments even the mean free path of
neutrinos is such that they rescatter and diffuse out before carrying
away energy. This picture was confirmed by the observed duration of
about $10~\seconds$ of the neutrino burst from the supernova SN~1987A.
Since axion emission would be an additional energy drain, it would
reduce the cooling time and this burst duration.
The observed burst duration thereby implies $f_a\gtrsim 4\times
10^8\,\GeV$~\cite{Raffelt:2006cw} as indicated by the upper bar in the
rightmost column of Fig.~\ref{Fig:AxionLimits}.  For $f_a \lesssim
6\times 10^5\,\GeV$, axions cannot affect the neutrino burst duration
significantly since their coupling becomes so strong that they
rescatter within the dense
medium~\cite{Burrows:1988ah,Burrows:1990pk}. For $f_a \lesssim 3\times
10^5\,\GeV$, the axion coupling would have been strong enough for
axion induced events that have not been observed for the supernova
SN~1987A~\cite{Engel:1990zd}. This is indicated by the lower bar in
the rightmost column of Fig.~\ref{Fig:AxionLimits}. The gap between
the two bars is known as the ``hadronic axion window'' which was
thought to be an allowed region for KSVZ models with strongly
suppressed $\gagg$~\cite{Moroi:1998qs}.  However, this window is
closed by the hot dark matter constraint from LSS
observations~\cite{Hannestad:2005df,Hannestad:2007dd,Hannestad:2008js}
discussed in Sect.~\ref{sec:AxionProduction} and indicated by the
second column from the left in Fig.~\ref{Fig:AxionLimits}.

\subsection{Experimental Searches and Prospects}
\label{sec:AxionExperiments}

Axion searches started already more than 30 years ago and excluded
soon $f_a$ values close to the weak scale, e.g., from studies of the
branching ratio BR$(K^+\to\pi^+ + \mathrm{nothing})$, as indicated by
the bar at the bottom (``Laboratory'') of the leftmost column in
Fig.~\ref{Fig:AxionLimits}. Present day axion searches are probing
much larger values of $f_a/N$ including those of invisible axion
models and those in which axions can provide the dominant component of
dark matter~\cite{Sikivie:1999sy,Battesti:2007um,Amsler:2008zz}.
Depending on the expected origin of the axions, these searches can be
classified into those for cosmic axions, solar axions (or more
generally axions from astrophysical sources), and axions produced in
the laboratory.

\begin{figure}[t!]
  \centerline{\includegraphics[width=0.495\textwidth,clip=true]{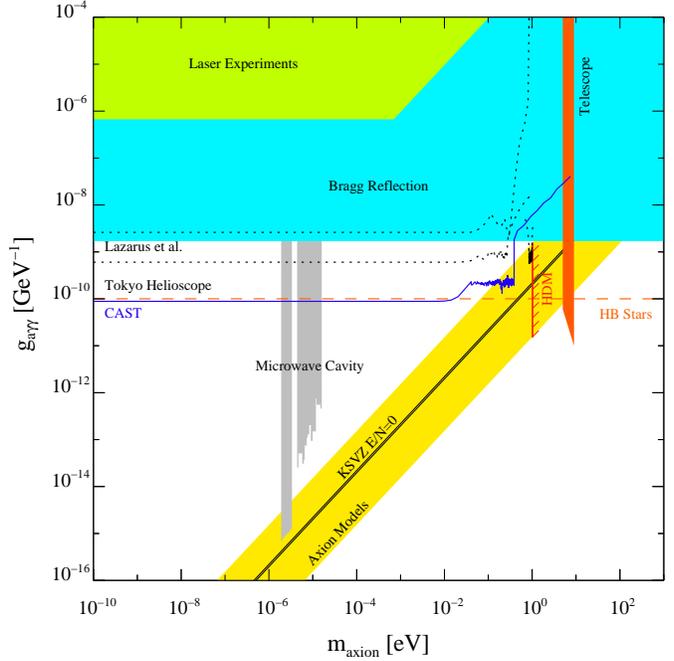}}
  \caption{Axion exclusion limits in the plane spanned by $m_a$ and
    $\gagg$. The diagonal (yellow) bar (``Axion Models'') and the
    solid diagonal line (``KSVZ E/N=0'') show respectively the region
    of typical invisible axion models and $\gagg$ in the KSVZ model
    with $E/N=0$.  The cosmological hot dark matter
    constraint~\cite{Hannestad:2005df,Hannestad:2007dd,Hannestad:2008js}
    is shown by the short vertical (orange) line (``HDM'') within this
    region and the astrophysical limit from studies of stars on the
    horizontal branch (HB) in GBs by the vertical dashed (orange) line
    (``HB
    stars'')~\cite{Raffelt:1996wa,Raffelt:1999tx,Raffelt:2006cw}.  The
    telescope search for $a\to\gamma\gamma$ decays reported in
    Ref.~\cite{Grin:2006aw} excludes the pencil-like (orange) region
    (``Telescope'') and direct axion searches with microwave cavities
    the pencil-like gray region (``Microwave
    Cavity'')~\cite{Asztalos:2003px,Duffy:2006aa,Battesti:2007um}.
    Solar axion searches with axion helioscopes exclude the regions
    above the upper (``Lazarus et al.'') and lower (``Tokyo
    Helioscope'') dotted lines and the one above the solid (blue) line
    (``CAST'')~\cite{Zioutas:2004hi,Andriamonje:2007ew,Ruz:2008zz,Minowa:2008uj}.
    The medium-shaded (cyan) region (``Bragg Reflection'') is excluded
    by searches for axion conversion in the Coulomb field of nuclei in
    a crystal
    lattice~\cite{Morales:2001we,Bernabei:2001ny,Avignone:1997th} and
    the light-shaded (green) region (``Laser Experiments'') by
    searches for a ``light-shining through a wall''
    event~\cite{Cameron:1993mr}. I thank M.~Kuster for providing this
    update of Fig.~10.26 of Ref.~\cite{Battesti:2007um}.}
  \label{Fig:AxionLimitsSearches}
\end{figure}

Most axion searches rely on the axion-photon coupling $\gagg$.%
\footnote{Fifth-force experiments allow for $\gagg$-independent axion
  searches~\cite{Moody:1984ba}; see~\cite{Heckel:2006ww} and
  references therein.}
Since the axion has not been discovered so far, negative searches can
be translated into $m_a$-dependent (or $f_a$-dependent) limits on
$\gagg$. A summary of those exclusion limits is given in
Fig.~\ref{Fig:AxionLimitsSearches} which is an update of Fig.~10.26 of
Ref.~\cite{Battesti:2007um} by courtesy of M.~Kuster.
%
%
Here the diagonal (yellow) bar and the solid diagonal line
indicate the region of typical invisible axion models and $\gagg$ in
the KSVZ model with $E/N=0$, respectively. The cosmological hot dark
matter constraint $m_a\lesssim
1~\eV$~\cite{Hannestad:2005df,Hannestad:2007dd,Hannestad:2008js} is
shown by the short vertical (orange) line (``HDM'') within this region
and the astrophysical limit from studies of stars on the horizontal
branch (HB) in GBs $\gagg\lesssim 10^{-10}~\GeV^{-1}$ by the vertical
dashed (orange) line (``HB
stars'')~\cite{Raffelt:1996wa,Raffelt:1999tx,Raffelt:2006cw}. Let us
recall that axions can provide a significant contribution to
$\OmegaDM$ for $m_a\lesssim 3\times 10^{-4}\,\eV$.

Sizable cosmic axion densities are expected in galaxy clusters and
galaxies and thereby also on Earth. There are indirect and direct
searches for those cosmic axions. In the indirect ones, telescopes are
used to look for photons from $a\to\gamma\gamma$ decays, for example,
in the Abell
clusters~\cite{Bershady:1990sw,Ressell:1991zv,Grin:2006aw} and in
nearby dwarf galaxies~\cite{Blout:2000uc}. After correcting for the
Doppler shift due to the motion of the host, one expects a basically
mono-chromatic spectrum of the resulting $\gamma$'s at
$E_{\gamma}=m_a/2$. In addition to possible insights into $m_a$, also
$\tau_a$ is probed in these telescope searches. No such $\gamma$'s
have been identified unambiguously so far which implies that $\tau_a$
can respect~(\ref{Eq:lifetimeDM}). The associated exclusion limits
(labeled as ``Telescope'') are shown by the corresponding bar in the
leftmost column in Fig.~\ref{Fig:AxionLimits} and by the vertical
pencil-like (orange) region in Fig.~\ref{Fig:AxionLimitsSearches}.

In direct searches for cosmic axions, microwave cavities are used to
look for resonant conversion of those axions---pervading the
Earth---into
photons~\cite{DePanfilis:1987dk,Wuensch:1989sa,Hagmann:1990tj,Bradley:2003kg,Asztalos:2003px,Duffy:2006aa,Carosi:2007uc}
along the haloscope technique proposed in~\cite{Sikivie:1983ip}.
Through the cavity frequency at which the resonance would appear and
throught the width of the resonance, these experiments are sensitive
to $m_a$ and to the virial distribution of thermalized axions and
thereby to the axion distribution in the galactic halo. In fact,
microwave resonant cavity experiments probe exactly the $m_a$ range in
which axions can provide a sizable contribution to $\OmegaDM$; cf.\
bar labeled as ``ADMX'' in the leftmost column in
Fig.~\ref{Fig:AxionLimits}. Moreover, the ADMX experiment has achieved
a sensitivity such that realistic axion models have already been
probed and excluded at 90\% CL in a narrow $m_a$
range~\cite{Asztalos:2003px,Duffy:2006aa}.  Indeed, no axion signal
has been observed so far. The associated exclusion limits are shown by
the horizontal (brown) bar in Fig.~\ref{Fig:AxionLimitsfaEinflaton}
and by the vertical pencil-like gray regions labeled as ``Microwave
Cavity'' in Fig.~\ref{Fig:AxionLimitsSearches}.  An upgrade of ADMX is
underway that should allow one to probe realistic axion models over a
much larger $m_a$ range~\cite{Bradley:2003kg,Carosi:2007uc}.  Relying
on Rydberg-atom detectors, an upgrade of the cavity experiment CARRACK
is aiming at a similar sensitivity and search
range~\cite{Tada:1999tu,Bradley:2003kg}.

As already addressed in the previous section, axions could be produced
in astrophysical sources such as our sun. Searching for solar axions
means to look for the conversion of such axions into $\gamma$'s in an
electromagnetic field via the inverse Primakoff process. In axion
helioscopes~\cite{Sikivie:1983ip,Lazarus:1992ry,Moriyama:1998kd,Inoue:2002qy,Zioutas:2004hi,Andriamonje:2007ew},
this field is provided by a strong magnet that is pointed at the sun.
Using Bragg diffraction at crystal
detectors~\cite{Avignone:1997th,Morales:2001we,Bernabei:2001ny}, it is
the Coulomb field of the nuclei in the crystal lattice that provides
the field.
Also the geomagnetic field of the Earth can allow in principle for the
conversion of solar axions into photons which could be detected by a
satellite on the dark side of the Earth~\cite{Davoudiasl:2005nh}.
Existing solar axion searches provide so far only $m_a$-dependent
exclusion limits on $\gagg$.  The ones from axion helioscopes are
indicated by the upper (``Lazarus et al.'') and lower (``Tokyo
Helioscope'') dotted lines and by the solid (blue) line (``CAST'')
line in Fig.~\ref{Fig:AxionLimitsSearches}.
Among those limits, the most restrictive one is provided by the CERN
Axion Solar Telescope (CAST): $\gagg<8.8\times 10^{-11}\,\GeV^{-1}$
for $m_a<0.02~\eV$~\cite{Zioutas:2004hi,Andriamonje:2007ew}.
Currently, the Tokyo Helioscope and CAST are probing already the
parameter region with realistic axion
models~\cite{Ruz:2008zz,Minowa:2008uj} but for $m_a>10^{-2}\,\eV$
which is associated with warm/hot axion dark matter scenarios. Indeed,
an axion discovery in that region would imply that axions cannot be
the dominant component of $\OmegaDM$ for a standard cosmological
history%
\footnote{For axion cosmology with a non-standard cosmological
  history, see e.g.~\cite{Grin:2007yg}.}
so that there is room, for example, for the LSP to take over that
part. 
The $m_a$-independent exclusion limit from Bragg reflection
$\gagg\lesssim 1.7\times 10^{-9}\,\GeV^{-1}$ obtained by the DAMA
collaboration~\cite{Bernabei:2001ny} is indicated by the medium-shaded
(cyan) region (labeled accordingly) in
Fig.~\ref{Fig:AxionLimitsSearches}; similar limits were obtained by
COSME~\cite{Morales:2001we} and SOLAX~\cite{Avignone:1997th}.
Not shown is the limit $\gagg\lesssim 10^{-11}\,\GeV$ for $m_a\lesssim
10^{-9}\,\eV$ that has been inferred from the absence of $\gamma$-ray
bursts in coincidence with the SN~1987A neutrino
burst~\cite{Brockway:1996yr,Grifols:1996id}.  Such bursts could have
originated from axions produced in the SN~1987A that had been
converted subsequently into photons in the galactic magnetic field.

Axion searches are also performed purely in the laboratory. Using the
Primakoff process, one should in principle be able to convert a small
fraction of the photons of a laser beam in a strong transverse
magnetic field into axions. In contrast to the photons, those axions
should be able to traverse a wall. In a second strong magnetic field
behind the wall, there should also be a non-zero probability for the
inverse Primakoff effect in which the axion is reconverted into a
detectable photon. The overall probability of such a ``light-shining
through a wall'' event is proportional to $\gagg^4$.  No event of this
sort has been observed so far
and the Brookhaven-Fermilab-Rutherford-Trieste (BFRT) collaboration
has inferred $\gagg<6.7\times 10^{-7}\,\GeV$ (95\%~CL) for
$m_a<10^{-3}\,\eV$~\cite{Cameron:1993mr} as indicated by the
light-shaded (green) region labeled as ``Laser Experiments'' in
Fig.~\ref{Fig:AxionLimitsSearches}.

Another way to search for axions in the laboratory relies on the
prediction that axions can affect the polarization of light that
propagates in vacuum through a transverse magnetic field. In fact,
because of their coupling to photons $\gagg$, axions can induce
dispersive and absorptive processes and thereby the following two
phenomena~\cite{Maiani:1986md}: (i)~Linear dichroism, which refers to
a rotation of the polarization vector by a finite angle, and
(ii)~birefringence, which refers to the introduction of an ellipticity
and an rotation of an initially linearly polarized beam. Searching for
these phenomena, the BFRT experiment has extracted the exclusion limit
$\gagg<3.6\times 10^{-7}\,\GeV$ (95\%~CL) for $m_a<5\times
10^{-4}\,\eV$. Note that the evidence for vacuum dichroism claimed by
the PVLAS collaboration in the year 2006~\cite{Zavattini:2005tm} has
been retracted recently~\cite{Zavattini:2007ee}. Originally, they had
interpreted their findings in terms of the presence of an axion-like
particle (ALP) with a mass of $(1\!-\!1.5)\times 10^{-3}~\eV$ and a
coupling to photons in the range $(1.7\!-\!5)\times
10^{-6}\,\GeV^{-1}$~\cite{Zavattini:2005tm}.
The retraction reassures the validity of the astrophysical constraints
that had already excluded the region in which this signal was
reported.
 
With new experiments and updates underway, the next years will become
very exciting for axion searches. Both helioscope and haloscope
experiments are about to probe significant parts of complementary
parameter regions of realistic axion models. For example, an axion
signal at ADMX would support the hypothesis of $\OmegaDM$ provided by
axions from the misalignment mechanism. In contrast, an axion
observation at CAST would point to axion hot/warm dark matter which
can only provide a minor fraction of $\OmegaDM$ so that the dominant
contribution can still be provided, e.g., by the LSP if SUSY is
realized in nature. In fact, if the axino---the fermionic superpartner
of the axion---is the LSP, collider signatures predicted for the axino
LSP~\cite{Brandenburg:2005he} could become an additional hint towards
the existence of the axion and the solution of the strong CP problem
proposed by Peccei and Quinn.

\section{Neutralino Dark Matter}
\label{sec:NeutralinoDM}

In this section we consider SUSY scenarios in which the lightest
neutralino $\neutralino$ is the LSP. This hypothetical particle is
probably the most studied and most popular SUSY dark matter candidate
and a concrete example for a weakly interacting massive particle
(WIMP).  Before discussing $\neutralino$ dark matter scenarios, let us
review some generic properties of SUSY extensions of the Standard
Model. For details, we refer to dedicated reviews on
SUSY~\cite{Wess:1992cp,Nilles:1983ge,Haber:1984rc,Martin:1997ns,Drees:2004jm,Baer:2006rs}.

Extending the Standard Model with SUSY, there is a superpartner of
each Standard Model particle and an extended Higgs sector with a least
two Higgs doublets.
The couplings of these superpartners arise by supersymmetrizing the
Standard Model couplings and are thus fixed by symmetry. This allows
for model independent SUSY predictions that can be tested in collider
experiments~\cite{Brandenburg:2002ff,Freitas:2007fd,Brandenburg:2008gd}.
The masses of the Standard Model superpartners are governed by the
Higgs-higgsino mass parameter $\mu$ and by soft SUSY breaking
parameters which depend on the SUSY breaking mechanism and thereby on
physics at high energy scales such as the one of grand unification
$\mgut\simeq 2\times 10^{16}\,\GeV$. The experimental determination of
the SUSY mass spectrum and of the Higgs masses at colliders can thus
provide insights into high scale physics and into the SUSY breaking
mechanism~\cite{Baer:2000gf,Blair:2002pg,Lafaye:2004cn,Bechtle:2004pc}.

Assuming that SUSY is realized not only as global but as a local
symmetry~\cite{Wess:1992cp}, the gravitino $\gravitino$ appears as the
spin-3/2 superpartner of the graviton in addition to the Standard
Model superpartners.
The gravitino is the gauge field associated with local SUSY
transformations and a singlet with respect to the gauge groups of the
Standard Model. Its interactions---given by the supergravity
Lagrangian~\cite{Cremmer:1982en,Wess:1992cp}---are suppressed by the
(reduced) Planck scale~\cite{Amsler:2008zz}
\begin{align}
       \MPl=2.4\times 10^{18}\,\GeV \,.
\label{Eq:MPLmacro}
\end{align}
Once SUSY is broken, the extremely weak gravitino interactions are
enhanced through the super-Higgs mechanism, in particular, at
energy/mass scales that are large with respect to the gravitino mass
$\mgr$.  Nevertheless, the gravitino can be classified as an extremely
weakly interacting particle (EWIP).
Since the gravitino $\gravitino$ is the gauge field of local SUSY, its
mass~$m_{\gravitino}$ is governed by the scale of SUSY breaking and
can range from the eV scale to scales beyond the TeV
region~\cite{Nilles:1983ge,Martin:1997ns,Dine:1994vc,Dine:1995ag,Giudice:1998bp,Randall:1998uk,Giudice:1998xp,Buchmuller:2005rt}.
For example, in gauge-mediated SUSY breaking
schemes~\cite{Dine:1994vc,Dine:1995ag,Giudice:1998bp}, the mass of the
gravitino is typically less than 1~GeV, while in gravity-mediated
sche\-mes~\cite{Nilles:1983ge,Martin:1997ns} it is expected to be in
the GeV to TeV range.
In fact, SUSY scenarios in which the gravitino is the stable LSP are
well-motivated and and will be discussed in the next section.
In this section, $m_{\gravitino}$ is assumed to be above the
neutralino mass $m_{\neutralino}$. This implies an unstable
$\gravitino$ which can be associated with an additional source of
$\neutralino$ dark matter (cf.\ Sect.~\ref{sec:NeutralinoProduction})
and with restrictive BBN constraints on the reheating temperature
$\TR$ (cf.\ Sect.~\ref{sec:NeutralinoConstraints}).

The lightest neutralino $\neutralino$ appears in the minimal
supersymmetric Standard Model (MSSM) as the lightest mass eigenstate
among the four neutralinos being mixtures of the bino $\bino$, the
wino $\wino$, and the neutral higgsinos $\HiggsinoUp$ and
$\HiggsinoDown$. Accordingly, $\neutralino$ is a spin 1/2 fermion with
weak interactions only. Its mass $\mneu$ depends on the gaugino mass
parameters $M_{1}$ and $M_{2}$, on the ratio of the two MSSM Higgs
doublet vacuum expectation values $\tanb$, and on the Higgs-higgsino
mass parameter $\mu$. Expecting $\mneu=\Order(100~\GeV)$,
$\neutralino$ is classified as a WIMP.

Motivated by theories of grand unification and
supergravity~\cite{Brignole:1997dp} and by experimental constraints on
flavor mixing and CP violation~\cite{Amsler:2008zz}, one often assumes
universal soft SUSY breaking parameters at the scale of grand
unification $\mgut$;
cf.~\cite{Martin:1997ns,Drees:2004jm,Baer:2006rs,Olive:2008uf}
and references therein.  For example, in the framework of the
constrained MSSM (CMSSM), the gaugino masses, the scalar masses, and
the trilinear scalar interactions are assumed to take on the
respective universal values $\monetwo$, $\mzero$, and $A_0$ at
$\mgut$. Specifying $\monetwo$, $\mzero$, $A_0$, $\tanb$, and the sign
of $\mu$, the low-energy mass spectrum is given by the renormalization
group running from $\mgut$ downwards.

For $A_0=0$, for example, the lightest Standard Model
superpartner---or lightest ordinary superpartner (LOSP)---is either
the lightest neutralino $\neutralino$ or the lighter stau $\stau$,
whose mass is denoted by $m_{\stau}$. If the LSP is assumed to be the
LOSP, the parameter region in which $m_{\stau}<m_{\neutralino}$ is
usually not considered because of severe upper limits on the abundance
of stable charged particles~\cite{Amsler:2008zz}. However, in
gravitino/axino LSP scenarios, in which the LOSP is the
next-to-lightest supersymmetric particle (NLSP), the $\stau$ LOSP case
is viable and particularly promising for collider phenomenology as
will be discussed in Sects.~\ref{sec:GravitinoDM}
and~\ref{sec:AxinoDM}. 

\begin{figure}[t!]
\includegraphics[width=0.45\textwidth]{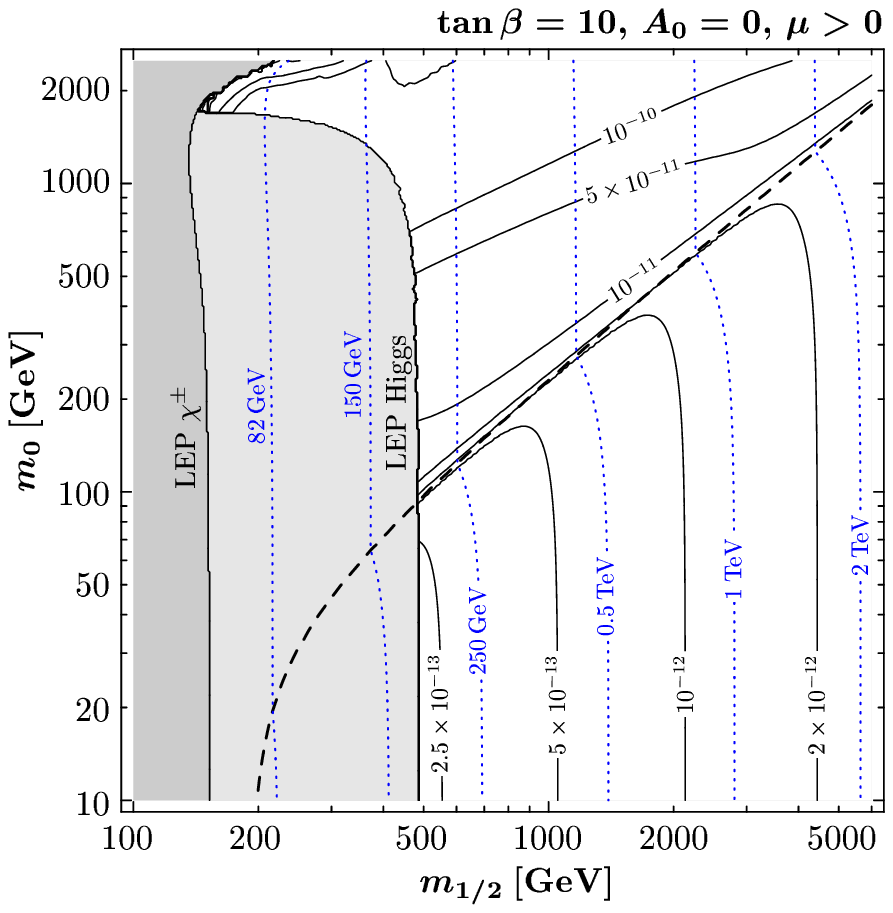} 
\caption{Contours of $m_{\LOSP}$ (dotted blue lines) and
  $Y_{\LOSP}^{\dec}$ (solid black lines) in the $(m_{1/2},m_0)$ plane
  for $A_0=0$, $\mu>0$, $\tan\beta=10$. Above (below) the dashed line,
  $m_{\neutralino}<m_{\stau}$ ($m_{\stau}<m_{\neutralino}$). The
  medium gray and the light gray regions show the LEP bounds
  $m_{\chargino}>94~\GeV$ and $m_{\Higgs}>114.4~\GeV$,
  respectively~\cite{Amsler:2008zz}. The contours are obtained with
  the spectrum generator {\tt SuSpect~2.34}~\cite{Djouadi:2002ze}
  using $m_t=172.5~\GeV$ and
  $m_{\mathrm{b}}(m_{\mathrm{b}})^{\mathrm{\overline{MS}}} = 4.23\
  \GeV$, and with {\tt
    micrOMEGAs~1.3.7}~\cite{Belanger:2001fz,Belanger:2004yn}.
  From~\cite{Pradler:2006hh}.}
\label{Fig:YLOSP}
\end{figure}

In Fig.~\ref{Fig:YLOSP} (from~\cite{Pradler:2006hh})
%
%
the dotted (blue) lines show contours of $m_{\LOSP}$ in the
$(m_{1/2},m_0)$ plane for $A_0=0$, $\mu>0$, $\tan\beta=10$. Above
(below) the dashed line, $m_{\neutralino}<m_{\stau}$
($m_{\stau}<m_{\neutralino}$). The medium gray and the light gray
regions at small $m_{1/2}$ are excluded respectively by the mass
bounds $m_{\chargino}>94~\GeV$ and $m_{\Higgs}>114.4~\GeV$ from
chargino and Higgs searches at LEP~\cite{Amsler:2008zz}. It can be
seen that $\mneu=\Order(100~\GeV)$ appears naturally within the CMSSM.

Before proceeding, let us comment on other potential LOSP/LSP/NLSP
candidates.
For $A_0\neq 0$ and in less constrained frameworks such as models with
non-universal Higgs masses (NUHM), there are parameter regions in
which the LOSP is the lighter stop
$\scalartop$~\cite{Ellis:2001nx,DiazCruz:2007fc,Santoso:2007uw} or the
lightest sneutrino
$\sneutrino$~\cite{Ibanez:1983kw,Ellis:1983ew,Hagelin:1984wv,Buchmuller:2006nx,Covi:2007xj,Ellis:2008as}.
In fact, since the lightest sneutrino $\sneutrino$ is electrically
neutral and color neutral, the $\sneutrino$ LSP looks at first sight
like another promising WIMP dark matter candidate within the MSSM.
It turns out however that its couplings (and in particular the one to
the Z-boson) are ``too strong.'' From the invisible Z-boson width, we
know that the sneutrino must have a mass $m_{\sneutrino}>\mZ/2$, where
the its relic density is typically well below $\OmegaDM$ and/or its
interactions with nuclei are such that it should have already been
observed in direct dark matter searches (assuming a standard dark
matter halo profile); cf.\ Figs.~1 and~2 in Ref.~\cite{Arina:2007tm}.
Thus, the MSSM sneutrino LSP is not considered a viable dark matter
candidate~\cite{Falk:1994es} but it may well be the NLSP, e.g., in a
gravitino/axino LSP
scenario~\cite{Fujii:2003nr,Feng:2004mt,Buchmuller:2006nx,Kanzaki:2006hm,Covi:2007xj,Ellis:2008as}.%
\footnote{Variants of the MSSM sneutrinos are actively pursued as dark
  matter candidates; see~\cite{Arina:2007tm} and references therein.
  With an admixture of less strongly interacting ``right-handed''
  sneutrinos, sneutrino dark matter interactions can become compatible
  with $\Omega_{\sneutrino}\simeq\OmegaDM$ and with constraints from
  direct searches.}
The lighter stop $\scalartop$ is not viable as a stable LSP due to severe
constraints on exotic stable colored particles~\cite{Amsler:2008zz}
but another NLSP candidate.%
\footnote{Note that the mass of a long-lived $\scalartop$ NLSP has to
  respect the collider bound, $\mstop>250~\GeV$, inferred from SUSY
  searches at the Fermilab Tevatron~\cite{Santoso:2007uw}.}
While the NLSP governs cosmological constraints and experimental
prospects in gravitino/axino dark matter scenarios in a crucial way
(cf.\ Sects.~\ref{sec:GravitinoDM} and~\ref{sec:AxinoDM}), it can also
be important in a neutralino dark matter scenario (e.g., through
coannihilation processes) as will become clear below.

\subsection{Primordial Origin}
\label{sec:NeutralinoProduction}

The $\neutralino$'s were in thermal equilibrium for primordial
temperatures of $T>T_{\freezeout}\simeq\mneu/20$. At $T_{\freezeout}$,
the annihilation rate of the (by then) non-relativistic
$\neutralino$'s becomes smaller than the Hubble rate so that they
decouple from the thermal plasma.  Thus, for $T\lesssim
T_{\freezeout}$, their yield $Y_{\neutralino}\equiv n_{\neutralino}/s$
is given by
$Y_{\neutralino}^{\dec}\approx
Y^{\equil}_{\neutralino}(T_{\freezeout})$, 
where $n_{\neutralino}^{(\equil)}$ is the (equilibrium) number density
of $\neutralino$'s. Depending on details of the $\neutralino$
decoupling, $Y_{\neutralino}^{\dec}$ is very sensitive to the mass
spectrum and the couplings of the superparticles.  Indeed, convenient
computer programs such as {\tt DarkSUSY}~\cite{Gondolo:2004sc} or {\tt
  micrOMEGAs}~\cite{Belanger:2001fz,Belanger:2004yn,Belanger:2006is,Belanger:2007zz}
are available which allow for a numerical calculation of LOSP
decoupling and of the resulting thermal relic abundance in a given
SUSY model.

\begin{figure}[t!]
${m_0}$ [{\bf GeV}]
\begin{center}
\includegraphics[width=.45\textwidth,height=.4\textwidth]{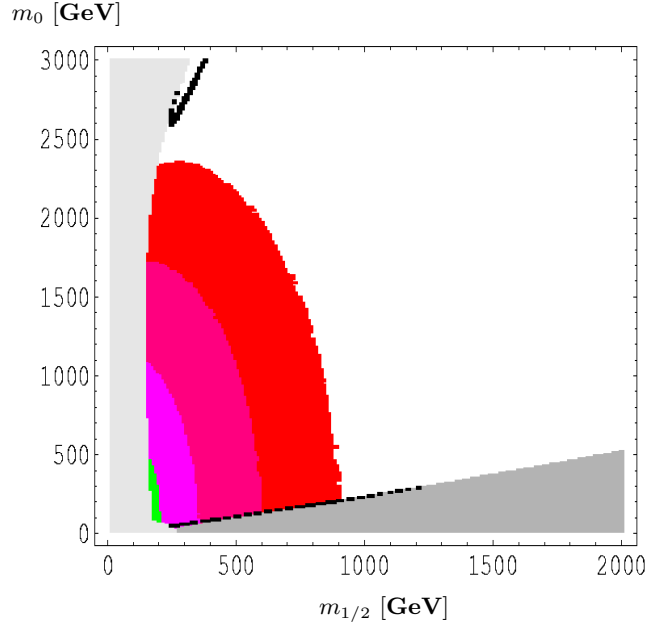}
\hspace*{1.cm} $m_{1/2}$ [{\bf GeV}]
\end{center}
\caption{Regions (black) with $0.087
  \leq\Omega_{\neutralino}^{\thermal}h^2\leq 0.138$ in the
  $(m_{1/2},m_0)$ plane for $A_0=0, \mu>0$, $\tan\beta=10$, and
  $m_t=172.7~\GeV$. In the dark gray triangular region,
  $\mneu>m_{\stau}$. The light gray region at small $m_{1/2}$ is
  excluded by the requirement of correct electroweak symmetry breaking
  or by sparticle search limits~\cite{Djouadi:2006be}, the two medium
  shaded (light pink) bands by the LEP bound $m_{\Higgs}>114~\GeV$,
  and the small light shaded (green) spot by the $b \rightarrow s
  \gamma$ constraint: $2.65\leq \mathrm{BR}(b \rightarrow s
  \gamma)/10^{-4} \leq 4.45$.  The dark shaded (red) band is
  compatible with having a Standard-Model-like Higgs boson near
  115~GeV.  From~\cite{Djouadi:2006be}.}
\label{Fig:NeutralinoDM}
\end{figure}

The $Y_{\LOSP}^{\dec}$ contours shown by the solid black lines in
Fig.~\ref{Fig:YLOSP} illustrate that the $\neutralino$ LSP yield can
easily vary by more than an order of magnitude. Because of this
sensitivity, the associated thermal relic density
\begin{equation}
  \Omega_{\neutralino}^{\thermal} h^2
  = \mneu\,Y_{\neutralino}^{\dec}\,s(T_0)\,h^2/\rho_c
\label{Eq:NeutralinoDensity}
\end{equation}
agrees with $\Omega_{\CDM}^{3\sigma}h^2$ only in narrow regions in the
parameter space; $\rho_c/[s(T_0)h^2]=3.6\times
10^{-9}\,\GeV$~\cite{Amsler:2008zz}.  This can be seen in
Fig.~\ref{Fig:NeutralinoDM} (from~\cite{Djouadi:2006be})
%
%
where the black strips indicate the region with $0.087 \leq
\Omega_{\neutralino}^{\thermal}h^2\leq 0.138$.

Remarkably, it is exactly the small width of the regions with
$\Omega_{\neutralino}^{\thermal}=\Omega_{\CDM}$, which could help us
to identify $\neutralino$ dark matter. If sparticles exist and if they
are produced at colliders, the data analysis will aim at determining
the SUSY model realized in nature~\cite{Lafaye:2004cn,Bechtle:2004pc}.
For the reconstructed model, a precise calculation of
$\Omega_{\neutralino}^{\thermal}$ is possible assuming a standard
thermal history of the Universe.  Because of the sensitivity of
$\Omega_{\neutralino}^{\thermal}$ with respect to the SUSY model, an
agreement of the obtained $\Omega_{\neutralino}^{\thermal}$ with
$\Omega_{\CDM}$ will then be a strong hint for the $\neutralino$ LSP
providing $\Omega_{\CDM}$ and for a standard thermal history up to the
$\neutralino$-decoupling temperature $T_{\freezeout}$. Since
$\neutralino$'s decouple already as a non-re\-la\-ti\-vis\-tic
species, it is also guaranteed that they are sufficiently cold to
allow for cosmic structure formation.

In fact, $\Omega_{\neutralino}^{\thermal}$ exceeds $\OmegaDM$ in most
of the parameter space. Thus, regions with
$\Omega_{\neutralino}^{\thermal}\simeq \OmegaDM$ are somewhat special
and associated with particularly efficient neutralino annihilation in
the early Universe. Depending on the origin for the efficient
annihilation, these regions can be classified as follows;
cf.~\cite{Drees:2004jm,Baer:2006rs,Djouadi:2006be,Baltz:2006fm,Olive:2008uf}
and references therein:

\begin{itemize}

\item Bulk region: This region is associated with light sleptons
  $\slepton$, $\mslepton\lesssim 200\,\GeV$, so that neutralinos can
  annihilate efficiently via slepton exchange into a lepton pair:
  $\neutralino\neutralino\to l^+ l^-$. This region is often in tension
  with the LEP Higgs bound. For example, in the CMSSM scenarios
  considered in Figs.~\ref{Fig:YLOSP} and \ref{Fig:NeutralinoDM}, the
  bulk region appears around
  $(\mzero,\monetwo)\simeq(60~\GeV,200~\GeV)$.
  
\item Focus point region/hyperbolic branch: This region is associated
  with a $\neutralino$ with a significant higgsino admixture so that
  $\neutralino\neutralino\to W^+ W^-,\,Z^0 Z^0$ become efficient.  (In
  fact, for a purely bino-like neutralino, $\neutralino=\Bino$,
  annihilation into these final states cannot occur.) In
  Fig.~\ref{Fig:NeutralinoDM}, the black region with
  $\Omega_{\neutralino}^{\thermal}\simeq\OmegaDM$ at large $m_0$ is
  associated with these neutralino annihilation channels being
  efficient.

\item Coannihilation region: In this region, the NLSP has a mass very
  close to the one of the neutralino LSP. Thereby, the number density
  of the NLSP during the freeze out of the $\neutralino$ LSP is
  sizable and $\neutralino$--NLSP coannihilation processes can enhance
  the efficiency of neutralino annihilation. In
  Fig.~\ref{Fig:NeutralinoDM}, the black region with
  $\Omega_{\neutralino}^{\thermal}\simeq\OmegaDM$ just above the dark
  gray stau LOSP region is associated with efficient
  $\neutralino$--$\stau$-coannihilation processes.  Moreover, in
  regions in which $m_{\scalartop}$ is close to $m_{\neutralino}$,
  also $\neutralino$--$\scalartop$ coannihilation can allow for
  $\Omega_{\neutralino}^{\thermal}\simeq\OmegaDM$.
  
\item Higgs funnel: In this region, $2\,m_{\neutralino}$ is close to
  the mass of the CP odd Higgs boson $\Ahiggs$, $\mA \sim
  2\,m_{\neutralino}$, so that neutralino annihilation proceeds very
  efficiently via the $\Ahiggs$ resonance. For large $\tanb$, the
  following annihilation channel becomes particularly efficient due to
  an $\tanb$-enhanced $\Ahiggs$ coupling of the $\bquark$ quark:
  $\neutralino\neutralino\to\Ahiggs\to\bquark\antibquark$.

\end{itemize}

In addition to $\Omega_{\neutralino}^{\thermal}$ from thermal freeze
out, $\neutralino$'s can also be produced non-thermally in late decays
of gravitinos. Because of their extremely weak interactions, unstable
gravitinos with $m_{\neutralino}<\mgr\lesssim 5~\TeV$ have long
lifetimes, $\tau_{\gravitino}\gtrsim 100~\seconds$ (cf.\ Fig.~1 of
Ref.~\cite{Kawasaki:2008qe}), and decay typically during or after BBN
into the LSP and into Standard Model particles.
While the decay into the LSP can proceed either directly or via a
cascade, each gravitino decays into one LSP. Thus, the resulting
non-thermally produced (NTP) neutralino density is given by
\begin{equation}
  \Omega_{\neutralino}^{\NTP} h^2
  = \mneu\,Y_{\gravitino}\,s(T_0)\,h^2/\rho_c
  \ ,
\label{Eq:NeutralinoDensityNTP}
\end{equation}
where $Y_{\gravitino}=n_{\gravitino}/s$ denotes the gravitino yield
prior to decay. Although gravitinos with
$\mgr>m_{\neutralino}=\Order(100~\GeV)$ are extremely weakly
interacting and not in thermal equilibrium with the primordial plasma,
they can be produced efficiently in thermal scattering of particles in
the hot plasma.%
\footnote{In this review I do not discuss gravitino production from
  inflaton decays which can be sub\-stan\-tial depending on the
  inflation model; see, e.g.,~\cite{Asaka:2006bv,Endo:2007sz}.}
Derived in a gauge-invariant treatment, the resulting thermally
produced (TP) gravitino yield at a temperature $\TL\ll\TR$
reads~\cite{Bolz:2000fu,Pradler:2006qh,Pradler:2006hh,Pradler:2007ne}
\begin{eqnarray}
        Y_{\gravitino}^{\TP}(\TL)
        &\equiv&
        \sum_{i=1}^3
        y_i\, g_i^2(\TR)
        \left(1+\frac{M^2_{i}(\TR)}{3\mgr^2}\right) 
\nonumber\\
        &&\quad\times\ln\left(\frac{k_i}{g_i(\TR)}\right)
        \left(\frac{\TR}{10^{10}\,\GeV} \right)
        \ ,
\label{Eq:YgravitinoTP}
\end{eqnarray}
with $y_i$, the gauge couplings $g_i$, the gaugino mass parameters
$M_i$, and $k_i$ as given in Table~\ref{Tab:Constants}.
%
%
\begin{table}[t]
  \caption{Assignments of the index $i$, the gauge coupling $g_i$, and 
    the gaugino mass parameter $M_i$, to the gauge groups
    U(1)$_\Hypercharge$, SU(2)$_\Weak$, and SU(3)$_\Color$,
    and the constants $k_i$, $y_i$, and $\omega_i$.}
  \label{Tab:Constants}
\begin{center}
\renewcommand{\arraystretch}{1.25}
\begin{tabular*}{3.25in}{@{\extracolsep\fill}cccccccc}
\hline
gauge group         & $i$ & $g_i$ & $M_i$  &  $k_i$ &  $(y_i/10^{-12})$ & $\omega_i$ 
\\ \hline
U(1)$_\Hypercharge$ & 1 & $g'$    & $M_1$  & 1.266  & 0.653 & 0.018 
\\
SU(2)$_\Weak$       & 2 & $g$     & $M_2$   & 1.312  & 1.604 & 0.044 
\\
SU(3)$_\Color$ & 3 & $g_\mathrm{s}$ & $M_3$ & 1.271  & 4.276 & 0.117 
\\
\hline
\end{tabular*}
\end{center}
\end{table}
%
%
Here $M_i$ and $g_i$ are understood to be evaluated at the reheating
temperature %
after inflation $T_{\Reheating}$~\cite{Pradler:2006qh}.%
\footnote{Note that the field-theoretical methods applied in the
  derivation of (\ref{Eq:YgravitinoTP})
  \cite{Bolz:2000fu,Pradler:2006qh,Pradler:2006hh,Pradler:2007ne}
  require weak couplings $g_i\ll 1$ and thus $T \gg 10^6~\GeV$. For an
  alternative approach, see~\cite{Rychkov:2007uq}.}
Using $Y_{\gravitino}=Y_{\gravitino}^{\TP}(\TL)$ as given
by~(\ref{Eq:YgravitinoTP}), one finds that
$\Omega_{\neutralino}^{\NTP}$ is sensitive to the gravitino mass
$\mgravitino$ and to the reheating temperature $\TR$. Thus, the dark
matter constraint
$\Omega_{\neutralino}^{\thermal}+\Omega_{\neutralino}^{\NTP}\leq\OmegaDM$
does imply an upper limit on $\TR$.
This limit is particularly restrictive in scenarios with
$\Omega_{\neutralino}^{\thermal}\simeq\OmegaDM$~\cite{Kohri:2005wn}
and/or $\mgravitino^2\ll M_i^2(\TR)$.
Since non-thermally produced $\neutralino$'s can be hot/warm dark
matter, additional constraints from LSS and potential solutions to
small scale structure problems can occur, in particular, for
$\Omega_{\neutralino}^{\NTP}\simeq
\OmegaDM$~\cite{Hisano:2000dz,Lin:2000qq}.


Let us comment at this point on the $\TR$ definition as discussed in
Ref.~\cite{Pradler:2006hh}. The analytic
expression~(\ref{Eq:YgravitinoTP}) is derived by assuming a
radiation-dominated epoch with an initial temperature of
$\TR$~\cite{Bolz:2000fu,Pradler:2006qh}.
In a numerical treatment, the epoch in which the coherent oscillations
of the inflaton field dominate the energy density of the Universe can
also be taken into account, where one usually defines $\TR$ in terms
of the decay width $\Gamma_{\phi}$ of the inflaton
field~\cite{Kawasaki:2004qu,Pradler:2006hh}. In fact, the numerical
result for $Y_{\gravitino}^{\TP}(\TL)$ agrees with the analytic
expression~(\ref{Eq:YgravitinoTP}) for~\cite{Pradler:2006hh}
\begin{equation}
  \TR 
  \simeq
  \left[\frac{90}{g_*(\TR)\pi^2}\right]^{1/4}
  \sqrt{\frac{\Gamma_{\phi}\MPl}{1.8}}
\label{Eq:TR_definition}
\end{equation}
which satisfies $\Gamma_{\phi}\simeq 1.8 H_{\mathrm{rad}}(\TR)$ with
the Hubble parameter
$H_{\mathrm{rad}}(T)=\sqrt{g_*(T)\pi^2/90}\, T^2/\MPl$ 
and an effective number of relativistic degrees of freedom of
$g_*(\TR)=228.75$.
Thus, (\ref{Eq:TR_definition}) provides the $\TR$ definition to which
the yield~(\ref{Eq:YgravitinoTP}) applies. For an alternative
definition
$\TR^{[\xi]} \equiv [90/(g_*(\TR^{[\xi]})\,\pi^2)]^{1/4}
\sqrt{\Gamma_{\phi}\MPl/\xi}$
given by $\Gamma_{\phi}=\xi H_{\mathrm{rad}}(\TR^{[\xi]})$,
the associated numerically obtained $Y_{\gravitino}^{\TP}(\TL)$ is
reproduced by the analytical expression~(\ref{Eq:YgravitinoTP}) after
substituting $\TR$ with
$\sqrt{\xi/1.8}\,\TR^{[\xi]}$~\cite{Pradler:2006hh}.

\begin{figure}
  \centerline{\includegraphics[width=0.45\textwidth]{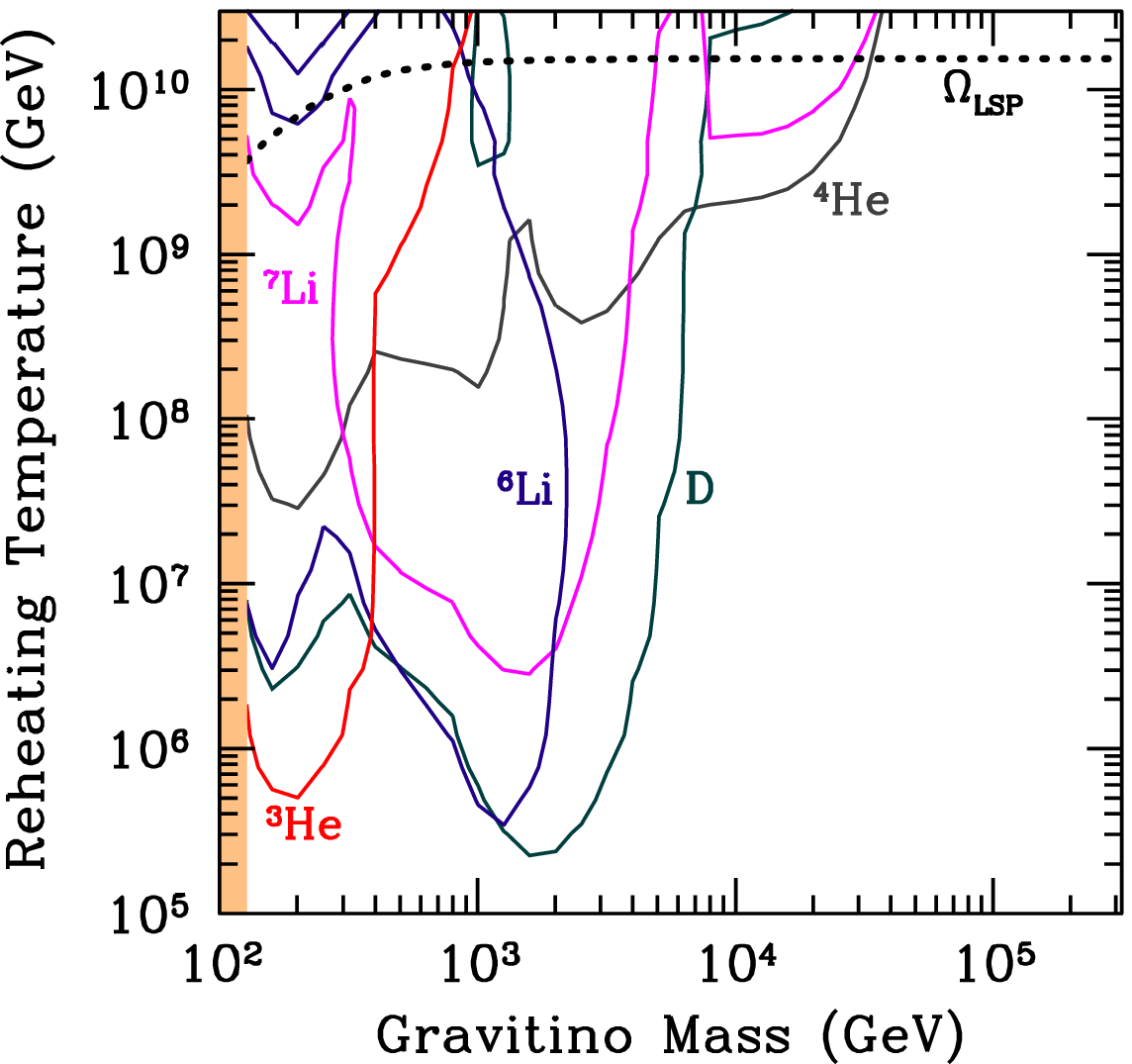}}
  \caption{Upper limits on the reheating temperature $\TR$ after
    inflation as a function of the gravitino mass $\mgr$ in the CMSSM
    with $\neutralino$ dark matter for $\monetwo=300~\GeV$,
    $\mzero=141~\GeV$, $A_0=0$, $\tanb=30$ and for the $\TR$
    definition associated with $\Gamma_{\phi}\simeq 3
    H_{\mathrm{rad}}(\TR)$.  The associated $\neutralino$ LSP mass is
    $m_{\neutralino}=117~\GeV$---as indicated by the shaded
    (light-orange) region in which $\mgravitino\leq
    m_{\neutralino}$---and the thermal relic density
    $\Omega_{\neutralino}^{\thermal}h^2=0.111$. Above the dotted line
    labeled as ``$\Omega_{\LSP}$'', the $\neutralino$ density from
    decays of thermally produced gravitinos exceeds
    $\Omega_{\neutralino}^{\NTP}h^2=0.118$. The lines labeled as
    $\dm$, $\hetm$, $\hefm$, $\lisxm$, and $\lisvm$ are upper limits
    on $\TR$ (95\% CL) inferred from observationally inferred
    primordial abundances of the corresponding light elements. Above
    those limits, BBN is reprocessed in an intolerable way by the
    Standard Model particles emitted in late decays of thermally
    produced gravitinos.  From~\cite{Kawasaki:2008qe}.}
  \label{Fig:NeutralinoConstraints}
\end{figure}

Figure~\ref{Fig:NeutralinoConstraints} (from~\cite{Kawasaki:2008qe})
%
%
%
shows the $\TR$ limit imposed by $\Omega_{\neutralino}^{\NTP}h^2 \leq
0.118$ (dotted line labeled as ``$\Omega_{\LSP}$'') as a function of
$\mgravitino$ for the CMSSM scenario with $\monetwo=300~\GeV$,
$\mzero=141~\GeV$, $A_0=0$, and $\tanb=30$.
Here the $\TR$ limits apply to the $\TR$ definition associated with
$\xi=3$, i.e., $\Gamma_{\phi}\simeq 3 H_{\mathrm{rad}}(\TR)$.
In the shaded (light-orange) region, $\mgravitino\leq
m_{\neutralino}=117~\GeV$.
The $\TR$ limit becomes more restrictive for $\mgravitino\to
m_{\neutralino}$ due to the $\mgr$-dependent goldstino component
in~(\ref{Eq:YgravitinoTP}).
For large $\mgravitino$, this spin-1/2 component becomes negligible
and the $\TR$ limit is governed by the $\mgravitino$-independent
spin-3/2 contribution in~(\ref{Eq:YgravitinoTP}).
Note that the shown limit is conservative since the thermal relic
density associated with this SUSY scenario,
$\Omega_{\neutralino}^{\thermal}h^2=0.111$, is not taken into account.

\subsection{Cosmological Constraints}
\label{sec:NeutralinoConstraints}

Late decaying gravitinos are not only associated with the
contribution~(\ref{Eq:NeutralinoDensityNTP}) to
$\Omega_{\neutralino}h^2$ but also with the injection of energetic
Standard Model particles. Because of the long $\gravitino$ lifetime of
$\tau_{\gravitino}>1~\seconds$ for $\mgr\lesssim
20~\TeV$~\cite{Kawasaki:2008qe}, those decay products are emitted
during/after BBN and can thus affect the abundances of the primordial
light
elements~\cite{Ellis:1984er,Cyburt:2002uv,Jedamzik:2004er,Kawasaki:2004qu,Kohri:2005wn,Jedamzik:2006xz,Kawasaki:2008qe}.
In fact, this is a concrete (and probably the most prominent) example
for the non-thermal BBN-affecting processes mentioned in the
Introduction.

The dominant mechanism affecting BBN depends on $\tau_{\gravitino}$,
or, more generally, on the time $t$ at which the electromagnetic or
hadronic energy is injected into the Universe.  For $1\,\seconds
\lesssim t \lesssim 100\,\seconds$, energetic hadrons are stopped
efficiently through electromagnetic interactions so that the direct
destruction of light elements is subdominant. The presence of
additional slow hadrons still can change the ratio of protons to
neutrons through interconversion processes and thus affect the
abundance of the light elements. For $100\,\seconds \lesssim t
\lesssim 10^{7}\,\seconds$, energetic hadrons and, in particular,
neutrons cannot be slowed down significantly.  Accordingly, they can
reprocess efficiently the produced light elements through
hadrodissociation processes. The effect of electromagnetic energy
release is negligible for $t \lesssim 10^{4}\,\seconds$ as the
interaction with the background particles thermalizes quickly any
high-energy photons or leptons emitted in the gravitino decay. Towards
later times, electromagnetic energy release becomes important. For
$10^{7}\,\seconds \lesssim t \lesssim 10^{12}\,\seconds$, the
reprocessing of light elements through energetic electromagnetic
showers, i.e., photodissociation, can become more significant than
hadrodissociation. (For more details, see,
e.g.,~\cite{Reno:1987qw,Kawasaki:2004qu} and references therein.)

Including these mechanisms in calculations of BBN, observationally
inferred abundances of primordial $\dm$, $\hetm$, $\hefm$, $\lisxm$,
and $\lisvm$ have been used to provide limits on quantities such
as~\cite{Cyburt:2002uv,Kawasaki:2004qu,Jedamzik:2006xz}
\begin{align}
        \xi_{\EM,\HAD} \equiv \epsilon_{\EM,\HAD}\, Y_{\gravitino}
        \ ,
\label{Eq:EnergyRelease}
\end{align}
where $\epsilon_{\EM,\HAD}$ is the (average) electromagnetic/hadronic
energy emitted in a single $\gravitino$ decay.
While the yield prior to decay $Y_{\gravitino}$ is given for thermally
produced gravitinos by~(\ref{Eq:YgravitinoTP}), $\epsilon_{\EM,\HAD}$
depends strongly on the sparticle spectrum. Once $\epsilon_{\EM,\HAD}$
is calculated for a given SUSY model, $\xi_{\EM,\HAD}$ limits can
basically be translated into $\mgravitino$-depen\-dent upper limits on
the reheating temperature~\cite{Kohri:2005wn,Kawasaki:2008qe}.

For the exemplary CMSSM point $\monetwo=300~\GeV$, $\mzero=141~\GeV$,
$A_0=0$, and $\tanb=30$, the $\mgravitino$-dependent BBN constraints
on $\TR$ (95\% CL) are shown in Fig.~\ref{Fig:NeutralinoConstraints}
(from~\cite{Kawasaki:2008qe})
for the $\TR$ definition given by $\Gamma_{\phi}\simeq 3
H_{\mathrm{rad}}(\TR)$.
The curves are inferred from observationally inferred abundances of
$\dm$, $\hetm$, $\hefm$, $\lisxm$, and $\lisvm$ (as labeled).
The $\hefm$ limit governs the BBN constraints for $\mgravitino\gtrsim
7~\TeV$ ($\tau_{\gravitino}\lesssim 10^2\,\seconds$) where
proton-neutron interconversion can affect BBN.
The $\dm$ ($\hetm$) limit is the most restrictive one in the region in
which the constraints from hadrodissociation (photodissociation) are
most relevant, $0.3~\TeV\lesssim\mgravitino\lesssim 7~\TeV$
($\mgravitino\lesssim 0.3~\TeV$) or $10^2\,\seconds
\lesssim\tau_{\gravitino}\lesssim 10^{7}\,\seconds$
($\tau_{\gravitino}\gtrsim 10^{7}\,\seconds$).
For $\mgravitino\gtrsim 40~\TeV$, the BBN bounds disappear since the
$\gravitino$'s decay well before the onset of BBN, i.e.,
$\tau_{\gravitino}\ll 1~\seconds$.

The range of allowed values of the reheating temperature is crucial
for our understanding of inflation and for the viability of potential
explanations of the matter-antimatter asymmetry in our Universe.
For example, thermal leptogenesis with hierarchical heavy right-handed
Majorana neutrinos---which provides an attractive explanation of this
asymmetry---requires very high reheating temperatures of $\TR\gtrsim
10^9\,\GeV$~\cite{Fukugita:1986hr,Davidson:2002qv,Buchmuller:2004nz,Blanchet:2006be,Antusch:2006gy}
and thus $\mgravitino\gtrsim 7~\TeV$ for the SUSY model considered in
Fig.~\ref{Fig:NeutralinoConstraints}. For smaller $\mgravitino$, the
$\TR$ limit can be as restrictive as $\TR<10^6\,\GeV$ which is known
as the gravitino problem.
Note that flavor
effects~\cite{Nardi:2006fx,Abada:2006fw,Blanchet:2006be,Antusch:2006gy}
do not change the lower bound $T_{\Reheating} > 10^9\,\GeV$ required
by successful thermal leptogenesis with hierarchical right-handed
neutrinos~\cite{Blanchet:2006be,Antusch:2006gy}. However, there are
special settings~\cite{Davidson:2003yk,Hambye:2003rt,Raidal:2004vt}
that allow for a CP asymmetry above the Davidson--Ibarra
bound~\cite{Davidson:2002qv} and for a relaxed lower $\TR$ bound.
Moreover, for (nearly) mass-degenerate heavy right-handed Majorana
neutrinos, resonant leptogenesis can explain the mat\-ter-antimat\-ter
asymmetry at smaller values of
$T_{\Reheating}$~\cite{Flanz:1994yx,Covi:1996fm,Pilaftsis:1997jf,Anisimov:2005hr}.
Another example for a framework in which the limit $T_{\Reheating} >
10^9\,\GeV$ is relaxed is non-thermal leptogenesis; see, e.g.,
\cite{HahnWoernle:2008pq} and references therein.

Before proceeding one should stress that the $\TR$ limits inferred
from $\Omega_{\neutralino}^{\NTP}\leq\OmegaDM$ and from BBN rely
crucially on assumptions on the cosmological history and the evolution
of physical parameters.
For example, for a non-standard thermal history with late-time entropy
production, the thermally produced gravitino yield can be diluted
$Y_{\gravitino}\to Y_{\gravitino}/\delta$ by a factor
$\delta>1$~\cite{Pradler:2006hh} so that $\TRmax\to\delta\,\TRmax$.
Moreover, $\TRmax$ can be relaxed if, e.g., the strong coupling $g_s$
levels off in a non-standard way at high
temperatures~\cite{Buchmuller:2003is}.
This emphasizes that the $\TR$ limits discussed above rely on the
assumptions of a standard cosmological history and of gauge couplings
that behave at high temperatures as described by the renormalization
group equation in the MSSM.
While tests of these assumptions seem inaccessible to terrestrial
accelerator experiments, the futuristic space-based gravitational-wave
detectors BBO or DECIGO~\cite{Seto:2001qf}---mentioned already in the
Introduction---could allow for tests of the thermal history after
inflation and could even probe
$\TR$~\cite{Nakayama:2008ip,Nakayama:2008wy} in a complementary way.

\subsection{Experimental Searches and Prospects}
\label{sec:NeutralinoExperiments}

For experimental tests of the $\neutralino$ dark matter hypothesis,
three complementary techniques exist: indirect, direct, and collider
searches.  While there is an enormous activity in each of those
fields, I will summarize only the main ideas. For more detailed
discussions,
see~\cite{Baltz:2006fm,Bertone:2007ki,Baudis:2007dq,Hooper:2007vy,deBoer:2008iu}
and references therein.

Let us first turn to indirect searches. Since dark matter clumps, one
expects regions with an increased $\neutralino$ density such as galaxy
halos, the center of galaxies, and the center of stars.  While
$\neutralino$ pair annihilation well after $\neutralino$ decoupling is
basically negligible for calculations of $\Omega_{\neutralino}$, it
should occur at a significant rate in these regions.  The resulting
Standard Model particles should then lead to energetic cosmic rays and
thereby to an excess of photons, neutrinos, positrons, and antiprotons
over backgrounds expected from standard cosmic ray models without dark
matter annihilation.  In fact, for example, data from the Energetic
Gamma Ray Experiment Telescope (EGRET) has already been interpreted as
evidence for $\neutralino$ annihilation~\cite{deBoer:2005bd} within
SUSY models that will be testable in direct and collider searches.
For a discussion of these and other potential hints,
see~\cite{Hooper:2007vy,Bertone:2007ki,deBoer:2008iu} and references
therein.

In direct searches, one looks for signals of $\neutralino$'s---or more
generally WIMPs---passing through Earth that scatter elastically off
nuclei. Being located in environments deep underground that are well
shielded against unwanted background, an enormous sensitivity has been
reached by a number of
experiments~\cite{Angloher:2004tr,Sanglard:2005we,Akerib:2005kh,Benetti:2007cd,Alner:2007ja,Angle:2007uj,Ahmed:2008eu,Lang:2008fa,Angloher:2008jj}.
Since no unambiguous signal of a $\neutralino$--nucleus scattering
event has been observed so far, $\mneu$-dependent upper limits on the
respective $\neutralino$ cross section are obtained.

\begin{figure}[t!]
  \centerline{\includegraphics[width=0.45\textwidth,clip=true,angle=270]{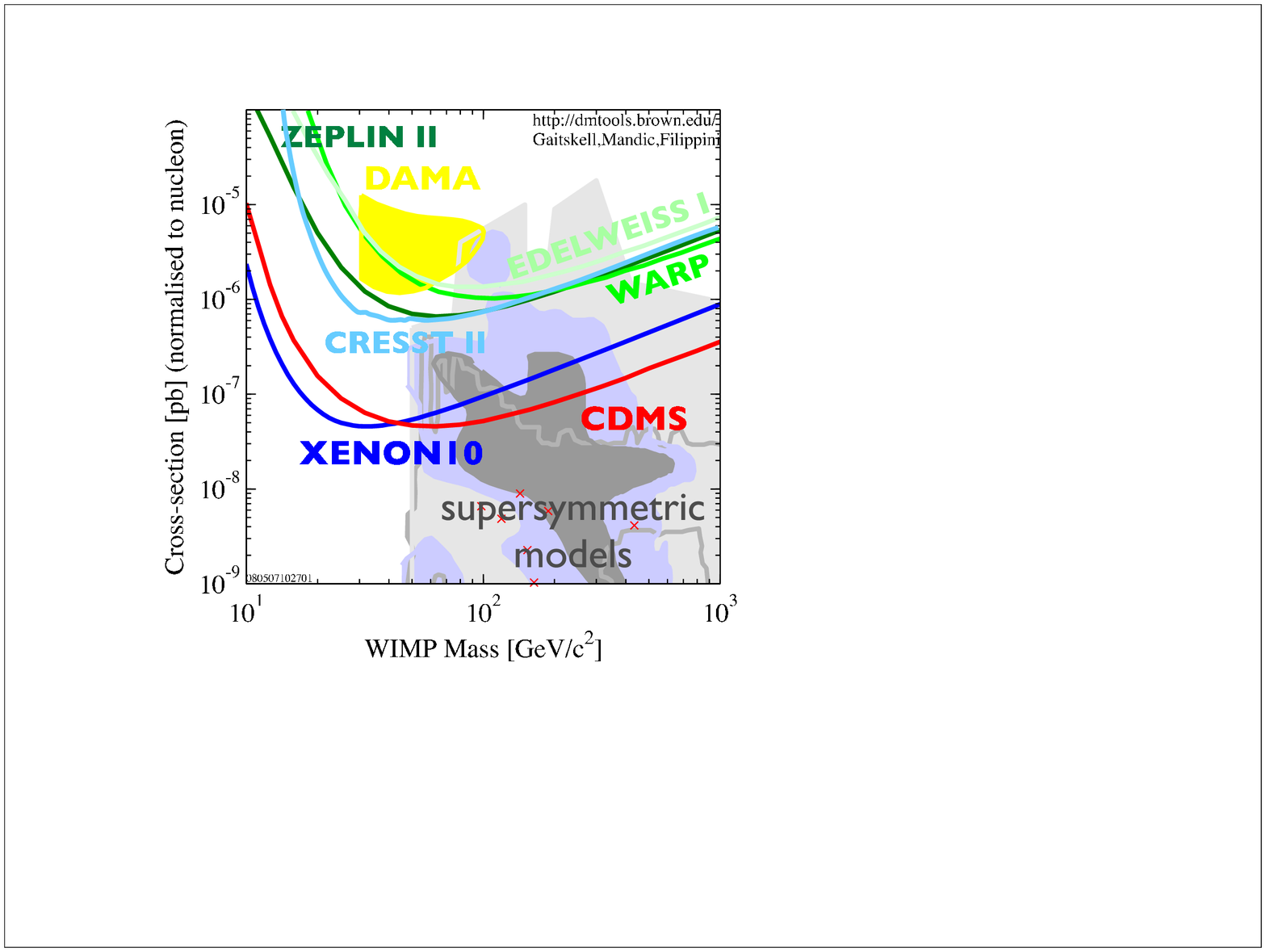}}
  \caption{Limits on the spin-independent neutralino--nucleon cross
    section from direct searches as a function of the neutralino mass
    (or WIMP mass). The shown limits are provided by
    CDMS~\cite{Ahmed:2008eu} (red), XENON10~\cite{Angle:2007uj} (dark
    blue), CRESST~II~\cite{Lang:2008fa,Angloher:2008jj} (light blue),
    ZEPLIN~II~\cite{Alner:2007ja} (dark green),
    WARP~\cite{Benetti:2007cd} (medium green), and
    EDELWEISS~I~\cite{Sanglard:2005we} (light green) (from bottom to
    top at a WIMP mass of $80~\GeV$, as labeled). The DAMA signal
    region~\cite{Bernabei:2000qi,Bernabei:2008yi} is indicated by the
    light-shaded patch (yellow, as labeled) and the (favored)
    parameter space of SUSY models with a $\neutralino$
    LSP~\cite{Baltz:2002ei,Baltz:2004aw,Ellis:2001hv,Battaglia:2003ab,Roszkowski:2007fd}
    by the shadings/contours/crosses at WIMP masses
    $m_{\neutralino}>40~\GeV$ (gray/purple/red, as labeled).  Standard
    galactic halo parameter are assumed, i.e., a local halo density of
    dark matter of $\rho_0=0.3~\GeV/\cm^3$ and a characteristic halo
    velocity of $v_0=220\!-\!240~\km/\seconds$. The figure was build
    by using the dark matter plotter available at {\tt
      http://dmtools.berkeley.edu/limitplots/} and maintained by
    R.~Gaitskell and J.~Filippini.}
  \label{Fig:NeutralinoDirectSearches}
\end{figure}

Figure~\ref{Fig:NeutralinoDirectSearches} shows current limits on the
spin-independent neutralino--nucleon cross section provided by
CDMS~\cite{Ahmed:2008eu} (red), XENON10~\cite{Angle:2007uj} (dark
blue), CRESST~II~\cite{Lang:2008fa,Angloher:2008jj} (light blue),
ZEPLIN~II~\cite{Alner:2007ja} (dark green), WARP~\cite{Benetti:2007cd}
(medium green), and EDELWEISS~I~\cite{Sanglard:2005we} (light green)
(from bottom to top at a WIMP mass of $80~\GeV$, as labeled).  The
light (yellow) patch indicates the region in which the DAMA experiment
is reporting the observation of a signal with the expected annual
modulation~\cite{Bernabei:2000qi,Bernabei:2008yi}. As can be seen,
this DAMA signal region is in a region in which null events were
observed by ZEPLIN~II, CRESST~II, CDMS, and XENON10. The DAMA signals
might thus be interpreted as signals of a ``non-standard'' dark matter
candidate (see e.g.~\cite{Bernabei:2005ca}), where cosmological and
astrophysical constraints can allow for crucial viability tests (see
e.g.~\cite{Pospelov:2008jk}); see also
~\cite{Savage:2008er,Feng:2008dz,Feng:2008qn} and references therein.
%
%
The various patches, crosses, and contours within the light gray
region labeled as ``supersymmetric models'' indicate the (favored)
parameter space of SUSY models with a $\neutralino$
LSP~\cite{Baltz:2002ei,Baltz:2004aw,Ellis:2001hv,Battaglia:2003ab,Roszkowski:2007fd}.
%
%
As can be seen, the current best limits given by the
CDMS~\cite{Akerib:2005kh,Ahmed:2008eu} and the
XENON10~\cite{Angle:2007uj} experiments disfavor already a sizable
part of the SUSY parameter space; see, for
example,~\cite{Bertone:2007ki,Baudis:2007dq,Olive:2008uf} and
references therein.  These limits, however, depend on the assumed
$\neutralino$ flux at the detector location. Standard galactic halo
parameter are assumed, i.e., a local halo density of dark matter of
$\rho_0=0.3~\GeV/\cm^3$ and a characteristic halo velocity of
$v_0=220\!-\!240~\km/\seconds$. Indeed, those assumptions are subject
to significant uncertainties due to possible inhomogeneities in the
dark matter distribution in galaxies. Such inhomogeneities should
manifest themselves also in indirect searches which can help to reduce
those uncertainties. Once $\neutralino$ events are observed in direct
searches, one can succeed in reconstructing the $\neutralino$ velocity
distribution~\cite{Drees:2007hr}. By analyzing the recoil spectra,
$\mneu$ can even be estimated in a way that is independent of the dark
matter density on Earth~\cite{Shan:2007vn}.

In most searches for SUSY at colliders, it is assumed that R-parity is
conserved. Accordingly, one expects that superpartners are produced in
pairs before decaying via cascades into the LSP and energetic
fermions. As a weakly-interacting particle, every $\neutralino$ LSP
produced will escape the detector without leaving a track. Thus, the
existence of SUSY and the $\neutralino$ LSP has to be inferred from
studies of missing transverse energy $\MET$ and of energetic jets and
leptons emitted along the cascades. Along these lines, ongoing
investigations are pursued based on data from $\proton\antiproton$
collisions with a center-of-mass energy of $\sqrt{s}=2~\TeV$ at the
Fermilab Tevatron Collider. While lower limits on the masses of
squarks and gluinos have been extracted, no evidence for SUSY or the
$\neutralino$ LSP has been reported so
far~\cite{Duperrin:2007uy,Shamim:2007yy}.
With the first $\proton\proton$ collisions with $\sqrt{s}=14~\TeV$ at
the CERN Large Hadron Collider (LHC) expected in the year 2009, there
are high hopes that the new energy range will allow for a copious
production of superpartners. Here large $\MET$ will be the key
quantity for early SUSY searches~\cite{Tytgat:2007gj,Yamamoto:2007it}.
Despite an enormous potential for mass and spin measurements of SUSY
particles at the LHC~\cite{Ozturk:2007ap}, additional precision
studies at the planned International Linear Collider
(ILC)~\cite{Weiglein:2004hn,Choi:2008hh} appear to be crucial for the
identification of the $\neutralino$
LSP~\cite{Baltz:2006fm,Choi:2006mr}.

\section{Gravitino Dark Matter}
\label{sec:GravitinoDM}

The gravitino $\gravitino$ has already been introduced in the previous
section and its appearance is an unavoidable implication of SUSY
theories including gravity.
In this section we consider the possibility of the gravitino LSP which
is well motivated, for example, in gauge-mediated and gravity-mediated
SUSY breaking schemes~\cite{Dine:1994vc,Dine:1995ag,Giudice:1998bp}.
Indeed, without consensus on the SUSY breaking mechanism and the SUSY
breaking scale, one may well consider the gravitino mass $\mgr$ as a
free parameter to be constrained by cosmological considerations (cf.\
Sects.~\ref{sec:GravitinoProduction}
and~\ref{sec:GravitinoConstraints}) and by collider experiments (cf.\
Sect.~\ref{sec:GravitinoExperiments}).%
\footnote{In scenarios with R-parity violation, the gravitino LSP can
  decay into Standard Model particles. Then, one may be able to infer
  its mass $\mgr$ from the decay spectra possibly observed in indirect
  dark matter searches~\cite{Ibarra:2008sa}.}

Being a singlet with respect to the gauge groups of the Standard
Model, the gravitino LSP is a promising dark matter candidate that can
be classified as an EWIP as mentioned above. In fact, it must not be
massive since even a light gravitino (e.g., $\mgr=1~\keV$) can evade
its production at colliders because of its tiny interaction strength.

Let us recall important characteristics of gravitino interactions:
(i)~Gravitino interactions are suppressed by inverse powers of $\MPl$.
Indeed, for example, gravitino-gau\-gino-gauge boson couplings are
described by dimension five operators and an energy scale appears in
the numerator of the respective vertex. Gravitino interactions can
thereby be enhanced at very high energies. (ii)~Through the
super-Higgs mechanism, the interactions of the spin-1/2 components of
the gravitino (i.e., the interactions of the goldstino components) are
enhanced at energy/mass scales that are large with respect to the
gravitino mass $\mgr$, i.e., a light gravitino interacts more strongly
than a heavy gravitino.%
\footnote{Note that this interaction-strength dependence on the mass
  is different than in the axion case, i.e., a light axion is less
  strongly interacting than a heavy axion; cf.\
  Sect.~\ref{sec:AxionDM}.}

Considering the case of the $\gravitino$ LSP, in which the LOSP is the
unstable NLSP that decays eventually into the $\gravitino$ LSP, both
cases $\mneu<\mst$ and $\mst<\mneu$ (cf.~Fig.~\ref{Fig:YLOSP}) are
viable as already mentioned. %
In less constrained frameworks such as models with non-universal Higgs
masses (NUHM), also other LOSP/NLSP candidates are still possible such
as the lighter stop $\scalartop$~\cite{DiazCruz:2007fc,Santoso:2007uw}%
\footnote{A long-lived stop $\widetilde{t}_1$ NLSP is not feasible in
  the CMSSM~\cite{DiazCruz:2007fc,Santoso:2007uw}.}
 or the lightest sneutrino
$\sneutrino$~\cite{Feng:2004mt,Buchmuller:2006nx,Kanzaki:2006hm,Covi:2007xj,Ellis:2008as}.
However, the $\stau$ NLSP case is probably the most promising one from
the phenomenological point view and is discussed more extensively than
the other NLSP cases in this review.

\subsection{Primordial Origin}
\label{sec:GravitinoProduction}

The potential primordial origin of gravitino dark matter depends on
the mass $\mgr$ that governs its interaction strength, on the SUSY
model, on the cosmological history, on the inflation model, and on the
reheating temperature after inflation.
The gravitino LSP can be a thermal relic~\cite{Pagels:1981ke} or be
produced in thermal scattering of particles in the primordial
plasma~\cite{Moroi:1993mb,Bolz:1998ek,Bolz:2000fu,Pradler:2006qh,Pradler:2006hh,Rychkov:2007uq}.
Additional more model dependent gravitino sources are NLSP
decays~\cite{Borgani:1996ag,Asaka:2000zh,Ellis:2003dn,Feng:2004mt} and
decays of scalar fields such as the
inflaton~\cite{Asaka:2006bv,Endo:2007sz}. The latter production
mechanism is not discussed in this review but can be substantial
depending on the inflation model.

Light gravitinos can have sufficiently strong interactions for being
in thermal equilibrium with the primordial plasma. For example, for
$\mgr\lesssim 2~\keV$, the gravitino decoupling temperature is
$T_{\freezeout}\lesssim 1~\TeV$; cf.~\cite{Drees:2004jm}. Before
gravitinos decouple as a relativistic species at $T_{\freezeout}$,
i.e., for $\TR>T>T_{\freezeout}$, the spin-1/2 components of the
gravitino were in thermal equilibrium. Thus, their ``hot'' thermal
relic density is the one of a spin-1/2 Majorana fermion
\begin{align}
        \Omega_{\gravitino}^{\thermal} h^2 
        = 0.115\,
        \left(\frac{100}{g_{*S}(T_{\freezeout})}\right)\,
        \left(\frac{m_{\gravitino}}{100~\eV}\right)
\label{Eq:GravitinoDensityEq}
\end{align}
and the right amount of dark matter,
$\Omega_{\gravitino}^{\thermal}\simeq\OmegaDM$, 
is provided for
$m_{\gravitino}\simeq 100~\eV$ and $g_{*S}(T_{\freezeout})\simeq 100$.
However, the present root mean square velocity of such gravitinos
\begin{align}
        (v_{\FS}^{\rms,0})_{\gravitino}
        = 0.77\,\frac{\km}{\seconds}\,
        \left(\frac{\Omega_{\gravitino}^{\thermal}h^2}{0.113}\right)^{1/3}\!\!
        \left(\frac{100~\eV}{m_{\gravitino}}\right)^{4/3}
\label{Eq:v0WDM}
\end{align}
%
exceeds significantly the constraints from observations and
simulations of cosmic structures listed, e.g., in Table~1 of
Ref.~\cite{Steffen:2006hw}. 
In fact, for $\Omega_{\gravitino}^{\thermal}\simeq\OmegaDM$, these
constraints imply $m_{\gravitino}> 500~\eV$ and
$g_{*S}(T_{\freezeout})>500$ which is well above the $228.75$ degrees
of freedom of the MSSM. Indeed, for $g_{*S}(T_{\freezeout})=228.75$, this
scenario is excluded by the dark matter constraint
$\Omega_{\gravitino}\leq\OmegaDM$ once a standard cosmological history
is assumed.
With a non-standard thermal history, light gravitinos can still be
viable thermal relics if their abundance is diluted by entropy
production, which can result, for example, from decays of messenger
fields in gauge-mediated SUSY breaking
scenarios~\cite{Baltz:2001rq,Fujii:2002fv,Fujii:2003iw,Lemoine:2005hu,Jedamzik:2005ir,Moultaka:2007pv}.

Gravitinos with $\mgravitino\gtrsim 0.1~\GeV$ ($1~\GeV$) have a high
decoupling temperature of $T_{\freezeout}> 10^{11}\,\GeV$
($10^{13}\,\GeV$)~\cite{Pradler:2006hh} because of their extremely
weak interactions.
Thus, those gravitinos have never been in thermal equilibrium with the
primordial plasma even for a reheating temperature as high as
$\TR=10^{10}\,\GeV$.
At high temperatures, however, those gravitinos can be produced
efficiently in thermal scattering of particles in the primordial
plasma, as already discussed for the case of an unstable $\gravitino$
in Sect.~\ref{sec:NeutralinoProduction}.
In the case of the stable $\gravitino$ LSP, the thermally produced
gravitino yield $Y_{\gravitino}^{\TP}(T_0)=Y_{\gravitino}^{\TP}(\TL)$
given in~(\ref{Eq:YgravitinoTP}) leads to the following thermally
produced (TP) gravitino
density~\cite{Bolz:2000fu,Pradler:2006qh,Pradler:2006hh}
\bea
        \Omega_{\gravitino}^{\TP}h^2
        &=&
        \mgravitino\,Y_{\gravitino}^{\TP}(T_0)\,s(T_0)\,h^2/\rho_c 
\nonumber\\
        &=&
        \sum_{i=1}^{3}
        \omega_i\, g_i^2(\TR)
        \left(1+\frac{M_i^2(\TR)}{3\mgr^2}\right)
        \ln\left(\frac{k_i}{g_i(\TR)}\right)
\nonumber\\
        &&
        \times
        \left(\frac{\mgr}{100~\GeV}\right)
        \left(\frac{\TR}{10^{10}\,\GeV}\right)
\label{Eq:GravitinoDensityTP}
\eea
with $\omega_i$ as given in Table~\ref{Tab:Constants}.
For the case of universal gaugino masses $M_{1,2,3}=\monetwo$ at
$\mgut$ and $\mgravitino \ll M_i$, i.e., $(1+M_i^2/3\mgr^2)\simeq
M_i^2/3\mgr^2$, $\Omegatp h^2$ can be approximated by the convenient
expression~\cite{Pradler:2007is}%
\footnote{Here the one-loop evolution described by the renormalization
  group equation in the MSSM is used to evaluate $g_i(\TR)$ at a
  representative scale of $\TR=10^8\,\GeV$ and to express $M_i(\TR)$
  at that scale in terms of $M_{1,2,3}(\mgut)=\monetwo$.  Note that
  the use of the two-loop evolution will lead to a somewhat smaller
  prefactor. Going in the other direction, i.e., using the two-loop
  evolution to express $M_3(\TR)$ in terms of the physical gluino
  mass, a pre\-factor has been found that is about twice as
  large~\cite{Buchmuller:2008vw} as the one obtained when the one-loop
  evolution is used to express $M_3(\TR)$ in terms of
  $M_3(1~\TeV)$~\cite{Bolz:2000fu}.}
\begin{align}
  \Omegatp h^2 \simeq 0.32 \Big( \frac{10\ \GeV}{\mgr} \Big) 
  \Big( \frac{\monetwo}{1\ \TeV} \Big)^2 \Big(
    \frac{\TR}{10^{8}\ \GeV} \Big) .
  \label{eq:omega-tp}
\end{align}
The thermally produced gravitinos do not affect the thermal evolution
of the LOSP (or NLSP) prior to its decay which occurs typically after
decoupling from the thermal plasma. Moreover, since each NLSP decays
into one $\gravitino$ LSP, the NLSP decay leads to a non-thermally
produced (NTP) gravitino
density~\cite{Borgani:1996ag,Asaka:2000zh,Feng:2003xh,Feng:2004mt}
\bea
        \Omega_{\gravitino}^{\NTP} h^2
        &=& 
        \mgravitino\, Y_{\NLSP}^{\dec}\, s(T_0) h^2 / \rho_{\mathrm{c}}
\label{Eq:GravitinoDensityNTP}
\eea
so that
the guaranteed gravitino density is given by %
 \begin{align}
   \Omega_{\gravitino}h^2=\Omega_{\gravitino}^{\TP}h^2+\Omega_{\gravitino}^{\NTP}h^2 
   \ .
   \label{Eq:GravitinoDensity}
 \end{align}

While $\Omegatp$ is sensitive to $M_i$ and $\TR$ for a given
$\mgravitino$, $\Omegantp$ depends on
$Y_{\NLSP}^{\dec}=Y_{\LOSP}^{\dec}$ and thereby on details of the SUSY
model realized in nature; cf.\ Sect.~\ref{sec:NeutralinoProduction}.
For the case of a charged slepton $\slepton$---such as the lighter
stau $\stau$---being the NLSP,%
\footnote{For $Y_{\NLSP}^{\dec}$ in the sneutrino and stop NLSP cases,
  see Refs.~\cite{Fujii:2003nr,Feng:2004mt,Arina:2007tm,Ellis:2008as}
  and~\cite{Fujii:2003nr,Kang:2006yd,DiazCruz:2007fc,Berger:2008ti},
  respectively.  Reference~\cite{Berger:2008ti} covers also the stau
  NLSP case for $\stau\simeq\stauR$ and the effects of Sommerfeld
  enhancement on $Y_{\NLSP}^{\dec}$.}
simple approximations have been used such
as~\cite{Asaka:2000zh,Fujii:2003nr,Steffen:2006hw,Steffen:2006wx,Pradler:2007is,Pospelov:2008ta}
\begin{align}
        Y_{\slepton}^{\dec}
        \simeq
        0.7 \times 10^{-12}
        \left(\frac{m_{\slepton}}{1~\TeV}\right)
        \ ,
\label{Eq:Yslepton}
\end{align}
where $Y_{\slepton}\equiv n_{\slepton}/s$ and $n_{\slepton}$ denotes
the total $\slepton$ number density assuming an equal number density
of positively and negatively charged $\slepton$'s.
Note that the yield~(\ref{Eq:Yslepton}) is in good agreement with the
curve in Fig.~1 of Ref.~\cite{Asaka:2000zh} that has been derived for
the case of a purely `right-handed' $\stau\simeq\stauR$ NLSP with a
mass that is significantly below the masses of the lighter selectron
and the lighter smuon, $m_{\stau} \ll m_{\sel,\smu}$, and with a
bino-like lightest neutralino, $\neutralino\simeq\Bino$, that has a
mass of $m_{\Bino}=1.1\,m_{\stau}$.
In the case of an approximate slepton mass degeneracy, $m_{\stau}
\lesssim m_{\sel,\smu} \lesssim 1.1\,m_{\stau}$, the $\stau$ NLSP
yield~(\ref{Eq:Yslepton}) can become twice as large due to slepton
coannihilation processes~\cite{Asaka:2000zh,Pradler:2006hh}.
Approaching the $\neutralino$--$\stau$ coannihilation region, $\mneu
\approx \mst$, even larger enhancement factors occur; cf.\
Fig.~\ref{Fig:YLOSP}.
On the other hand, while existing studies of $Y_{\slepton}^{\dec}$
focus mainly on the $\slepton\simeq\sleptonR$
case~\cite{Asaka:2000zh,Fujii:2003nr,Pradler:2006hh,Berger:2008ti}, it
has recently been found that a sizable left--right mixing of the stau
NLSP can be associated with an increase of its MSSM couplings and thus
with a significant reduction of
$Y_{\stau}^{\dec}$~\cite{Ratz:2008qh,Pradler:2008qc}.
The study~\cite{Pradler:2008qc} shows also explicitly that stau
annihilation at the resonance of the heavy CP even Higgs $\Hhiggs$ can
be particularly efficient via
$\stau\stau\to\Hhiggs\to\bquark\antibquark$ and thus be associated
with exceptionally small values of $Y_{\stau}^{\dec}$.

\begin{figure}[t!]
  \centerline{\includegraphics[width=0.45\textwidth]{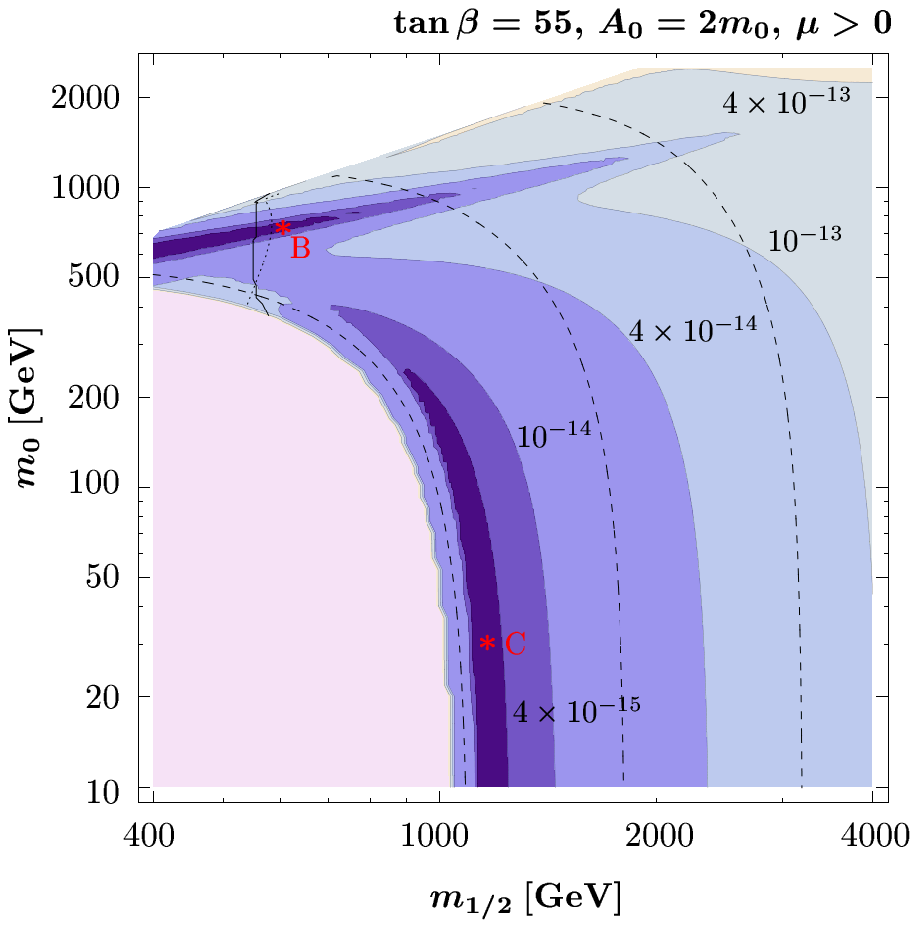}}
  \caption{Contours of $Y_{\stau}^{\dec}$ (as labeled) in the
    $(\monetwo,\,\mzero)$ plane for $\tanb =55$, $A_0 = 2m_0$, and
    $\mu>0$. Darker shadings imply smaller $Y_{\stau}^{\dec}$ values.
    The dashed lines are contours of $m_{\stau}=100,\, 300,$ and $600\
    \GeV$ (from left to right).  The large light-shaded region in the
    lower left corner is excluded by bounds from direct Higgs and SUSY
    searches (or by the appearance of a tachyonic spectrum).  In the
    region to the left of the vertical solid and dotted lines, $\mh
    \le 114.4\ \GeV$~\cite{Amsler:2008zz} and $B(b \rightarrow s
    \gamma)\ge 4.84\times 10^{-4}$~\cite{Mahmoudi:2007gd},
    respectively.  In the white area, $m_{\neutralino} < m_{\stau}$.
    From~\cite{Pradler:2008qc}.}
  \label{Fig:ExceptionalStauYields}
\end{figure}

Figure~\ref{Fig:ExceptionalStauYields} (from~\cite{Pradler:2008qc})
shows that $Y_{\stau}^{\dec}<4\times 10^{-15}$ occurs in special
regions even within the CMSSM.
%
%
The points B and C mark the regions in which $Y_{\stau}^{\dec}$ is
exceptionally small due to stau annihilation at the $\Hhiggs$
resonance and due to enhanced stau-Higgs couplings leading to
efficient annihilation into Higgs bosons, respectively.
Within these regions, $\Omega_{\gravitino}^{\NTP}$ is negligible and
otherwise restrictive cosmological constraints can be evaded as will
be explained in more detail in the next section.

An exceptional reduction of $Y_{\slepton}^{\dec}$ can occur also in a
non-standard thermal history with late-time entropy production after
the decoupling of the $\slepton$ NLSP and before
BBN~\cite{Buchmuller:2006tt,Pradler:2006hh,Hamaguchi:2007mp} or in low
$\TR$ scenarios~\cite{Takayama:2007du}.
Focussing on a standard cosmological history, we disregard such
possibilities and consider in the following mainly the more generic
$Y_{\slepton}^{\dec}$ values described approximately
by~(\ref{Eq:Yslepton}).

\begin{figure}[t!]
\includegraphics[width=.45\textwidth]{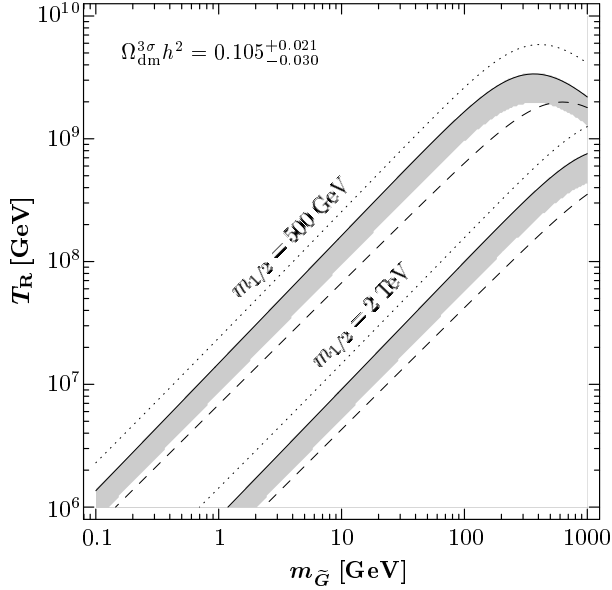} 
\caption{Upper limits on the reheating temperature $T_{\Reheating}$ in
  the $\gravitino$ LSP case, which are associated with the $\TR$
  definition~(\ref{Eq:TR_definition}).  On the upper (lower) gray
  band, $\Omega_{\widetilde{G}}^{\TP}\in\Omega_{\CDM}^{3\sigma}$ for
  $M_{1,2,3}=m_{1/2}=500~\GeV$ ($2~\TeV$) at $M_{\GUT}$. The
  corresponding limits from $\Omega_{\widetilde{G}}^{\TP}h^2\leq
  0.126$ shown by the dashed and dotted lines are obtained
  respectively with~(\ref{Eq:GravitinoDensityTP}) for
  $M_1/10=M_2/2=M_3=m_{1/2}$ at $M_{\GUT}$ and with the result of
  Ref.~\cite{Bolz:2000fu} for $M_3=m_{1/2}$ at $M_{\GUT}$.
  From~\cite{Pradler:2006hh}.}
\label{Fig:UpperLimitTR}
\end{figure}

\begin{figure}[t!]
  \includegraphics[width=0.45\textwidth]{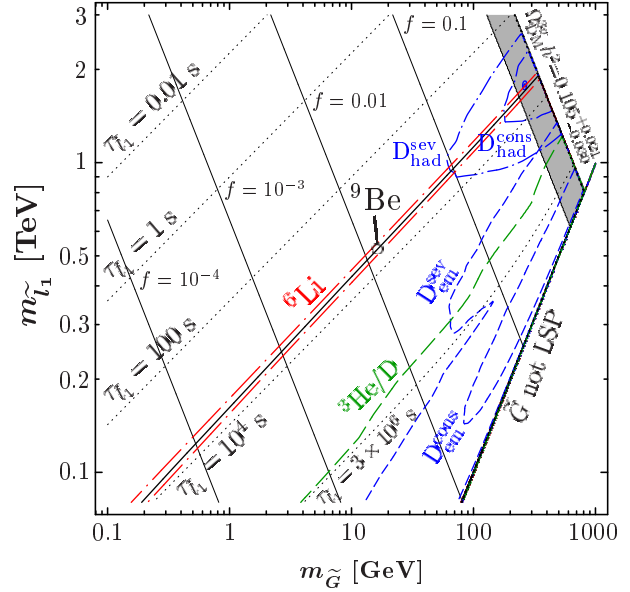}
  \caption{Cosmological constraints on the masses of the gravitino LSP
    and a purely `right-handed' $\slepton$ NLSP
    with~(\ref{Eq:Yslepton}). The gray band indicates
    $\Omega_{\gravitino}^{\NTP}\!\!\in\Omega_{\CDM}^{3\sigma}$. Above
    this band, $\Omega_{\gravitino}h^2 > 0.126$. On the thin solid
    lines labeled with $f$ values only $f\,\OmegaDM$ is provided by
    $\Omega_{\gravitino}^{\NTP}$.  The dotted lines show contours of
    $\tau_{\slepton}$. Due to CBBN, the region below the solid and the
    long-dash-dotted (red) lines is disfavored by observationally
    inferred abundances of $^9$Be and $^6$Li,
    respectively~\cite{Pospelov:2008ta}. The effect of electromagnetic
    and hadronic energy injection on primordial D disfavors the
    regions inside the short-dash-dotted (blue) curves and to the
    right or inside of the short-dashed (blue) curves, respectively.
    Those curves are obtained from the severe and conservative upper
    limits defined in Sect.~4.1 of~\cite{Steffen:2006hw} based on
    results of~\cite{Cyburt:2002uv,Kawasaki:2004qu}.  The region below
    the dashed (green) line is disfavored by the effect of
    electromagnetic energy injection
    on~$^3\mathrm{He}/\mathrm{D}$~\cite{Kawasaki:2004qu}.  While the
    constraints from hadronic energy injection are obtained for a
    purely `right-handed' $\slepton\simeq\sleptonR$ NLSP, the ones
    from electromagnetic energy injection are valid for the $\stau$
    NLSP case with a visible electromagnetic energy of
    $E_{\mathrm{vis}}=\epsilon_{\EM}=0.3 E_{\tau}$ released in
    $\stau\to\gravitino\tau$. From \cite{Pospelov:2008ta}.}
\label{Fig:GravitinoMassBounds}
\end{figure}

\begin{figure}[t!]
\includegraphics[width=.45\textwidth]{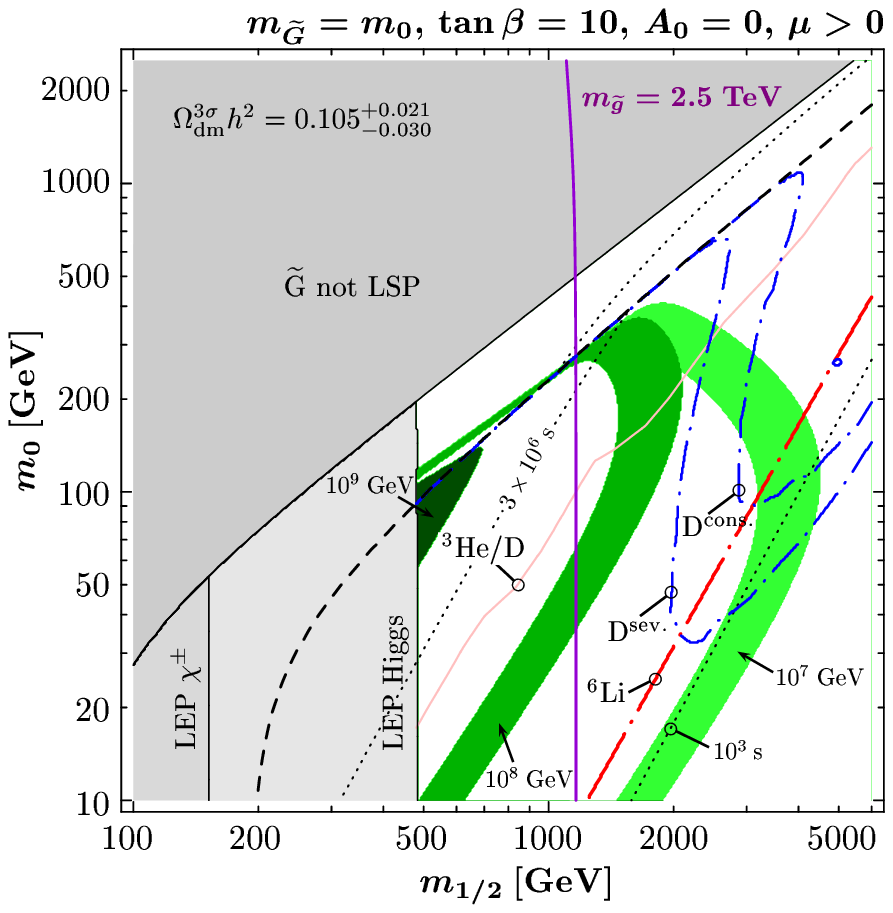} 
\caption{CMSSM regions with
  $\Omega_{\gravitino}h^2\in\Omega_{\CDM}^{3\sigma}$ for $\TR=10^7$,
  $10^8$, and $10^9~\GeV$ indicated respectively by the light, medium,
  and dark shaded (green) bands in the $(m_{1/2},m_0)$ planes for
  $\tan\beta=10$, $A_0=0$, $\mu>0$, and $\mgr=m_0$. The $\TR$ values
  are associated with the $\TR$ definition given
  in~(\ref{Eq:TR_definition}). The regions excluded by the chargino
  and Higgs mass bounds and the line indicating
  $m_{\neutralino}=m_{\stau}$ are identical to the ones shown in
  Fig.~\ref{Fig:YLOSP}.  In the dark gray region, the gravitino is not
  the LSP. The dotted lines show contours of the NLSP lifetime.  The
  region to the left of the long-dash-dotted (red) line and to the
  left of the thin gray (pink) line is disfavored by the
  observationally inferred abundances of primordial
  $^6$Li~\cite{Pradler:2007is}
  and~$^3\mathrm{He}/\mathrm{D}$~\cite{Kawasaki:2004qu}.  The effect
  of hadronic energy injection on primordial D~\cite{Steffen:2006hw}
  disfavors the $\stau$ NLSP region above the short-dash-dotted (blue)
  lines.  The $\neutralino$ NLSP region is disfavored by BBN
  constraints from energy
  injection~\cite{Ellis:2003dn,Feng:2004mt,Cerdeno:2005eu,Cyburt:2006uv}.
  On the solid vertical line (violet) $m_{\widetilde{g}}=2.5\ \TeV$.
  From~\cite{Pradler:2007ar}.}
\label{Fig:CMSSMtB10}
\end{figure}

Let us proceed with the discussion of the relic gravitino density.
Indeed, scenarios with $\Omega_{\gravitino}=\Omega_{\CDM}$ are found
for natural mass spectra and for a wide range of $\mgr$--$\TR$
combinations.
This is illustrated in Figs.~\ref{Fig:UpperLimitTR},
\ref{Fig:GravitinoMassBounds}, and~\ref{Fig:CMSSMtB10}.

In $\gravitino$ LSP scenarios, upper limits on $T_{\Reheating}$ can be
derived since $\Omega_{\gravitino}^{\TP}\leq \Omega_{\CDM}$%
~\cite{Moroi:1993mb,Asaka:2000zh,Cerdeno:2005eu,Steffen:2006hw,Pradler:2006hh,Choi:2007rh,Steffen:2008bt}.
These $\mgr$-dependent limits are shown in Fig.~\ref{Fig:UpperLimitTR}
(from \cite{Pradler:2006hh}) for the $\TR$
definition~(\ref{Eq:TR_definition}).
They can be confronted with inflation models and with potential
explanations of the cosmic matter-antimatter asymmetry in the same way
as in the case of the $\neutralino$ LSP (cf.\
Fig.~\ref{Fig:NeutralinoConstraints}) in
Sect.~\ref{sec:NeutralinoConstraints}.
%
%
For a given $\Omega_{\gravitino}^{\TP}$ and a $\slepton$ NLSP
with~(\ref{Eq:Yslepton}), the bound
$\Omega_{\gravitino}^{\NTP}\leq\Omega_{\CDM}-\Omega_{\gravitino}^{\TP}$
gives upper limits on $\mgr$ and $m_\slepton$. Those limits are shown
by the thin solid lines in Fig.~\ref{Fig:GravitinoMassBounds} (from
\cite{Pospelov:2008ta}).
%
%
In Fig.~\ref{Fig:CMSSMtB10} (from~\cite{Pradler:2007ar}) regions with
$\Omega_{\gravitino}\in\Omega_{\CDM}^{3\sigma}$ are shown for
$\TR=10^7$, $10^8$, and $10^9\,\GeV$, where $\TR$ is defined as
in~(\ref{Eq:TR_definition}). Here both $\Omega_{\gravitino}^{\TP}$ and
$\Omega_{\gravitino}^{\NTP}$ are taken into account for~$\mgr=m_0$
within the CMSSM.
%

While thermally produced gravitinos have a negligible free--streaming
velocity today, gravitinos from NLSP decays can be warm/hot dark
matter.  In the $\stau$ NLSP case, for example, upper limits on the
free--streaming velocity from simulations and observations of cosmic
structures exclude
$\mst\lesssim 0.7~\TeV$
for
$\Omega_{\gravitino}^{\NTP}\!\!\simeq\Omega_{\CDM}$~\cite{Steffen:2006hw}.
Such scenarios (gray band in Fig.~\ref{Fig:GravitinoMassBounds}),
however, require%
\footnote{Note that the BBN constraints discussed below can point to a
  $\stau$ NLSP mass of $\mst\gtrsim 2~\TeV$ for
  $\Omega_{\gravitino}^{\NTP}\!\!\simeq\Omega_{\CDM}$ as shown in
  Fig.~\ref{Fig:GravitinoMassBounds}. In the part of the gray band in
  which $\mst\lesssim 3~\TeV$ and in which the BBN constraints are
  respected, the present velocity of gravitinos emitted in $\stau$
  decays is between $0.004$ and $0.01~\km/\seconds$ which is
  comparable to that of a thermal relic warm dark matter species with
  a mass between $1$ and $5~\keV$; cf.\ Fig.~14 of
  Ref.~\cite{Steffen:2006hw}.}
$\mst\gtrsim 0.7~\TeV$ anyhow and could even resolve the small scale
structure problems inherent to cold dark
matter~\cite{Cembranos:2005us,Kaplinghat:2005sy,Jedamzik:2005sx}.
%

\subsection{Cosmological Constraints}
\label{sec:GravitinoConstraints}

In the $\gravitino$ LSP case with conserved R-parity, the NLSP can
have a long lifetime $\tau_{\NLSP}$.%
\footnote{For the case of broken R-parity, see, e.g.,~\cite{Takayama:2000uz,Buchmuller:2007ui,Ibarra:2007jz}.}
This is illustrated by the dotted $\tau_{\NLSP}$ contours in
Figs.~\ref{Fig:GravitinoMassBounds} and~\ref{Fig:CMSSMtB10}.
In particular, for a $\slepton$ NLSP, one finds in the limit $m_{l}\to
0$,
\begin{equation}
  \tau_{\slepton} 
  \simeq 
  \Gamma^{-1}(\slepton\to\gravitino l)
  = 
  \frac{48 \pi \mgr^2 \MPl^2}{m_{\slepton}^5}\!
  \left(\!1-\frac{\mgravitino^2}{m_{\slepton}^2}\right)^{\!\!\!\!-4},
\label{Eq:StauLifetime}
\end{equation}
which holds not only for a charged slepton NLSP but also for the
sneutrino NLSP $\sneutrino$. Expressions for the lifetimes of the
$\neutralino$ NLSP and the $\scalartop$ NLSP are given, e.g., in
Sect.~IIC of Ref.~\cite{Feng:2004mt} and in Sect.~2.2 of
Ref.~\cite{DiazCruz:2007fc}, respectively.

If the NLSP decays into the $\gravitino$ LSP occur during or after
BBN, the Standard Model particles emitted in addition to the gravitino
can affect the abundances of the primordial light elements as
explained in Sect.~\ref{sec:NeutralinoConstraints} for scenarios with
late-decaying $\gravitino$'s.
Indeed, these BBN constraints disfavor the $\neutralino$ NLSP for
$\mgr\gtrsim
100~\MeV$~\cite{Feng:2004mt,Cerdeno:2005eu,Cyburt:2006uv}.
For the charged slepton NLSP case, the BBN constraints associated with
electromagnetic/hadronic energy injection have also been considered
and found to be much weaker but still significant in much of the
parameter
space~\cite{Feng:2004mt,Cerdeno:2005eu,Steffen:2006hw,Cyburt:2006uv,Pradler:2006hh}.
This can be seen for a $\slepton\simeq\sleptonR$ NLSP
with~(\ref{Eq:Yslepton}) in Fig.~\ref{Fig:GravitinoMassBounds}, where
the constraints from electromagnetic and hadronic energy release are
shown respectively by the short-dashed (blue, labeled as
$\dm_{\EM}^{\mathrm{sev,cons}}$) and long-dashed (green, labeled as
$\hetm/\dm$) lines and by the short-dash-dotted (blue, labeled as
$\dm_{\HAD}^{\mathrm{sev,cons}}$) lines.
For the $\slepton$ NLSP within the CMSSM with $Y_{\slepton}^{\dec}$
calculated by {\tt micrOMEGAs 1.3.7}
\cite{Belanger:2001fz,Belanger:2004yn}, the electromagnetic
$\hetm/\dm$ and the hadronic $\dm_{\HAD}^{\mathrm{sev,cons}}$
constraints are also shown respectively by the thin gray (pink) and
the short-dash-dotted (blue) lines in Fig.~\ref{Fig:CMSSMtB10}.%
\footnote{The electromagnetic constraints shown in
  Figs.~\ref{Fig:GravitinoMassBounds} and~\ref{Fig:CMSSMtB10} apply to
  the $\stau$ NLSP and are obtained from Fig.~42 of
  Ref.~\cite{Kawasaki:2004qu} for a `visible' electromagnetic energy
  of $E_{\mathrm{vis}}=\epsilon_{\EM}= 0.3\,E_{\tau}$ of the $\tau$
  energy $E_{\tau}=(\mst^2-\mgr^2+m_{\tau}^2)/2\mst$ released in
  $\stau\to\gravitino\tau$.}
In the $\sneutrino$ NLSP case, there are basically no electromagnetic
constraints while the hadronic ones are similar to those in the
charged slepton NLSP case; see
Refs.~\cite{Feng:2004mt,Kanzaki:2006hm,Kawasaki:2008qe,Ellis:2008as}
for details.
In the $\scalartop$ NLSP case, there can be electromagnetic and
ha\-dron\-ic energy injection. Indeed, the (average) hadronic energy
emitted in a single $\scalartop$ decay $\epsilon_{\HAD}$ can be
relatively large.  However, since BBN constraints apply often to
combinations such as $\epsilon_{\HAD}Y_{\scalartop}^{\dec}$, this can
be compensated to some extend by a relatively small
$Y_{\scalartop}^{\dec}$, which results from efficient annihilation due
to the strong coupling of the colored
$\scalartop$~\cite{Fujii:2003nr,DiazCruz:2007fc,Berger:2008ti}.
In fact, a $\scalartop$ NLSP can experience color confinement due to
its color charge and thus allows for intriguing non-trivial
scenarios~\cite{Kang:2006yd,DiazCruz:2007fc,Santoso:2007uw}.

An additional constraint on electromagnetic energy injection can be
inferred from the observed Planck spectrum of the
CMB~\cite{Hu:1992dc,Lamon:2005jc}. This CMB constraint is not shown.
According to the results of~\cite{Lamon:2005jc} for the case of a
$\stauR$ NLSP, this limit is everywhere less severe than the severe
electromagnetic limit $D_{\EM}^{\mathrm{sev}}$ given by the
short-dashed (blue) line in Fig.~\ref{Fig:GravitinoMassBounds}.

As already mentioned in the Introduction, also the mere presence of a
long-lived negatively charged particle $\champ$---such as a long-lived
$\slepton^-$---can lead to bound states that catalyze BBN reactions
and can thereby be associated with BBN constraints.
In fact, the possibility that BBN can be affected by bound-state
formation of an $\champ$ with primordial nuclei had already been
realized almost twenty years
ago~\cite{Dimopoulos:1989hk,DeRujula:1989fe,Rafelski:1989pz}.
It was however only less than three years ago~\cite{Pospelov:2006sc}
when it was realized that bound-state formation of $\champ$ with
$\Hefour$ can lead to a substantial production of primordial $\Lisix$
via the CBBN reaction
\begin{align}
(\Hefour\champ)+\deuterium \rightarrow \Lisix + \champ
\ .
\label{Eq:CBBNLiSix}
\end{align}
Since then, there has been a considerable effort to refine various
aspects of CBBN and to understand its implications in the framework of
specific
models~\cite{Kohri:2006cn,Kaplinghat:2006qr,Cyburt:2006uv,Steffen:2006wx,Pradler:2006hh,Hamaguchi:2007mp,Bird:2007ge,Kawasaki:2007xb,Takayama:2007du,Jittoh:2007fr,Jedamzik:2007cp,Pradler:2007is,Kersten:2007ab,Pradler:2007ar,Jedamzik:2007qk,Kusakabe:2007fu,Kusakabe:2007fv,Pospelov:2007js,Kawasaki:2008qe,Steffen:2008bt,Pospelov:2008ta,Kamimura:2008fx}.
In particular, it has been pointed out in~\cite{Pospelov:2007js} that
there is also the possibility of efficient $\ben$ production via a
radiative fusion of $\hef$ and $(\hef\champ)$ leading to
$(\beetm\champ)$, which can capture a neutron resonantly,%
\footnote{The large $\ben$-production cross section reported and used
  in Refs.~\cite{Pospelov:2007js,Pospelov:2008ta} has very recently
  been questioned by Ref.~\cite{Kamimura:2008fx}, in which a study
  based on a four-body model is announced as work in progress to
  clarify the efficiency of $\ben$ production.}
\bea
\hef + (\hef\champ) & \rightarrow & (\beetm\champ)+\gamma
\label{Eq:RadFusion}
\\
(\beetm\champ)+n & \rightarrow & \ben+\champ \ .
\label{Eq:ResNCapture}
\eea
The efficiency of CBBN of both $\Lisix$ and $\ben$ depends strongly on
the abundance of $\champ$ at the relevant times, which is given by
$Y_{\slepton^-}^{\dec}$ and $\tau_{\slepton}$ for $\champ=\slepton^-$,
i.e., in the slepton NLSP case.
Observationally inferred limits on the primordial abundances of both
\lisx\ and \ben\ can thus be used to extract
$\tau_{\slepton}$-dependent upper limits on $Y_{\slepton^-}^{\dec}$.
Indeed, from the limits
$\Lisix/\mathrm{H}|_{\mathrm{obs}} \leq 10^{-11}\!-\!10^{-10}$ (cf.~\cite{Cyburt:2002uv,Asplund:2005yt,Jedamzik:2007qk})
and 
$\ben/\mathrm{H}|_{\mathrm{obs}} \leq 2.1\times 10^{-13}$~\cite{Pospelov:2008ta},
the $\tau_{\slepton}$-dependent bounds on $Y_{\slepton^-}^{\dec}$
shown in Fig.~5 of Ref.~\cite{Pospelov:2008ta} have been obtained.
Confronting~(\ref{Eq:Yslepton}) with those bounds, one finds the
$\ben$ and $\Lisix$ constraints shown respectively by the solid and by
the long-dash-dotted (red) lines in
Fig.~\ref{Fig:GravitinoMassBounds}.
The long-dash-dotted (red) line in Fig.~\ref{Fig:CMSSMtB10} shows the
constraint associated with $\Lisix/\mathrm{H} |_{\mathrm{obs}}
\lesssim 2\times 10^{-11}$ \cite{Cyburt:2002uv} as obtained from the
CBBN treatment of~\cite{Pradler:2007is} with $Y_{\slepton}^{\dec}$
from {\tt micrOMEGAs}.

For a typical yield~(\ref{Eq:Yslepton}), the CBBN constraints
associated with $\Lisix$ and $\ben$
imply~\cite{Pospelov:2006sc,Hamaguchi:2007mp,Bird:2007ge,Takayama:2007du,Pradler:2007is,Pospelov:2008ta}
\begin{align}
\tau_\slepton \lesssim 5\times 10^3\;\seconds .
\label{Eq:tauslepton}
\end{align}
While numerous other CBBN reactions can also affect the abundances of
$\Lisix$, $\ben$, and other primordial
elements~\cite{Cyburt:2006uv,Bird:2007ge,Jedamzik:2007cp,Jedamzik:2007qk,Pospelov:2007js,Pospelov:2008ta,Kamimura:2008fx},
the approximate $\tau_\slepton$ bound is relatively robust. 
In particular, the possibility of allowed islands in the parameter
region with large $Y_{\slepton}^{\dec}$/large
$\tau_{\slepton}$---which was advocated to remain viable in
Ref.~\cite{Jedamzik:2007cp}---does not exist~\cite{Pospelov:2008ta}.
The finding of relaxed $Y_{\slepton^-}^{\dec}$ limits at long
lifetimes $\tau_\slepton$ in Ref.~\cite{Jedamzik:2007cp} did rely on
the presence of $(p\,\slepton^-)$-bound states and on a claimed
significant reprocessing of $\Lisix$ by nuclear reactions such as
$(p\,\slepton^-)+\Lisix\to\slepton^-+\hefm+\hetm$.
However, it is clarified in Ref.~\cite{Pospelov:2008ta} that the
presence of $(p\,\slepton^-)$-bound states cannot relax the
$Y_{\slepton^-}^{\dec}$ limits at long lifetimes $\tau_\slepton$ in
any substantial way because of charge exchange reactions such as
$(p\,\slepton^-)+\hefm\to(\hefm\,\slepton^-)+p$ and
$(p\,\slepton^-)+\Lisix\to(\Lisix\,\slepton^-)+p$ that can be very
efficient~\cite{Pospelov:2008ta}.
This finding has recently been confirmed in quantum three-body
calculations~\cite{Kamimura:2008fx}.

As can be seen in Fig.~\ref{Fig:GravitinoMassBounds}, the cosmological
constraints provide an upper bound on $\mgravitino$ once $m_{\st}$ is
measured. This bound implies upper bounds on the SUSY breaking scale,
on $\Omega_{\gravitino}^{\NTP}$, and---as can be seen in
Fig.~\ref{Fig:UpperLimitTR}---on $\TR$.
Figure~\ref{Fig:CMSSMtB10} shows that the cosmological constraints
imply not only an upper limit on $\TR$~\cite{Pradler:2006hh} but also
a lower limit on $\monetwo$~\cite{Cyburt:2006uv,Pradler:2006hh}.
Indeed, $\mgr$-dependent limits on
the reheating temperature,
\begin{align}
  \TR &\le 4.9\times 10^7 \ \GeV \left( \frac{\mgr}{10\ \GeV}
  \right)^{1/5},
    \label{eq:UpperLimitTR}
\end{align}
and on the gaugino mass parameter,
 \begin{align}
  \monetwo &\ge 0.9\, \TeV \left( \frac{\mgr}{ 10\ \GeV}
  \right)^{2/5},
  \label{eq:LowerLimitm12}
\end{align}
have been derived within the CMSSM~\cite{Pradler:2007is} from the
limit (\ref{Eq:tauslepton}), i.e., under the assumption that
$Y_{\slepton}^{\dec}$ is described approximately
by~(\ref{Eq:Yslepton}).%
\footnote{Similar limits have also been discussed in models where the
  ratio $\mgr/\monetwo$ is bounded from below~\cite{Kersten:2007ab}.}
While the $\TR$ bound can be restrictive for models of inflation and
of baryogenesis, the $\monetwo$ bound can have implications for SUSY
searches at the LHC. Depending on $\mgr$, (\ref{eq:LowerLimitm12})
implies sparticle masses which can be associated with a mass range
that will be difficult to probe at the LHC.  This is illustrated by
the vertical (violet) line in Fig.~\ref{Fig:CMSSMtB10} which indicates
the gluino mass $m_{\widetilde{g}} = 2.5\
\TeV$~\cite{Pradler:2007ar}.%
\footnote{Note that the mass of the lighter stop is
  $m_{\widetilde{t}_1} \simeq 0.7 m_{\widetilde{g}}$ in the considered
  $\stau$ NLSP region with $m_{\mathrm{h}}>114.4\ \GeV$.}

\begin{figure}
  \includegraphics[width=3.25in]{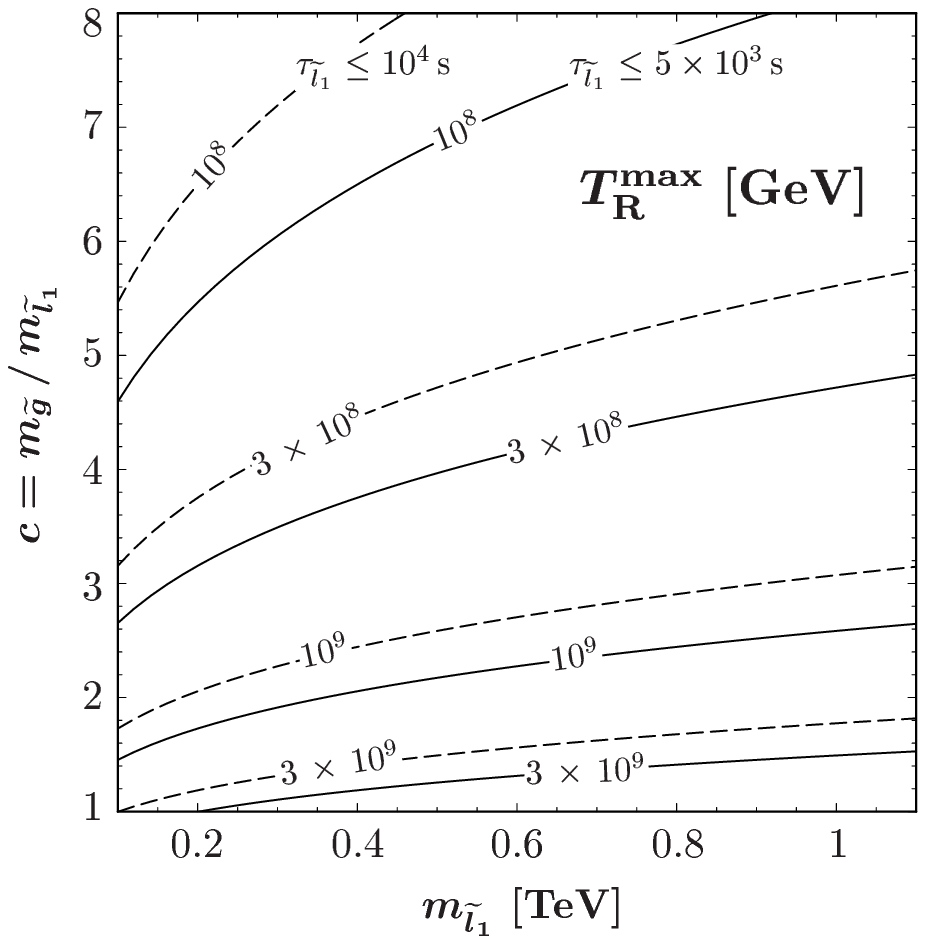}
  \caption{Upper limits on the mass ratio $c=m_{\gluino}/\mslepton$
    imposed by $\Omega_\gravitino^{\TP} h^2\leq\OmegaDM h^2\leq 0.126$
    and $\tau_\slepton\leq5\times 10^3\,\seconds$ ($10^4\,\seconds$)
    are shown as a function of $\mslepton$ by the solid (dashed) lines
    for values of $\TRmax$ ranging from $10^8\,\GeV$ up to $3\times
    10^9\,\GeV$. These values refer to the $\TR$ definition given
    in~(\ref{Eq:TR_definition}).  From~\cite{Steffen:2008bt}.}
\label{Fig:ProbingC}
\end{figure}

Here one should emphasize that the $\TR$ limit~(\ref{eq:UpperLimitTR})
relies on the CMSSM-specific minimal splitting~\cite{Pradler:2007is}
$m_\stau^2\leq 0.21 \monetwo^2$. In a less restrictive model, higher
$\TR$ values can be viable for a smaller splitting between the masses
of the gluino and the $\slepton$ NLSP~\cite{Steffen:2008bt}. This is
shown in Fig.~\ref{Fig:ProbingC} (from~\cite{Steffen:2008bt}),
%
%
where the maximum reheating temperature $\TRmax$ imposed by
$\Omega_{\gravitino}^{\TP}\leq\OmegaDM$ and $\tau_\slepton\leq 5\times
10^3\,\GeV$ $(10^4\,\GeV)$ is shown by the solid (dashed) lines in the
plane spanned by $m_\slepton$ and the ratio $c=m_\gluino/m_\slepton$
at the weak scale. Note that the shown $\TRmax$ values are given
by~\cite{Steffen:2008bt}
\begin{eqnarray}
  \TR 
  &\leq& 
  \frac{2.37\times 10^9~\GeV}{c^{2}} 
  \left(\frac{\Omega_{\CDM}h^2}{0.1}\right)
  \nonumber\\
  &&\times
  \left(\frac{\tau_{\slepton}}{10^4~\seconds}\right)^{\!\!\frac{1}{2}}\!\!
  \left(\frac{\mslepton}{100~\GeV}\right)^{\!\!\frac{1}{2}}\!\!
  \equiv
  \TR^{\max}
\label{Eq:TR_max}
\end{eqnarray}
which has been derived in a very conservative way taking into account
only the SUSY QCD contribution to $\Omega_{\gravitino}^{\TP}$.

The BBN constraints shown in Figs.~\ref{Fig:GravitinoMassBounds},
\ref{Fig:CMSSMtB10}, and~\ref{Fig:ProbingC} and~(\ref{Eq:tauslepton}),
(\ref{eq:UpperLimitTR}), and~(\ref{eq:LowerLimitm12}) rely on typical
$Y_{\slepton}^{\dec}$ values such as the ones given
by~(\ref{Eq:Yslepton}). However, $Y_{\stau}^{\dec}<4\times 10^{-15}$
can be found even within the CMSSM as is shown in
Fig.~\ref{Fig:ExceptionalStauYields}. Indeed, for a $\stau$ NLSP with
a sizable left-right mixing, exceptional values of
$Y_{\stau^-}^{\dec}\lesssim10^{-15}$ can
occur~\cite{Ratz:2008qh,Pradler:2008qc} for which even the restrictive
CBBN bounds associated with $\Lisix/\mathrm{H}|_{\mathrm{obs}} \leq
10^{-10}$ and $\ben/\mathrm{H}|_{\mathrm{obs}} \leq 2.1\times
10^{-13}$ (cf.\ Fig.~5 of~\cite{Pospelov:2008ta}) can be evaded. In
these exceptional cases, the $\tau_\slepton$
limit~(\ref{Eq:tauslepton}) does not exist and the $\TR$ limit is
governed by $\Omega_{\gravitino}^{\TP}\leq\OmegaDM$ only (cf.\
Fig.~\ref{Fig:UpperLimitTR}) since $\Omega_{\gravitino}^{\NTP}$ is
negligible.

\subsection{Experimental Prospects}
\label{sec:GravitinoExperiments}

Because of its extremely weak couplings, gravitino dark matter is
inaccessible to direct and indirect searches if R-parity is
conserved.%
\footnote{For broken R-parity, the $\gravitino$ LSP is unstable and
  its decay products can lead to signals in indirect
  searches~\cite{Buchmuller:2007ui,Bertone:2007aw,Ibarra:2007wg,Ibarra:2008qg,Ishiwata:2008cu,Covi:2008jy,Hisano:2008ti}.}
Also the direct production of gravitinos with $\mgravitino\gtrsim
0.1~\keV$ at colliders is extremely suppressed.  Instead, one expects
a large sample of (quasi-) stable NLSPs if the NLSP belongs to the
MSSM spectrum and if its mass is within the kinematical reach.

The $\neutralino$ and the $\sneutrino$ NLSP cases will be associated
with an excess of missing transverse energy and it might become a
major challenge to distinguish such cases from the $\neutralino$ LSP
case; cf.~\cite{Covi:2007xj,Ellis:2008as}.
Indeed, the phenomenological prospects will be much more promising in
scenarios with the $\slepton$ NLSP or the $\scalartop$ NLSP. These
particles can appear respectively as (quasi-) stable electrically
charged elementary particles or as stop hadrons in the collider
detectors.
Because of possibly non-trivial hadronization properties of the
$\scalartop$ NLSP, we focus on the simpler $\stau$ NLSP case in the
remainder of this section.

In the $\stau$ NLSP case, each heavier superpartner produced will
cascade down to the $\stau$ which will appear as a (quasi-) stable
particle in the detector. Such a heavy charged particle would
penetrate the collider detector in a way similar to
muons~\cite{Drees:1990yw,Nisati:1997gb,Feng:1997zr}.  If the produced
staus are slow, the associated highly ionizing tracks and
time--of--flight measurements will allow one to distinguish the
$\stau$ from a
muon~\cite{Drees:1990yw,Nisati:1997gb,Feng:1997zr,Ambrosanio:2000ik}.
With measurements of the $\stau$ velocity $\beta_{\stau} \equiv
v_{\stau}/c$ and the slepton momentum $p_{\stau}\equiv
|\vec{p}_{\stau}|$, $\mst$ can be determined:
$\mst=p_{\stau}(1-\beta_{\stau}^2)^{1/2}/\beta_{\stau}$~\cite{Ambrosanio:2000ik}.
For the upcoming LHC experiments, studies of hypothetical scenarios
with long-lived charged particles are actively pursued; see,
e.g.,~\cite{Ellis:2006vu,Ellis:2007mc,Bressler:2007gk,Zalewski:2007up}.
For example, it has been found that one should be able to measure the
mass $\mst$ of a (quasi-) stable $\stau$ quite
accurately~\cite{Ellis:2006vu,Ellis:2007mc}.%
\footnote{(Quasi-) stable $\stau$'s could also be pair-produced in
  interactions of cosmic neutrinos in the Earth matter and be detected
  in a neutrino telescope such as IceCube~\cite{Ahlers:2006pf}.}

If some of the staus decay already in the collider detectors, the
statistical method proposed in~\cite{Ambrosanio:2000ik} could allow
one to measure the $\stau$ lifetime. With~(\ref{Eq:StauLifetime}) and
the measured value of $m_{\stau}$, one will then be able to determine
also the gravitino mass $\mgravitino$ and thereby the scale of SUSY
breaking. As a test of our understanding of the early Universe, it
will also be interesting to confront the experimentally determined
$(\mgravitino,m_{\stau})$ point with the cosmological constraints in
Fig.~\ref{Fig:GravitinoMassBounds}. 

Ways to stop and collect charged long-lived particles for an analysis
of their decays have also been
proposed~\cite{Goity:1993ih,Hamaguchi:2004df,Feng:2004yi,DeRoeck:2005bw,Hamaguchi:2006vu,Cakir:2007xa}.
It was found that up to $\Order(10^3$--$10^4)$ and
$\Order(10^3$--$10^5)$ $\stau$'s could be trapped per year at the LHC
and the ILC, respectively, by placing 1--10~kt of massive additional
material around existing or planned collider
detectors~\cite{Hamaguchi:2004df,Feng:2004yi}. A measurement of
$\tau_{\stau}$ can then be used to determine $\mgravitino$ as already
described above.
If $\mgr$ can be determined independently from the kinematics of the
2-body decay $\stau\to\gravitino\tau$,
\begin{align}
      \mgr = \sqrt{{m_{\st}^2}+{m_\tau^2}-2{m_{\st} E_\tau}}  
      \, ,
\label{Eq:mgrKinematics}
\end{align}
the lifetime $\tau_{\st}$ can allow for a measurement of the Planck
scale~\cite{Buchmuller:2004rq,Martyn:2006as,Hamaguchi:2006vu,Martyn:2007mj}
\begin{align}
  \MPl^2 =  
  \frac{\tau_{\stau}}{48\pi} 
  \frac{\mst^5}{\mgr^2}
  \left(
  1 - \frac{\mgr^2}{\mst^2}
  \right)^4
  .
\label{Eq:Planck_Scale}
\end{align}
An agreement with~(\ref{Eq:MPLmacro}), which is inferred from Newton's
constant~\cite{Amsler:2008zz}
$G_{\rm N} = 6.709\times 10^{-39}\,\GeV^{-2}$,
would then provide a strong experimental hint for the existence of
supergravity in nature~\cite{Buchmuller:2004rq}. In fact, this
agreement would be a striking signature of the gravitino LSP.
Unfortunately, the required kinematical determination of $\mgravitino$
appears to be feasible only
for~\cite{Martyn:2006as,Hamaguchi:2006vu,Martyn:2007mj}
$\mgravitino/\mst \gtrsim 0.1$
which seems to be disfavored in most of the SUSY parameter space
according to our present understanding of the cosmological constraints
(see Fig.~\ref{Fig:GravitinoMassBounds}).%
\footnote{Note that the cosmological constraints described in
  Sect.~\ref{sec:GravitinoConstraints} assume a standard thermal
  history. In fact, entropy production after NLSP decoupling and
  before BBN can weaken the BBN constraints
  significantly~\cite{Buchmuller:2006tt,Pradler:2006hh}.}
Accordingly, alternative methods such as the ones proposed
in~\cite{Brandenburg:2005he,Steffen:2005cn} could become essential to
identify the gravitino as the LSP.
In the special regions with exceptionally small
$Y_{\stau^-}^{\dec}\lesssim
10^{-15}$~\cite{Ratz:2008qh,Pradler:2008qc}, however, the cosmological
constraints could be evaded even within a standard cosmological
history and $\mgr$ can still be sufficiently close to $m_\stau$ so
that the kinematical $m_\gravitino$ determination can still be viable.
Here, also the spin-3/2 character of the gravitino can become relevant
so that it could be probed in principle by analyzing the decays
$\stau\to\gravitino\tau\gamma$~\cite{Buchmuller:2004rq}.

With additional experimental insights into the masses of the gluino
and of the neutralinos (and thereby into $M_{1,2,3}$ and their
possibly universal value $\monetwo$ at $\mgut$), the determination of
$\mgravitino$ (or $\tau_\stau$) can allow one to probe also the upper
limit on the reheating temperature $\TRmax$ at
colliders~\cite{Moroi:1993mb,Bolz:1998ek,Bolz:2000fu,Fujii:2003nr,Steffen:2006hw,Pradler:2006qh,Takayama:2007du,Choi:2007rh,Steffen:2008bt};
cf.\ Figs.~\ref{Fig:UpperLimitTR}, \ref{Fig:CMSSMtB10},
and~\ref{Fig:ProbingC}.
This possibility results from the extremely weak gravitino couplings,
the associated $\TR$ dependence of $\Omega_{\gravitino}^{\TP}$, and
the limit $\Omega_{\gravitino}^{\TP}\leq\OmegaDM$.
Any $\TRmax$ value inferred from collider experiments will however
depend crucially on assumptions on the cosmological history and on the
evolution of physical parameters
as discussed already for $\TR$ bounds in $\neutralino$ LSP scenarios
in Sect.~\ref{sec:NeutralinoConstraints}.
It would still be very insightful to compare, e.g., the minimum
temperature required by thermal leptogenesis in its simplest setting,
$T\gtrsim 10^9\,\GeV$, with the $\TRmax$ value obtained under the
assumptions of a standard cosmological history and of couplings that
behave at high temperatures as described by the renormalization group
equations in the MSSM.

\section{Axino Dark Matter}
\label{sec:AxinoDM}

The axino
$\axino$~\cite{Nilles:1981py,Kim:1983dt,Tamvakis:1982mw,Kim:1983ia}
appears as the spin-1/2 superpartner of the axion once the MSSM is
extended with the PQ mechanism~\cite{Peccei:1977hh,Peccei:1977ur} in
order to solve the strong CP problem. Depending on the model and on
the SUSY breaking scheme, the axino mass $m_{\axino}$ can range
between the eV and the GeV
scale~\cite{Tamvakis:1982mw,Nieves:1985fq,Rajagopal:1990yx,Goto:1991gq,Chun:1992zk,Chun:1995hc}.
The axino is a singlet with respect to the gauge groups of the
Standard Model. It interacts extremely weakly since its couplings are
suppressed by the PQ scale $f_a\gtrsim 6\times
10^8\,\GeV$~\cite{Amsler:2008zz,Sikivie:2006ni,Raffelt:2006rj,Raffelt:2006cw}.
and thus can be classified as an EWIP.  The detailed form of the axino
interactions depends on the axion model under consideration; cf.\
Sect.~\ref{sec:AxionDM}.  We focus on hadronic (or KSVZ) axion
models~\cite{Kim:1979if,Shifman:1979if} in a SUSY setting, in which
the axino couples to the MSSM particles only indirectly through loops
of additional heavy KSVZ (s)quarks.  Considering $\axino$ LSP
scenarios in which the LOSP is the NLSP, $\mst<\mneu$ is again viable
as also the alternative $\neutralino$, $\sneutrino$, and $\scalartop$
NLSP cases.

Before proceeding, it should be stressed that the bo\-sonic partners
of the axino, the axion and the saxion, can have important
implications for cosmology: (i)~Those associated with the axion have
already been discussed in Sect.~\ref{sec:AxionDM} and most important
for this section is the possibly significant axion contribution to
$\OmegaDM$ which can tighten the constraints from
$\Omega_{\axino}^{\TP} < \Omega_{\CDM}$ discussed below. (ii)~Late
decays of the saxion can lead to significant entropy
production~\cite{Kim:1992eu,Lyth:1993zw,Chang:1996ih,Hashimoto:1998ua}
and can thereby affect the cosmological
constraints~\cite{Kawasaki:2007mk}. In this review, however, a
standard thermal history is assumed which implies that saxion effects
are negligible.

\subsection{Primordial Origin}
\label{sec:AxinoProduction}

Because of their extremely weak interactions, the temperature
$T_{\freezeout}$ at which axinos decouple from the thermal plasma in
the early Universe is very high. For example, an axino decoupling
temperature of $T_{\freezeout} \approx 10^9\,\GeV$ is obtained for
$f_a=10^{11}\,\GeV$~\cite{Rajagopal:1990yx,Brandenburg:2004du}.  For
$\TR>T_{\freezeout}$, axinos were in thermal equilibrium before
decoupling as a relativistic species so
that~\cite{Rajagopal:1990yx,Asaka:2000ew,Covi:2001nw,Brandenburg:2004du}
\begin{align}
  \Omega_{\ax}^{\thermal}h^2
  \approx 0.1 \left(\frac{m_{\ax}}{0.2\,\keV}\right)
  \ ,
  \label{Eq:AxinoDensityEq}
\end{align}
where $g_{*S}(T_{\freezeout})=228.75$ is assumed.

For $\TR<T_{\freezeout}$, axinos have never been in thermal
equilibrium with the primordial plasma but can be generated
efficiently in scattering processes of particles that are in thermal
equilibrium with in the hot MSSM
plasma~\cite{Asaka:2000ew,Covi:2001nw,Brandenburg:2004du,Gomez:2008js}.
Within SUSY QCD, the associated thermally produced (TP) axino density
can be calculated in a consistent gauge-invariant treatment that
requires weak couplings ($g_\mathrm{s}\ll
1$)~\cite{Brandenburg:2004du}:
\bea
        \Omega_{\axino}^{\TP}h^2
        &\simeq&
        5.5\,g_\mathrm{s}^6(\TR) \ln\left(\frac{1.108}{g_\mathrm{s}(\TR)}\right) 
        \left(\frac{10^{11}\,\GeV}{f_a/N}\right)^{\! 2}\!\!
\nonumber\\
        &&
        \times
        \bigg(\frac{m_{\ax}}{0.1~\GeV}\bigg)\!
        \left(\frac{T_R}{10^4\,\GeV}\right)
\label{Eq:AxinoDensityTP}
\eea
with the axion-model-dependent color anomaly $N$ of the PQ symmetry.
The thermally produced axinos do not affect the thermal evolution of
the LOSP (or NLSP) which decays after its decoupling into the $\axino$
LSP. Taking into account the non-thermally produced (NTP) density from
NLSP decays~\cite{Covi:1999ty,Covi:2001nw}
\bea
        \Omega_{\axino}^{\NTP} h^2
        &=& 
        m_{\axino}\, Y_{\NLSP}^{\dec}\, s(T_0) h^2 / \rho_{\mathrm{c}}
        \ ,
\label{Eq:AxinoDensityNTP}
\eea
the guaranteed axino density is%
\footnote{Axino production in inflaton decays is not considered.}
\begin{align}
\Omega_{\axino}=\Omega_{\axino}^{\,\,\thermal/\TP}+\Omega_{\axino}^{\NTP}
\ .
\end{align}

\begin{figure}[t]
\includegraphics*[width=.45\textwidth]{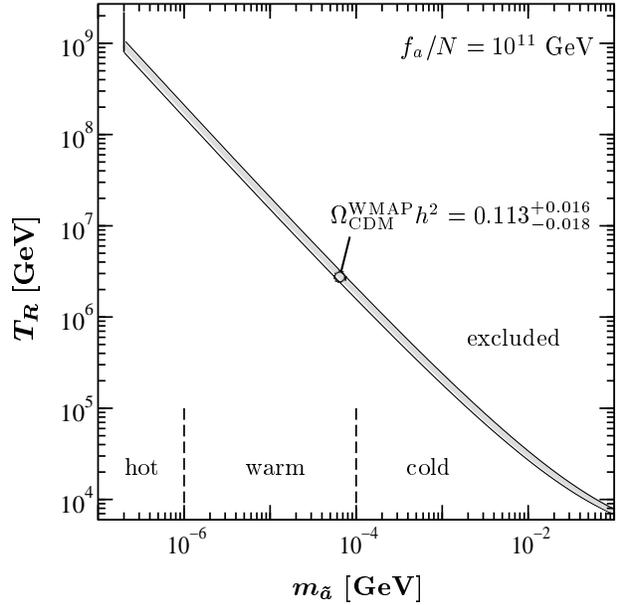}
\caption{Upper limits on the reheating temperature $T_{\Reheating}$ in
  the $\axino$ LSP case for $f_a/N = 10^{11}\,\GeV$. On (above) the
  gray band, $\Omega^{\TP}_{\ax}h^2\in 0.113^{+0.016}_{-0.018}$
  ($\Omega^{\TP}_{\ax}h^2> 0.129$).  Thermally produced axinos can be
  classified as hot, warm, and cold dark matter~\cite{Covi:2001nw} as
  indicated.  From~\cite{Brandenburg:2004du}.}
\label{Fig:axino_dark_matter_limits}
\end{figure}

In Fig.~\ref{Fig:axino_dark_matter_limits}
(from~\cite{Brandenburg:2004du}) the $(m_{\ax},\TR)$ region with
$0.097\leq\Omega_{\axino}^{\TP}\leq 0.129$ for $f_a/N = 10^{11}\,\GeV$
is shown by the gray band.  Note that~(\ref{Eq:AxinoDensityTP}) shows
a different dependence on the LSP mass than the corresponding
expression in the~$\gravitino$ LSP case~(\ref{Eq:GravitinoDensityTP}).
Accordingly, one finds the different $m_{\LSP}$ dependence of the
$T_{\Reheating}$ limits inferred from
$\Omega_{\axino/\gravitino}^{\TP} < \Omega_{\CDM}$.
%
%
Since thermally produced axinos are generated in kinetic equilibrium
with the primordial plasma, they have a thermal spectrum which allows
for the $m_{\ax}$-dependent classification into cold, warm, and hot
dark matter~\cite{Covi:2001nw} shown in
Fig.~\ref{Fig:axino_dark_matter_limits}.
As can be seen, the $\TR$ limit does not exist for $m_{\ax}\lesssim
0.2~\keV$ because of the equality of $\axino$ production and $\axino$
disappearance rates for $T > T_{\freezeout}\approx 10^9\,\GeV$.  With
a thermal relic density~(\ref{Eq:AxinoDensityEq}) in this regime,
there will be a limit on $m_{\ax}$ depending on the constraints
inferred from studies of warm/hot dark matter~\cite{Hannestad:2007dd}.

The non-thermally produced axino density $\Omega_{\axino}^{\NTP}$
differs from the corresponding expression in the $\gravitino$ LSP case
(\ref{Eq:GravitinoDensityNTP}) only by the obvious difference in
$m_{\axino/\gravitino}$. In particular, for given
$\Omega_{\axino}^{\TP}$, the bound
$\Omega_{\axino}^{\NTP}\leq\Omega_{\CDM}-\Omega_{\axino}^{\,\equil/\TP}$ 
as obtained with~(\ref{Eq:Yslepton}) implies limits on $m_{\ax}$ and
$\mst$ which can be read off directly from
Fig.~\ref{Fig:GravitinoMassBounds} after the replacement
$\mgr\rightarrow m_{\ax}$.  Note, however, that the $\taustau$
contours and the cosmological constraints are different in the axino
LSP case. For the $\stau$ NLSP, the following lifetime was
estimated~\cite{Brandenburg:2005he}
\bea
        \taustau 
        && \!\!\simeq 
        \Gamma^{-1}(\stau\to\tau\,\axino)
        \simeq
        25~\mathrm{s}\,\,  
        \xi^{-2}\,
        \left(1-\frac{m_{\ax}^2}{m_{\st}^2}\right)^{\!\!\!\!-1}
\nonumber\\
        &&
        \times
        \left(\frac{100\,\GeV}{m_{\st}}\right)\!\!
        \left(\frac{f_a/C_{\rm aYY}}{10^{11}\,\GeV}\right)^{\!\!2}\!\!
        \left(\frac{100\,\GeV}{m_{\tilde{B}}}\right)^{\!\!2}\!
\label{Eq:Axino2Body}
\eea
with the KSVZ-model dependence expressed by $C_{\rm aYY}\simeq
\mathcal{O}(1)$ and the uncertainty of the estimate absorbed into
$\xi\simeq\Order(1)$. One thus finds a $\stau$ lifetime in the
$\axino$ LSP case that cannot be as large as the one in the
$\gravitino$ LSP case~(\ref{Eq:StauLifetime}). Accordingly, the BBN
constraints are much weaker for the $\axino$ LSP. For discussions of
$\axino$ LSP scenarios with the $\neutralino$ NLSP,
see~\cite{Covi:1999ty,Covi:2001nw,Covi:2004rb}. For both the $\stau$
NLSP and the $\neutralino$ NLSP, it has been shown that non-thermally
produced axinos with $m_{\ax}\lesssim 10~\GeV$ would be warm/hot dark
matter~\cite{Jedamzik:2005sx}.

\subsection{Experimental Prospects}
\label{sec:AxinoExperiments}

Being an EWIP, the axino LSP is inaccessible in direct and indirect
dark matter searches if R-parity is conserved. Also direct $\axino$
production at colliders is strongly suppressed.  Nevertheless,
(quasi-) stable $\stau$'s could appear in collider detectors (and
neutrino telescopes~\cite{Ahlers:2006pf}) as a possible signature of
the $\axino$ LSP. However, since the $\MPl$ measurement at
colliders~\cite{Buchmuller:2004rq}, which would have been a decisive
test of the $\gravitino$ LSP, seems cosmologically disfavored in most
of the parameter space of R-parity conserving SUSY models, it may very
well be a challenge to distinguish between the $\axino$ LSP and the
$\gravitino$ LSP.

For $m_{\st}=100\,\GeV$ and $m_{\Bino}=110\,\GeV$, for example, the
$\stau$ lifetime in the $\axino$ LSP scenario~(\ref{Eq:Axino2Body})
can range from $\Order(10^{-4}\,\seconds)$ for $f_a=5\times
10^8\,\GeV$ to $\Order(10~{\mbox{h}})$ for $f_a=5\times
10^{12}\,\GeV$. In the $\gravitino$ LSP case, the corresponding
lifetime~(\ref{Eq:StauLifetime}) can vary over an even wider range,
e.g., from $6\times 10^{-8}\,{\rm s}$ for $\mgravitino = 1~\keV$ to
15~years for $\mgravitino = 50~\GeV$. Thus, both a very short
lifetime, $\taustau \lesssim$~ms, and a very long one, $\taustau
\gtrsim$~days, will point to the $\gravitino$ LSP. On the other hand,
if the LSP mass cannot be measured kinematically and if
$\taustau=\Order(0.01~{\mbox{s}})$--$\Order(10~{\mbox{h}})$, the stau
lifetime alone will not allow us to distinguish between the $\axino$
LSP and the $\gravitino$ LSP.

\begin{figure}[t]
\includegraphics*[width=0.45\textwidth]{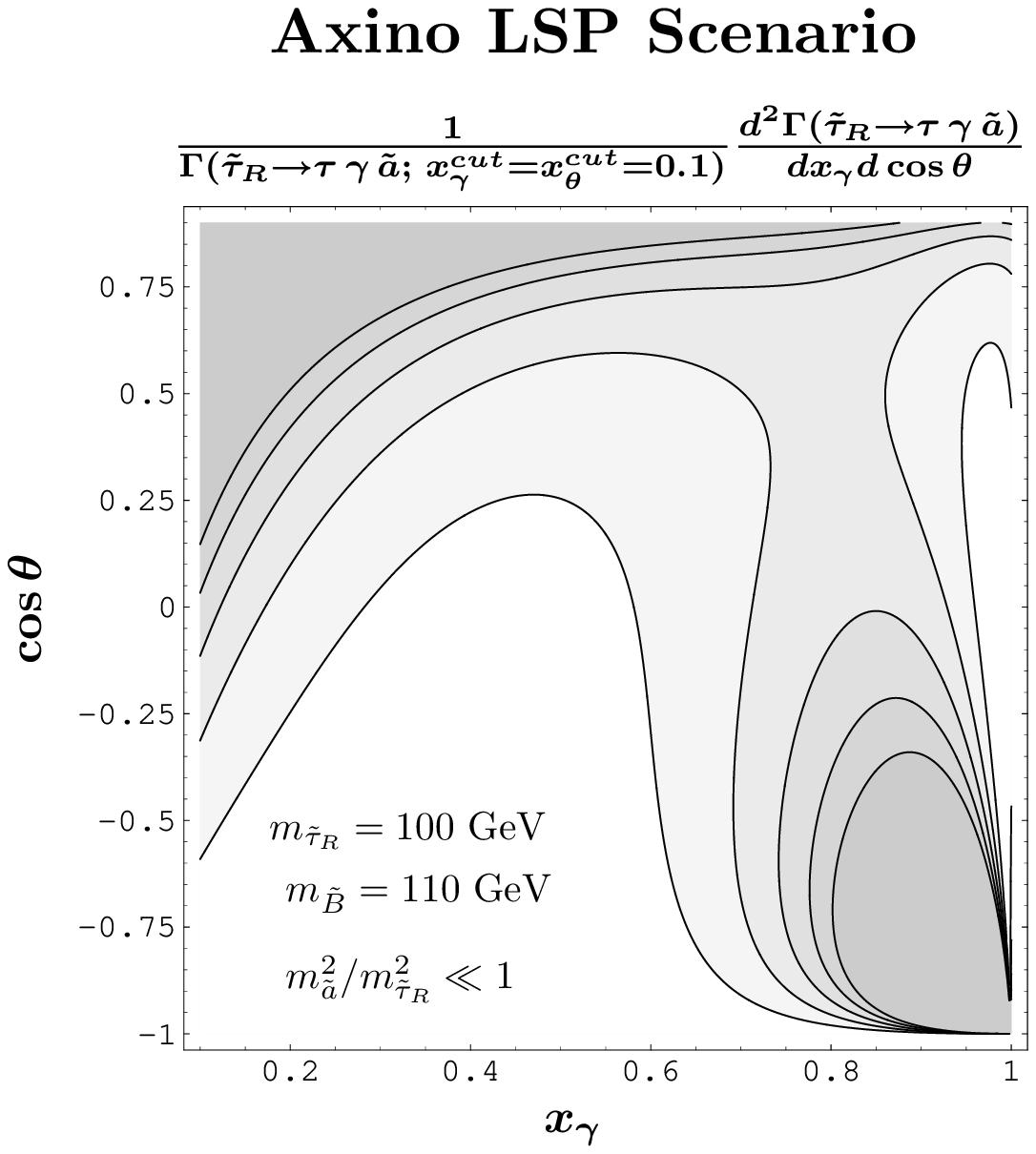}\\ \\
\includegraphics*[width=0.45\textwidth]{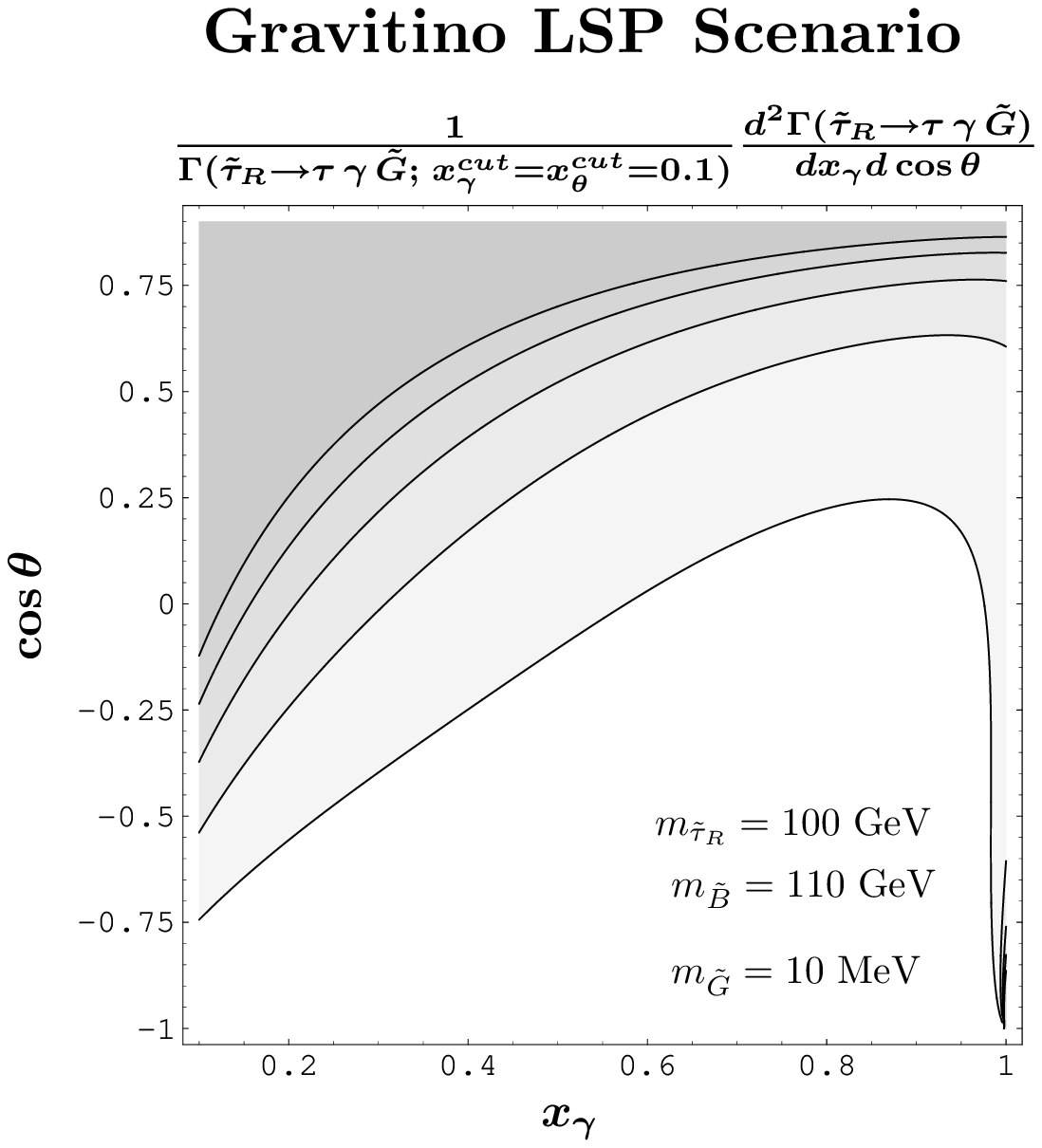}
\caption{ The normalized differential
  distributions~(\ref{Eq:Fingerprint}) of the visible decay products
  in the decays $\stau\to\tau+\gamma+\axino/\gravitino$ for the cases
  of the $\axino$ LSP (upper panel) and the $\gravitino$ LSP (lower
  panel) for $m_\stau = 100\,\GeV$, $\neutralino\simeq\Bino$, $m_\Bino
  = 110\,\GeV$, $m_{\axino}^2/m_{\stau}^2 \ll 1$, and $\mgravitino =
  10\,\MeV$.  The cut parameters are set to $x_\gamma^{\mathrm{cut}} =
  x_\theta^{\mathrm{cut}}=0.1$. The contour lines represent the values
  0.2, 0.4, 0.6, 0.8, and 1.0, where the darker shading implies a
  higher number of events. From~\cite{Brandenburg:2005he}.}
\label{Fig:Fingerprint}
\end{figure}

The situation is considerably improved when one considers the 3-body
decays $\stau\to \tau \gamma \axino/\gravitino$. From the
corresponding differential rates~\cite{Brandenburg:2005he}, one
obtains the differential distributions of the visible decay products.
These are illustrated in Fig.~\ref{Fig:Fingerprint}
(from~\cite{Brandenburg:2005he}) in terms of
\begin{align}
        {1 
        \over 
        \Gamma(\stau\to\tau\,\gamma\, i\,;
        x_{\gamma}^{\mathrm{cut}},x_{\theta}^{\mathrm{cut}})}
        \,\,
        {d^2\Gamma(\stau\to\tau\,\gamma\, i)
        \over
        d x_{\gamma}d \cos\theta}
\label{Eq:Fingerprint}
\end{align}
where $x_\gamma\equiv 2 E_\gamma/m_{\st}$ is the scaled photon energy,
$\theta$ the opening angle between the directions of $\gamma$ and
$\tau$,
\bea
        \Gamma(\stau\to\tau\,\gamma\,i\,;
        x_\gamma^{\mathrm{cut}},x_\theta^{\mathrm{cut}})
        &\equiv&
        \int^{1-A_{i}}_{x_\gamma^{\mathrm{cut}}}
        d x_\gamma
        \int^{1-x_\theta^{\mathrm{cut}}}_{-1}
        d \cos\theta \,\,
\nonumber\\
        &&\times
        \frac{d^2\Gamma(\stau\to\tau\,\gamma\,i)}{dx_\gamma d\cos\theta}
\eea
the respective integrated 3-body decay rate with the cuts $x_\gamma >
x_\gamma^{\mathrm{cut}}$ and $\cos\theta < 1-x_\theta^{\mathrm{cut}}$,
and $A_{i} \equiv m_{i}^2/m_{\st}^2$.
Note that~(\ref{Eq:Fingerprint}) is independent of the 2-body decay,
of the total NLSP decay rate, and of the PQ/Planck scale.
%
%
The figure shows~(\ref{Eq:Fingerprint}) for the axino LSP ($i =
\axino$) with $m_{\ax}^2/m_{\stau}^2 \ll 1$ (upper panel) and the
gravitino LSP ($i=\gravitino$) with $\mgravitino = 10~\MeV$ (lower
panel), where $m_{\st} = 100~\GeV$, $m_{\Bino} = 110~\GeV$, and
$x_\gamma^{\mathrm{cut}} = x_\theta^{\mathrm{cut}}=0.1$.
In the $\gravitino$ LSP case, the events are peaked only in the region
where photons are soft and emitted with a small opening angle with
respect to the tau ($\theta\simeq 0$).  In contrast, in the $\axino$
LSP case, the events are also peaked in the region where the photon
energy is large and the photon and the tau are emitted back-to-back
($\theta \simeq \pi$).
Thus, if the observed number of events peaks in both regions, this can
be evidence against the gravitino LSP and a hint towards the axino
LSP~\cite{Brandenburg:2005he}.%
\footnote{There is a caveat: If $\mgr < m_{\ax} < m_{\st}$ and
  $\Gamma(\stau \to \axino\,X) \gg \Gamma(\stau \to \gravitino\,X)$,
  one would still find the distribution shown in the upper panel of
  Fig.~\ref{Fig:Fingerprint}. The axino would then eventually decay
  into the gravitino LSP and the axion.}

To be specific, with $10^4$ analyzed stau NLSP decays, one expects
about 165$\pm$13 (stat.) events for the $\axino$ LSP and about
100$\pm$10 (stat.) events for the $\gravitino$
LSP~\cite{Brandenburg:2005he}, which will be distributed over the
respective ($x_\gamma$,\,$\cos\theta$)-planes shown in
Fig.~\ref{Fig:Fingerprint}. In particular, in the region of
$x_{\gamma}\gtrsim 0.8$ and $\cos\theta \lesssim -0.3$, we expect
about 28\% of the 165$\pm$13 (stat.) events in the $\axino$ LSP case
and about 1\% of the 100$\pm$10 (stat.) events in the $\gravitino$ LSP
case.  These numbers illustrate that $\Order(10^4)$ of analyzed stau
NLSP decays could be sufficient for the distinction based on the
differential distributions. To establish the feasibility of this
distinction, dedicated studies including details of the detectors and
the additional massive material will be
crucial~\cite{Hamaguchi:2006vu}.

\subsection{Probing the Peccei-Quinn Scale \boldmath$f_a$ and \boldmath$m_{\axino}$}
\label{sec:AxinoPQScale}

If $\axino$ is the LSP and $\stau$ the NLSP, 
the analysis of the 2-body decay $\stau\to\tau\axino$ will allow us to
probe the PQ scale $f_a$ and the axino mass $m_{\axino}$.
In fact, the measurement of $\taustau$~(\ref{Eq:Axino2Body}) with
methods described in Sect.~\ref{sec:GravitinoExperiments} leads to the
following estimate of the Peccei-Quinn scale
$f_a$~\cite{Brandenburg:2005he}:
\bea
  f_a^2 
  &\simeq&
  {\xi^2\,C_{\rm aYY}^2} 
  \left(10^{11}\,\GeV\right)^2
  \Big(1-\frac{{m_{\axino}^2}}{{m_{\stau}^2}}\Big)
  \left(\frac{\tau_{\stau}}{25~\mathrm{s}}\right)\, 
\nonumber\\
  && \times
  \left(\frac{m_{\stau}}{100\,\GeV}\right)
  \left(\frac{m_{\Bino}}{100\,\GeV}\right)^2
  \ ,
\label{Eq:PQ_Scale}
\eea
which can be confronted with $f_a$ limits from axion studies; cf.\
Sect.~\ref{sec:AxionDM}.
Indeed, we expect that $m_{\st}$ and $m_{\Bino}$ will already be known
from other processes when the $\stau$ NLSP decays are analyzed;
cf.~Sect.~\ref{sec:GravitinoExperiments}. The dependence on $m_{\ax}$
is negligible for $m_{\ax}/m_{\st}\lesssim 0.1$.  For larger values of
$m_{\ax}$, the $\stau$ NLSP decays can be used to determine $m_{\ax}$
from the kinematics of the 2-body decay, i.e., from a measurement of
the energy of the emitted tau $E_\tau$,
\begin{align}
  m_{\axino} =
  \sqrt{{m_{\st}^2}+{m_\tau^2}-2{m_{\st} E_\tau}} 
  \ ,
\label{Eq:Axino_Mass} 
\end{align}
with an error governed by the experimental uncertainties on $m_{\st}$
and $E_\tau$. As is evident from~(\ref{Eq:AxinoDensityTP})
and~(\ref{Eq:AxinoDensityNTP}), the determination of both the
PQ scale $f_a$ and the axino mass $m_{\ax}$ is crucial for
insights into the cosmological relevance of the axino LSP.

In principle, the determination of both $f_a$ and $m_{\ax}$ at
colliders would allow one to probe an upper limit on the reheating
temperature $\TRmax$ at colliders for $m_{\ax}\gtrsim 0.2~\keV$ as can
be seen in Fig.~\ref{Fig:axino_dark_matter_limits}; see
also~\cite{Choi:2007rh}.
Indeed, Ref.~\cite{Choi:2007rh} has outlined theoretical ways to probe
$\TR$ values as low as $100~\GeV$ with $\Omega_{\ax}^{\NTP}$ taken
into account but without addressing the difficulties to determine an
axino mass $m_{\ax}\lesssim 10~\GeV$ experimentally.
In analogy to the $\gravitino$ LSP case, the theoretical possibility
results from the extremely weak axino couplings, the associated $\TR$
dependence of $\Omega_{\ax}^{\TP}$, and the limit
$\Omega_{\ax}^{\TP}\leq\OmegaDM$.
However, the high $\TR$ values at which the gauge-invariant result for
$\Omega_{\ax}^{\TP}$ is reliable, $\TR\gtrsim 10^6\,\GeV$, are
associated with $m_{\ax}\ll 1~\GeV$ while the kinematical
determination~(\ref{Eq:Axino_Mass}) appears to be feasible only for
$m_{\ax}\gtrsim 0.1\mst\gtrsim 10~\GeV$,
as in the $\gravitino$ LSP
case~\cite{Martyn:2006as,Hamaguchi:2006vu,Martyn:2007mj}.
Since also $\taustau$ is practically $m_{\ax}$-independent for
$m_{\ax}\lesssim 10~\GeV \lesssim 0.1\mst$, it seems unfortunately to
be impossible to probe those $\TR$ values in the axino LSP case at
colliders.

\section{Conclusion}
\label{sec:Conclusion}
%
\begin{table*}
  \caption{Dark matter candidates, their identity, and key properties. With the listed production mechanisms, $\Omega_{X}=\Omega_{\CDM}$ is possible for each candidate $X$. The respective production mechanism leads typically to a cold, warm, or hot dark matter component as indicated. Quantities marked with `(?)' seem to be unaccessible in large parts of the parameter space in light of the current understanding of experimental feasibility and/or of cosmological constraints within a standard thermal history.}
\label{tab:DMCandidates}
\begin{tabular}{lllllll}
\hline\noalign{\smallskip}
candidate & 
identity &
mass &
interactions &
production &
constraints &
experiments\\
\noalign{\smallskip}\hline\noalign{\smallskip}
$a$ &
axion &
$<0.01~\eV$ &
$(p/f_a)^{n}$ &
misalign.\ mech. &
$\leftarrow$ cold &
direct searches with
\vspace{0.1cm}
\\
&
(spin 0)&
&
extremely weak &
&
& 
microwave cavities
\vspace{0.1cm}
\\
&
N.-Goldst.\ boson &
&
{\tiny $f_a\gtrsim 6\times 10^{8}\,\GeV$} &
&
& 
$\hookrightarrow$ $m_a$, $f_a$, $\gagg$
\\
&
PQ symm.\ break. &
&
&
&
CMB & 
\\
\noalign{\smallskip}\hline\noalign{\smallskip}
$\widetilde{\chi}^0_1$ LSP & 
lightest neutralino &
$\Order(100\,\GeV)$ &
g, g', $y_i$ &
therm.~relic &
$\leftarrow$ cold &
indirect searches
\vspace{0.1cm}
\\
&
(spin 1/2)&
& 
weak &
$\gravitino$ decay &
$\leftarrow$ warm/hot &
direct searches
\vspace{0.1cm}
\\
&
mixture of&
&
{\tiny $M_{\mathrm{W}}\sim 100~\GeV$} &
&
&
collider searches
\vspace{0.1cm}
\\
&
$\Bino$, $\Wino$, $\HiggsinoUp$, $\HiggsinoDown$&
&
&
&
BBN &
$\hookrightarrow$ $m_{\neutralino}$, $\neutralino$ coupl.
\\
\noalign{\smallskip}\hline\noalign{\smallskip}
$\widetilde{G}$ LSP &
gravitino &
eV--TeV &
$(p/\MPl)^{n}$ &
therm.~prod. &
$\leftarrow$ cold &
$\stau$ prod. at colliders
\vspace{0.1cm}
\\
&
(spin 3/2)&
&
extremely weak &
NLSP decay &
$\leftarrow$ warm &
+ $\stau$ collection 
\vspace{0.1cm}
\\
&
superpartner&
&
{\tiny $\MPl = 2.4\!\times\! 10^{18}\,\GeV$} &
&
&
+ $\stau$ decay analysis
\vspace{0.1cm}
\\
&
of the graviton&
&
&
&
BBN&
$\hookrightarrow$ $\mgravitino$, $\MPl$ (?), $\TR$
\\
\noalign{\smallskip}\hline\noalign{\smallskip}
$\widetilde{a}$ LSP &
axino &
eV--GeV &
$(p/f_a)^{n}$ &
therm.~prod.  &
$\leftarrow$ cold/warm &
$\stau$ prod. at colliders
\vspace{0.1cm}
\\
&
(spin 1/2)&
&
extremely weak &
NLSP decay &
$\leftarrow$ warm/hot &
+ $\stau$ collection
\vspace{0.1cm}
\\
&
superpartner&
&
{\tiny $f_a\gtrsim 6\times 10^{8}\,\GeV$} &
&
&
+ $\stau$ decay analysis 
\vspace{0.1cm}
\\
&
of the axion&
&
&
&
BBN&
$\hookrightarrow$ $m_{\ax}$ (?), $f_a$, $\TR$ (?)
\\
\noalign{\smallskip}\hline
\end{tabular}
\end{table*}

The existence of dark matter provides strong evidence for physics
beyond the Standard Model.  Extending the Standard Model with PQ
symmetry and/or SUSY, the axion and/or an electrically neutral and
color neutral LSP appear as promising dark matter candidates.
The axion is well motivated by the PQ solution to the strong CP
problem.
With and without SUSY being realized in nature, the axion can exist
and can contribute significantly to $\OmegaDM$.
In SUSY extensions of the Standard Model, the lightest neutralino
$\neutralino$, the gravitino $\gravitino$, or the axino $\axino$ can
be the LSP and as such explain the non-baryonic dark matter in our
Universe.  The neutralino $\neutralino$ is already part of the MSSM
which provides a solution of the hierarchy problem and allows for
gauge coupling unification. Being the superpartner of the graviton and
the gauge field associated with supergravity, the gravitino
$\gravitino$ is equally well motivated with a mass $\mgr$ that
reflects the SUSY breaking scale. As the superpartner of the axion,
also the axino $\axino$ appears naturally once the strong CP problem
is solved with the PQ mechanism in a SUSY setting.

While mass values and interactions can be very different for the $a$,
the $\neutralino$ LSP, the $\gravitino$ LSP, and the $\axino$ LSP, I
have illustrated for each of these dark matter candidates that there
are natural regions in the associated parameter space in which
$\Omega_{a/\LSP}=\Omega_{\CDM}$.
In the axion case, the $\Omega_{a}=\Omega_{\CDM}$ region can be
subject to constraints from limits on axionic isocurvature and
non-Gaussian perturbations inferred from the CMB anisotropies.
In the SUSY case, regions with $\Omega_{\LSP}=\Omega_{\CDM}$ are
limited most importantly by bounds from electroweak precision
observables, B-physics observables, Higgs and sparticle searches at
LEP, and by BBN constraints. The constraints from $\Omega_{\CDM}$ and
BBN also imply restrictive upper limits on the reheating temperature
after inflation $\TR$ which can be relevant for models of inflation
and of baryogenesis.

Most promising are the experimental prospects in the case of the
$\neutralino$ LSP.
Being a WIMP, the $\neutralino$ LSP should be accessible in direct and
indirect dark matter searches.  Indeed, first hints might have already
been found in the EGRET data~\cite{deBoer:2005bd}.  With ongoing
indirect searches, the increasing sensitivity of direct searches, and
the advent of the LHC at which $\neutralino$ dark matter could be
produced, we will be able to test whether these hints are indeed the
first evidence for the existence of SUSY dark matter. While an excess
in missing transverse energy could provide a first hint for SUSY at
the LHC already within the next three years, the identification of the
$\neutralino$ being the LSP will require the reconstruction of the
SUSY model realized in nature. If superparticles are within the
kinematical reach, precision studies at the ILC will be crucial for
this endeavor.

Also an axion discovery is conceivable to occur in the near future.
Indeed, a current upgrade of the microwave cavity experiment ADMX aims
at a significant experimental exploration of a ``natural'' part of the
$\Omega_{a}=\Omega_{\CDM}$ region within the upcoming years.
Complementary to that, axion searches with helioscopes such as CAST
and those that are performed purely in the laboratory will extend
their search ranges. However, a signal in those searches would support
the existence of axions as hot/warm dark matter which can only provide
a minor fraction of $\OmegaDM$. This would leave room for a
significant $\Omega_{\LSP}$ contribution to $\Omega_{\CDM}$.

In $\gravitino/\axino$ LSP scenarios with conserved R-parity, no dark
matter signal should appear in direct or indirect searches.  However,
since an electrically charged lightest Standard Model superpartner
such as the $\stau$ is viable in $\gravitino/\axino$ LSP scenarios,
(quasi-) stable $\stau$'s might occur as muon-like particles instead
of an excess in missing transverse energy. Indeed, an excess of
(quasi-) stable $\stau$'s could appear as an alternative first hint
for SUSY at the LHC in the next three years. Because of the severe
limits on the abundance of stable charged
particles~\cite{Amsler:2008zz}, one would then expect that the $\stau$
is the NLSP that decays eventually into the $\gravitino/\axino$ LSP or
that R-parity is broken. A distinction between those scenarios will
require the analysis of $\stau$ decays.  For this challenge, the ILC
with its tunable beam energy seems
crucial~\cite{Hamaguchi:2004df,Feng:2004yi,Martyn:2006as,Hamaguchi:2006vu,Cakir:2007xa,Martyn:2007mj}.

Table~\ref{tab:DMCandidates} presents an overview of the dark matter
candidates discussed in this review. The axion and each LSP
candidate---the lightest neutralino $\neutralino$, the gravitino
$\gravitino$, or the axino $\axino$---could provide $\Omega_{\CDM}$
and could be produced and identified experimentally in the near
future.

\bigskip

I am grateful to A.~Brandenburg, L.~Covi, A.~Frei\-tas, C.~Frugiuele,
P.~Graf, K.~Hamaguchi, J.~Hamann, M.~Kuster, J.~Lesgour\-gues,
A.~Mirizzi, G.~Panotopoulos, T.~Plehn, M.~Pospelov, J.~Prad\-ler,
G.G.\ Raffelt, A.~Ringwald, L.~Rosz\-kowski, S.~Schilling,
N.~Tajuddin, F.~Takahashi, Y.Y.Y. Wong, D.~Wyler, and M.~Zagermann for
valuable discussions and/or collaborations on the topics covered in
this review.
This research was supported by the DFG cluster of excellence `Origin
and Structure of the Universe' (www.universe-cluster.de).

%
\bibliographystyle{epj}
%
%

\end{document}